\documentclass[11pt]{article}

\usepackage{amssymb,amsmath}
\usepackage{graphicx}
\usepackage{graphics}
\usepackage{subfigure} 
\usepackage{paralist}
\usepackage{pifont}    
\usepackage[latin1]{inputenc}    
\usepackage{eepic,epic}
\usepackage{psfrag}

\newcommand{\OMIT}[1]{} 
\newcommand{\bfootnote}[1]{} 

\newtheorem{theorem}{Theorem}[section]
\newtheorem{corollary}[theorem]{Corollary}

\newtheorem{lemma}[theorem]{Lemma}

\newtheorem{definition}[theorem]{Definition}

\newcommand{\p}{\mbox{\rm P}}

\newcommand{\np}{\mbox{\rm NP}}

\newcommand{\DP}{\mbox{\rm DP}}
\newcommand{\sigmastar}{\mbox{$\Sigma^\ast$}}
\newcommand{\naturalnumber}{\ensuremath{{\mathbb{N}}}}
\def\integers{\mathbb{Z}}
\def\nats{\naturalnumber}

\newcommand{\condition}{\,|\:}
\newcommand{\seq}{\subseteq}

\newcommand{\sharpp}{{\rm \#P}}

\newcommand{\shape}{\mathit{shape}}

\newcommand{\solutionktrp}[2]{\mbox{\sc Sol}_{\scriptsize \ktrp{#1}}({#2})}

\newenvironment{desctight}
  {\begin{list}{}{\setlength\labelwidth{0pt}%
        \setlength{\itemsep}{0pt}%
        \setlength{\parsep}{0pt}%
        \setlength\itemindent{-\leftmargin}%
        }}
    {\end{list}}

\newcommand{\probbf}{\rm}
\newcommand{\sat}{{\probbf SAT}}
\newcommand{\usat}{{\probbf Unique\hbox{-}\allowbreak{SAT}}}

\newcommand{\sharpsat}{{\probbf \#{SAT}}}

\newcommand{\circuitandnotsat}{{\probbf Circuit_{\wedge,\neg}\hbox{-}\allowbreak{SAT}}}

\newcommand{\threetrp}{{\probbf \mbox{\rm{}$3$}\hbox{-}TRP}}
\newcommand{\ktrp}[1]{{\probbf \mbox{\rm{}$#1$}\hbox{-}TRP}}

\newcommand{\uktrp}[1]{{\probbf Unique\hbox{-}\allowbreak\mbox{\rm{}$#1$}\hbox{-}TRP}}

\newcommand{\sharpktrp}[1]{{\probbf \#\mbox{\rm{}$#1$}\hbox{-}TRP}}
\newcommand{\infktrp}[1]{{\probbf Inf\hbox{-}\mbox{\rm{}$#1$}\hbox{-}TRP}}
\newcommand{\asktrp}[1]{{\probbf AS\hbox{-}\mbox{\rm{}$#1$}\hbox{-}TRP}}

\newcommand{\x}{{$\bullet$}}

\newcommand{\littlep}{{p}}

\newcommand{\randomized}{\ensuremath{\leq_{\mathit{ran}}^{{\littlep}}}}
\newcommand{\manyone}{\ensuremath{\leq_{m}^{{\littlep}}}}

\newcommand{\aspred}{\ensuremath{\leq_{asp}^{{\littlep}}}}

\newcommand\qedblob{\ding{113}}
\def\literalqed{{\ \nolinebreak\hfill\mbox{\qedblob\quad}}}

\newenvironment{proofs}{\noindent{\bf Proof.}\hspace*{1em}}{\literalqed\bigskip}

\begin{document}
\title{The Three-Color and Two-Color Tantrix\texttrademark\ Rotation
Puzzle Problems are NP-Complete via Parsimonious
Reductions\thanks{Supported in part by DFG grants RO~\mbox{1202/9-3} 
and RO~\mbox{1202/11-1}, 
the European Science Foundation's EUROCORES program LogICCC,
and the Alexander von Humboldt Foundation's 
TransCoop program.  URL:
${\tt{}http\mbox{:}//ccc.cs.uni\mbox{-}duesseldorf.de/\mbox{\small$\sim$}rothe}$
(J. Rothe).}  }

\author{Dorothea Baumeister
\quad 
and
\quad 
J\"{o}rg Rothe
\\ Institut f\"ur Informatik
\\ Heinrich-Heine-Universit\"at D\"usseldorf
\\ 40225 D\"usseldorf, Germany
}

\date{June 9, 2008}

\maketitle

\begin{abstract}
Holzer and Holzer~\cite{hol-hol:j:tantrix} proved that the
Tantrix\texttrademark\ rotation puzzle problem with four colors is
$\np$-complete, and they showed that the infinite variant of this
problem is undecidable.
%
In this paper, we study the three-color and two-color
Tantrix\texttrademark\ rotation puzzle problems 
($\ktrp{3}$ and $\ktrp{2}$) and their variants.
Restricting the
number of allowed colors to three (respectively, to two) 
reduces the
set of available Tantrix\texttrademark\ tiles from 56 to 14
(respectively, to~8).
We prove that $\ktrp{3}$ and $\ktrp{2}$
are $\np$-complete,
which answers a question raised by Holzer and
Holzer~\cite{hol-hol:j:tantrix} in the affirmative.  Since
our reductions are parsimonious,
it follows that the problems $\uktrp{3}$
and $\uktrp{2}$
are $\DP$-complete under randomized reductions.
We also show that the another-solution problems
associated with $\ktrp{4}$, $\ktrp{3}$, and $\ktrp{2}$
are $\np$-complete.
Finally, we prove that the infinite variants of
$\ktrp{3}$ and $\ktrp{2}$
are undecidable.
\end{abstract}

\section{Introduction}
\label{sec:introduction}

The puzzle game Tantrix\texttrademark, invented by Mike McManaway
in 1991, is a domino-like strategy game played
with hexagonal tiles in the plane.  
Each tile contains three colored lines in different
patterns (see Figure~\ref{fig:TantrixTiles}).
We are here interested in the variant of the Tantrix\texttrademark\ rotation
puzzle game whose aim it
is to match the line colors of the joint edges for each pair of adjacent tiles,
just by rotating the tiles around their axes
while their locations remain fixed.
This paper continues the complexity-theoretic study of such problems that was
initiated by Holzer and Holzer~\cite{hol-hol:j:tantrix}.
Other results on the complexity of domino-like strategy
games can be found, e.g., in Gr{\"{a}}del's work~\cite{gra:j:domino}.
Ueda and Nagao~\cite{ued-nag:t:nonogramm} and Yato and
Seta~\cite{yat-set:j:finding-another-solution} provided a framework
for studying the problem of finding another solution of any given 
$\np$ problem when some solutions to this $\np$ problem are already
known---an approach particularly appropriate for puzzle games.
Tantrix\texttrademark\
puzzles have also 
been studied with regard to ``evolutionary computation,''
see Downing~\cite{dow:j:tantrix}.

Holzer and Holzer~\cite{hol-hol:j:tantrix} defined two decision problems 
associated with four-color Tantrix\texttrademark\ rotation puzzles.
The first problem's instances are restricted to a finite number of tiles,
and the second problem's instances are allowed to have infinitely
many tiles.  They proved that the finite variant of this problem is
$\np$-complete
and
that the infinite problem variant is undecidable.
The constructions
in~\cite{hol-hol:j:tantrix} use tiles
with four
colors, just as the original Tantrix\texttrademark\ tile set.
Holzer and Holzer posed
the question of whether the Tantrix\texttrademark\ rotation puzzle
problem remains $\np$-complete if
restricted to only three colors, or if restricted to otherwise reduced
tile sets.
In this paper, we answer this question in the affirmative for the
three-color and the two-color version of this problem.

For each~$k$, $1 \leq k \leq 4$, Table~\ref{tab:results}
summarizes the previously known and our new
results for $\ktrp{k}$,
the $k$-color Tantrix\texttrademark\ rotation puzzle
problem, and its variants.
(All problems are formally defined in
Section~\ref{sec:definitions}.)

\begin{table}[h!t]
\centering
\small
\begin{tabular}[c]{|@{\hspace*{1mm}}c@{\hspace*{1mm}}||@{\hspace*{1mm}}l@{\hspace*{1mm}}|@{\hspace*{1mm}}l@{\hspace*{1mm}}||@{\hspace*{1mm}}l@{\hspace*{1mm}}||@{\hspace*{1mm}}l@{\hspace*{1mm}}||@{\hspace*{1mm}}l@{\hspace*{1mm}}|}
\hline
$k$ & $\ktrp{k}$ is  & Parsimonious?
    & $\uktrp{k}$ is & $\asktrp{k}$ is & $\infktrp{k}$ is
 \\ \hline \hline
$1$ & in $\p$   & & in $\p$   & in $\p$   & decidable 
 \\
    & (trivial) & & (trivial) & (trivial) & (trivial)
 \\ \hline
$2$ & $\np$-complete 
    & yes 
    & $\DP$-$\randomized$-complete
    & $\np$-complete 
    & undecidable 
 \\
    & (see Cor.~\ref{cor:two-trp-is-np-complete})
    & (see Thm.~\ref{thm:two-trp-is-np-complete})
    & (see Cor.~\ref{cor:3-uniquetrp-is-DP-randomized-complete})
    & (see Cor.~\ref{cor:another-solution-trp-is-NP-complete})
    & (see Thm.~\ref{thm:inf-3-trp-and-inf-2-trp-are-undecidable})
 \\ \hline
$3$ & $\np$-complete 
    & yes 
    & $\DP$-$\randomized$-complete
    & $\np$-complete 
    & undecidable 
 \\
    & (see Cor.~\ref{cor:3trp-npc})
    & (see Thm.~\ref{thm:3trp-npc})
    & (see Cor.~\ref{cor:3-uniquetrp-is-DP-randomized-complete})
    & (see Cor.~\ref{cor:another-solution-trp-is-NP-complete})
    & (see Thm.~\ref{thm:inf-3-trp-and-inf-2-trp-are-undecidable})
 \\ \hline
$4$ & $\np$-complete
    & yes
    & $\DP$-$\randomized$-complete 
    & $\np$-complete 
    & undecidable
 \\
    & (see \cite{hol-hol:j:tantrix})
    & (see \cite{bau-rot:c:tantrix})
    & (see \cite{bau-rot:c:tantrix})     
    & (see Cor.~\ref{cor:another-solution-trp-is-NP-complete})
    & (see \cite{hol-hol:j:tantrix})
 \\ \hline
\end{tabular}
\caption{Overview of complexity and decidability 
results for $\ktrp{k}$ and its variants}
\label{tab:results}
\end{table}

Since the four-color Tantrix\texttrademark\ tile set contains 
the three-color Tantrix\texttrademark\ tile set,
our new complexity results for $\ktrp{3}$
imply the previous results for $\ktrp{4}$ (both its
$\np$-completeness~\cite{hol-hol:j:tantrix} and that satisfiability
\emph{parsimoniously} reduces to $\ktrp{4}$~\cite{bau-rot:c:tantrix}).
In contrast,
the three-color Tantrix\texttrademark\ tile set does not contain the
two-color Tantrix\texttrademark\ tile set (see
Figure~\ref{fig:tileset} in Section~\ref{sec:definitions}).  Thus,
$\ktrp{3}$ does not straightforwardly inherit its hardness results
from those of $\ktrp{2}$, which is why both reductions, the one to
$\ktrp{3}$ and the one to $\ktrp{2}$, have to be presented.  Note that
they each substantially differ---both regarding the subpuzzles
constructed and regarding the arguments showing that the constructions
are correct---from the previously known reductions presented
in~\cite{hol-hol:j:tantrix,bau-rot:c:tantrix}, and we will explicitly
illustrate the differences between our new and the original
subpuzzles.

Our reductions will be from a boolean circuit problem, and we
construct a Tantrix\texttrademark\ rotation puzzle that simulates the
computation of such a circuit, where suitable subpuzzles are used to
simulate the wires and gates of the circuit.  In particular, the
previous reductions presented
in~\cite{hol-hol:j:tantrix,bau-rot:c:tantrix,bau-rot:c-toappear:tantrix-three-two-color}
use McColl's planar ``cross-over'' circuit with AND and NOT gates to
simulate wire crossings~\cite{mcc:j:planar-crossovers} and they
employ Goldschlager's log-space transformation from general to planar
circuits~\cite{gol:j:monotone-planar-circuits}.  We take the same
approach in our construction for $\ktrp{2}$.  In contrast, we simulate
wire crossings in the circuit in the construction for $\ktrp{3}$
directly by a new subpuzzle called CROSS, which we will introduce in
Section~\ref{sec:sharpsat-parsimoniously-reduces-to-sharp3trp} and
which will make our reduction for $\ktrp{3}$ significantly more
efficient compared with the reduction for $\ktrp{3}$ presented in a
previous version of this
paper~\cite{bau-rot:c-toappear:tantrix-three-two-color}.  Note that
using the CROSS results in a puzzle with a considerably smaller total
number of tiles that are needed to simulate a given circuit.

Since we provide \emph{parsimonious} reductions from the
satisfiability problem to $\ktrp{3}$ and to $\ktrp{2}$, our
reductions preserve the uniqueness of the solution.
Thus, the unique variants of both
$\ktrp{3}$ and $\ktrp{2}$
are $\DP$-complete under polynomial-time randomized reductions, where
$\DP$ is the class of differences of $\np$ sets.
In addition, we will show that our parsimonious reductions
for $\ktrp{3}$ and $\ktrp{2}$
also provide ``another-solution problem reductions''
(i.e., $\aspred$-reductions, see Section~\ref{sec:definitions-complexity}),
and so the ``another-solution problems'' associated with $\ktrp{3}$ and
$\ktrp{2}$ are also $\np$-complete.\footnote{Informally stated, an
\emph{another-solution problem} associated with an $\np$ problem $A$ asks,
given an instance $x \in A$ and some solutions $y_1, y_2, \ldots , y_n$
for ``$x \in A$'' (i.e., the $y_i$'s encode accepting computation paths
of an $\np$ machine solving $A$ on input~$x$), whether or not there exists
\emph{another} solution, $y \not\in \{y_1, y_2, \ldots , y_n\}$,
for ``$x \in A$.''
See Ueda and Nagao~\cite{ued-nag:t:nonogramm} and Yato and
Seta~\cite{yat-set:j:finding-another-solution} for more details and
results, and also for a discussion of why these problems are particularly
important for puzzle games.
\label{foo:asp}}
Moreover, since $\ktrp{4}$ inherits
the hardness results for $\ktrp{3}$, the another-solution problem
associated with $\ktrp{4}$ is $\np$-complete as well.  Finally, we will
prove that the infinite variants of $\ktrp{3}$ and $\ktrp{2}$
are undecidable, via
a circuit construction similar to the one
Holzer and Holzer~\cite{hol-hol:j:tantrix} used
to show that the infinite
$\ktrp{4}$ problem is undecidable.
%

We mention in passing that the present paper differs from and extends
its preliminary
version~\cite{bau-rot:c-toappear:tantrix-three-two-color} in various
ways.  First, the proof of Theorem~\ref{thm:3trp-npc}, which was only
sketched in~\cite{bau-rot:c-toappear:tantrix-three-two-color}, is
given here in full length, where we also display the original
subpuzzles of Holzer and Holzer~\cite{hol-hol:j:tantrix} to allow
comparison and where we explicitly show the differences between the
subpuzzles used in the their original construction (that provides a
reduction for $\ktrp{4}$ that is not parsimonious;
see~\cite{bau-rot:c:tantrix} for a parsimonious reduction for
$\ktrp{4}$) and in our new reduction showing $\ktrp{3}$ $\np$-complete
via a parsimonious reduction.  Moreover, the proof of this result for
$\ktrp{3}$ presented here additionally differs from the one sketched
in~\cite{bau-rot:c-toappear:tantrix-three-two-color}, since the
reduction given here uses the CROSS subpuzzle, which---as explained
above---makes the reduction significantly more efficient.  Second, we
here provide the proof of Theorem~\ref{thm:two-trp-is-np-complete},
which was completely omitted
in~\cite{bau-rot:c-toappear:tantrix-three-two-color}.  Third,
Corollary~\ref{cor:another-solution-trp-is-NP-complete} and the
related discussion of the another-solution variants of $\ktrp{k}$, $k
\in \{2, 3, 4\}$, are completely new to the current version.

This paper is organized as follows.  Section~\ref{sec:definitions}
provides the complexity-theoretic definitions and notation used and
defines the $k$-color Tantrix\texttrademark\ rotation puzzle
problem and its variants.
Section~\ref{sec:sharpsat-parsimoniously-reduces-to-sharp3trp} shows
that the three-color Tantrix\texttrademark\ rotation puzzle problem is
$\np$-complete via a parsimonious reduction.
To allow comparison,
the original subpuzzles from Holzer and Holzer's
construction~\cite{hol-hol:j:tantrix} are also presented in this section.
Section~\ref{sec:two-trp-is-np-complete}
presents our result that $\ktrp{2}$ is $\np$-complete, again via a
parsimonious reduction.
Section~\ref{sec:unique-infinite-variants} is concerned with the
complexity of the unique and infinite variants of the
three-color and the two-color Tantrix\texttrademark\
rotation puzzle problem, and with the corresponding another-solution problems.

\section{Definitions and Notation}
\label{sec:definitions}

\subsection{Complexity-Theoretic Notions and Notation}
\label{sec:definitions-complexity}

We assume that the reader is familiar with the standard notions of
complexity theory, such as the complexity classes $\p$ (deterministic
polynomial time) and $\np$ (nondeterministic polynomial time); see,
e.g., the
textbooks~\cite{pap:b-1994:complexity,rot:b:cryptocomplexity}.  $\DP$
denotes the class of differences of any two $\np$
sets~\cite{pap-yan:j:dp}.  Note that $\DP$ is also known to be the
second level of the boolean hierarchy over~$\np$, see Cai et
al.~\cite{cai-gun-har-hem-sew-wag-wec:j:bh1,cai-gun-har-hem-sew-wag-wec:j:bh2}.

Let $\sigmastar$ denote the set of strings over the alphabet $\Sigma =
\{0,1\}$.  Given any language $L \seq \sigmastar$, $\|L\|$ denotes the
number of elements in~$L$.
We consider both decision problems and function problems.
The former are formalized as languages whose elements are those
strings in $\sigmastar$ that encode the yes-instances of the problem
at hand.  Regarding the latter, we focus on the counting problems
related to sets in~$\np$.
The counting version $\#A$ of
an $\np$ set $A$ maps each instance $x$ of $A$ to the number of
solutions of~$x$.  That is, counting problems are functions from
$\sigmastar$ to~$\nats$.  As an example, the counting version
$\sharpsat$ of $\sat$, the $\np$-complete satisfiability problem,
asks how many satisfying assignments a given boolean formula
has.  Solutions of $\np$ sets can be viewed as accepting paths of
$\np$ machines.  Valiant~\cite{val:j:permanent} defined the function
class $\sharpp$ to contain the functions that give the number of
accepting paths of some $\np$ machine.  In particular, $\sharpsat$ is
in~$\sharpp$.
Another class of problems we consider are the another-solution problems 
(see Footnote~\ref{foo:asp} for an informal definition and
Definition~\ref{def:sharpktrp-uniquektrp} for the 
another-solution problems associated with $\ktrp{k}$).

The complexity of two decision problems, $A$ and~$B$,
will here be compared 
via the \emph{polynomial-time many-one reducibility}: $A \manyone
B$ if there is a polynomial-time computable function $f$ such that for
each $x \in \sigmastar$, $x \in A$ if and only if $f(x) \in B$.
A set $B$ is said to be $\np$-complete if $B$ is
in $\np$ and every $\np$ set $\manyone$-reduces to~$B$.

Many-one reductions do not always preserve the number of solutions.  A
reduction that does preserve the number of solutions is said to be
\emph{parsimonious}.  Formally, if $A$ and $B$ are any two sets
in~$\np$, we say \emph{$A$ parsimoniously reduces to $B$} if there
exists a polynomial-time computable function $f$ such that for each~$x
\in \sigmastar$, $\#A(x) = \#B(f(x))$.

To compare two another-solution problems associated with two given
$\np$ problems, $A$ and~$B$, Ueda and Nagao~\cite{ued-nag:t:nonogramm}
introduced the following notion of reducibility.\footnote{They call
this notion ``parsimonious reduction with the property
($\ast$)''~\cite{ued-nag:t:nonogramm}.  Yato and
Seta~\cite{yat-set:j:finding-another-solution} introduce a similar
notion (albeit tailored to the case of function problems), which they
denote by ``polynomial-time ASP reduction.''}
We say that $A \aspred B$ if $A$ is parsimoniously reducible to $B$
and, in addition, there exists a polynomial-time computable bijective
function from the set of solutions of $A$ to the set of solutions of
$B$.  Let AS-$A$ and AS-$B$ be the another-solution problems
associated with $A$ and~$B$ (see Footnote~\ref{foo:asp} for an
informal definition and, specifically,
Definition~\ref{def:sharpktrp-uniquektrp} for the another-solution
problems associated with $\ktrp{k}$).
Ueda and Nagao~\cite{ued-nag:t:nonogramm} show that if
AS-$A$ is $\np$-complete and $A \aspred B$, then
$\mbox{AS-}B$ is also $\np$-complete~\cite{ued-nag:t:nonogramm}.  In
particular, AS-$\sat$ is known to be
$\np$-complete~\cite{yat-set:j:finding-another-solution}.

Valiant and Vazirani~\cite{val-vaz:j:np-unique} introduced the
following type of \emph{randomized polynomial-time many-one reducibility}:
$A \randomized B$ if there exists a polynomial-time randomized algorithm
$F$ and a polynomial $p$ such that for each~$x \in \sigmastar$, if $x
\in A$ then $F(x) \in B$ with probability at least $1/p(|x|)$, and if
$x \not\in A$ then $F(x) \not\in B$ with certainty.  In particular,
they proved that the unique version of the satisfiability problem,
$\usat$, is $\DP$-complete under randomized reductions; see also
Chang, Kadin, and
Rohatgi~\cite{cha-kad-roh:uniquesat-randomized-reductions} for 
further related results.

\subsection{Variants of the Tantrix\texttrademark\ Rotation Puzzle Problem}
\label{sec:definitions-tantrix}

\subsubsection{Tile Sets, Color Sequences, and Orientations}
\label{sec:tile-sets-color-sequences-orientations}

\begin{figure}[h!]
  \centering
  \subfigure[Sint]{
    \label{fig:Sint}
\quad
\includegraphics[width=1.5cm]{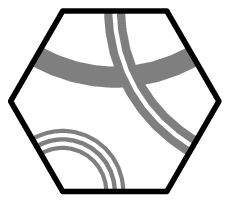}
\quad
  }
  \subfigure[Brid]{
    \label{fig:Brid}
\quad
\includegraphics[width=1.5cm]{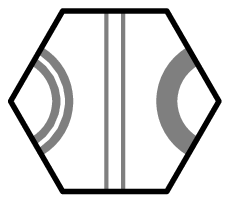}
\quad
  }
  \subfigure[Chin]{
    \label{fig:Chin}
\quad
\includegraphics[width=1.5cm]{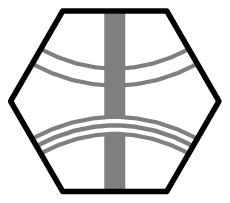}
\quad
  }      
  \subfigure[Rond]{
    \label{fig:Rond}
\quad
\includegraphics[width=1.5cm]{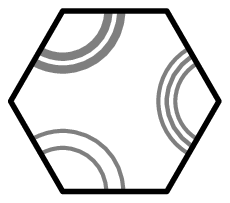}
\quad
  }

  \subfigure[red]{
    \label{fig:red}
\quad
\includegraphics[width=1.5cm]{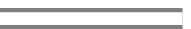}
\quad
  }
  \subfigure[yellow]{
    \label{fig:yellow}
\quad
\includegraphics[width=1.5cm]{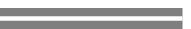}
\quad
  }
  \subfigure[blue]{
    \label{fig:blue}
\quad
\includegraphics[width=1.5cm]{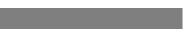}
\quad
  }      
  \subfigure[green]{
    \label{fig:green}
\quad
\includegraphics[width=1.5cm]{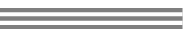}
\quad
  }
\caption{Tantrix\texttrademark\ tile types and the
encoding of Tantrix\texttrademark\ line colors}
\label{fig:TantrixTiles}
\end{figure}

The Tantrix\texttrademark\ rotation puzzle consists of four different
kinds of hexagonal tiles, named \emph{Sint}, \emph{Brid}, \emph{Chin},
and \emph{Rond}.  Each tile has three lines colored differently, where
the three colors of a tile are chosen among four possible colors, see
Figures~\ref{fig:TantrixTiles}(a)--(d). The original
Tantrix\texttrademark\ colors are \emph{red}, \emph{yellow},
\emph{blue}, and \emph{green}, which we encode here as shown in
Figures~\ref{fig:TantrixTiles}(e)--(h). The combination of four kinds
of tiles having three out of four colors each gives a total of 56
different tiles.

\begin{figure}[h!]
  \centering
  \subfigure[Tantrix\texttrademark\ tile set $T_2$]{
    \label{fig:tileset-two-color}
\quad
\includegraphics[width=4cm]{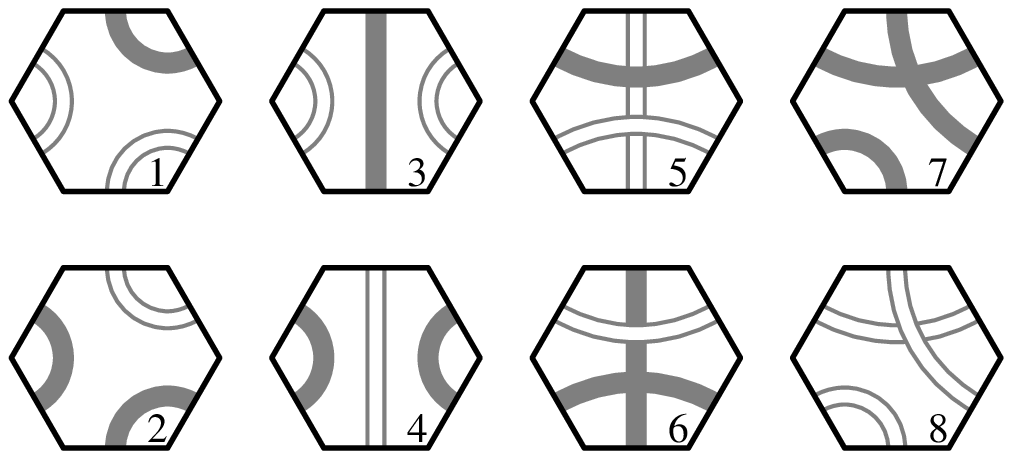}
\quad
  }
  \subfigure[Tantrix\texttrademark\ tile set $T_3$]{
    \label{fig:tileset-three-color}
\quad
\includegraphics[width=5cm]{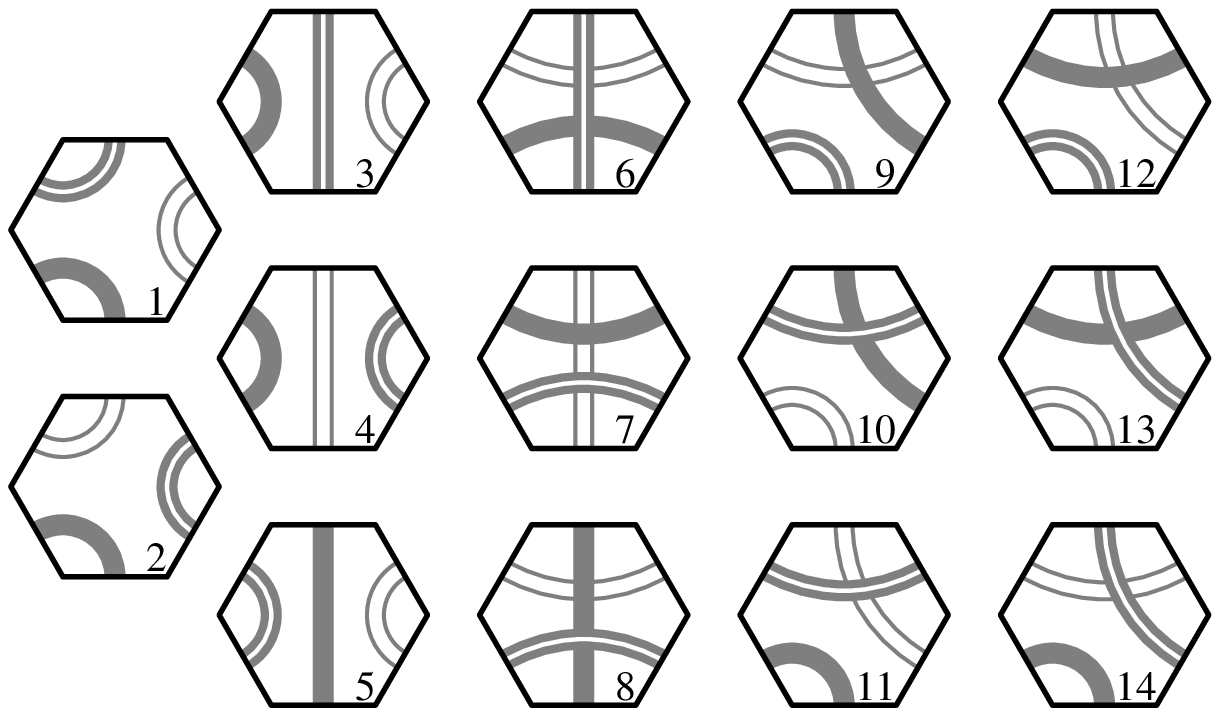}
\quad
  }
  \caption{Tantrix\texttrademark\ tile sets $T_2$ (for \emph{red} and
\emph{blue}) and $T_3$ (for
\emph{red}, \emph{yellow}, and \emph{blue})}
  \label{fig:tileset}
\end{figure}

Since we wish to study Tantrix\texttrademark\ rotation puzzle problems
for which the number of allowed colors is restricted, the set of
Tantrix\texttrademark\ tiles available in a given problem instance
depends on which variant of the Tantrix\texttrademark\ rotation puzzle
problem we are interested in.  Let $C$ be the set that contains the
four colors \emph{red}, \emph{yellow}, \emph{blue}, and \emph{green}.
For each~$i \in \{1, 2, 3, 4\}$, let $C_i \seq C$ be some fixed subset
of size~$i$, and let $T_i$ denote the set of Tantrix\texttrademark\
tiles available when the line colors for each tile are restricted
to~$C_i$.  For example, $T_4$ is the original
Tantrix\texttrademark\ tile set containing 56 tiles, and if
$C_3$ contains, say, the three colors \emph{red}, \emph{yellow}, and
\emph{blue}, then tile set $T_3$ contains the 14 tiles shown in
Figure~\ref{fig:tileset-three-color}.

%

Some more remarks on the tile sets are in order.  First, for $T_3$
and~$T_4$, we require the three lines on each tile to have distinct
colors, as in the original Tantrix\texttrademark\ tile set.
For $T_1$ and $T_2$, however, this is not possible, so we allow the
same color being used for more than one of the three lines of any
tile.  Second, note that we care only about the sequence of colors on
a tile,\footnote{The reason for this and the resulting conventions on
the tile sets stated in this paragraph is that our problems refer to
the variant of the Tantrix\texttrademark\ game that seeks, via
rotations, to make the line colors match on all joint edges of
adjacent tiles.  The objective
of other Tantrix\texttrademark\ games is to create lines and loops of
the same color as long as possible; for problems related to these
Tantrix\texttrademark\ game variants, other conventions on the sets of
allowed tiles would be reasonable.}  where we always use the clockwise
direction to represent color sequences.  However, since different
types of tiles can yield the same color sequence, we will use just one
such tile to represent the corresponding color sequence.  For example,
if $C_2$ contains, say, the two colors \emph{red} and \emph{blue},
then the color sequence \emph{red-red-blue-blue-blue-blue} (which we
abbreviate as ${\tt rrbbbb}$) can be represented by a \emph{Sint}, a
\emph{Brid}, or a \emph{Rond} each having one short \emph{red} arc and
two \emph{blue} additional lines, and we add only one such tile (say,
the \emph{Rond}) to the tile set~$T_2$.
That is, though there is some freedom in choosing a particular set of
tiles, to be specific we fix the tile set $T_2$ shown in
Figure~\ref{fig:tileset-two-color}.
Thus, we have $\|T_1\| = 1$,
$\|T_2\| = 8$, $\|T_3\| = 14$, and $\|T_4\| = 56$, regardless of which
colors are chosen to be in~$C_i$, $1 \leq i \leq 4$.

\begin{table}[h!]
\centering
\begin{tabular}[c]{|c|c||c|c||c|c||c|c|}
\hline
\multicolumn{2}{|c||}{\emph{Rond}} & 
\multicolumn{2}{|c||}{\emph{Brid}} & 
\multicolumn{2}{|c||}{\emph{Chin}} & 
\multicolumn{2}{|c|}{\emph{Sint}}  \\ \hline
$t_1$ & $t_2$ & $t_3$ & $t_4$ & $t_5$ & $t_6$ & $t_7$ & $t_8$ \\ \hline
${\tt bbrrrr}$ &
${\tt rrbbbb}$&
${\tt brrbrr}$ &
${\tt rbbrbb}$ &
${\tt rbrrrb}$ &
${\tt brbbbr}$ &
${\tt bbbbbb}$ &
${\tt rrrrrr}$ \\
\hline
\end{tabular}
\caption{Color sequences of the tiles in $T_2$}
\label{tab:colorsequences-two-colors}
\end{table}

\begin{table}[h!]
\centering
\begin{tabular}[c]{|c|c||c|c|c||c|c|c|}
\hline
\multicolumn{2}{|c||}{\emph{Rond}} & 
\multicolumn{3}{|c||}{\emph{Brid}} & 
\multicolumn{3}{|c|}{\emph{Chin}} \\ \hline
$t_1$ & $t_2$ & $t_3$ & $t_4$ & $t_5$ & $t_6$ & $t_7$ & $t_8$ \\ \hline
${\tt yrrbby}$ &
${\tt ryybbr}$ &
${\tt yrrybb}$ &
${\tt ryyrbb}$ &
${\tt brrbyy}$ &
${\tt yrbybr}$ &
${\tt rbyryb}$ &
${\tt brybyr}$ \\
\hline\hline
\multicolumn{6}{|c|}{\emph{Sint}} & 
 \multicolumn{1}{c}{} &
 \multicolumn{1}{c}{} \\ \cline{1-6}
 \multicolumn{1}{|c|}{$t_9$}  & \multicolumn{1}{|c|}{$t_{10}$} &
 \multicolumn{1}{|c|}{$t_{11}$} & \multicolumn{1}{|c|}{$t_{12}$} &
 \multicolumn{1}{|c|}{$t_{13}$} & \multicolumn{1}{|c|}{$t_{14}$} &
 \multicolumn{1}{c}{}  &
 \multicolumn{1}{c}{}  \\ \cline{1-6}
 \multicolumn{1}{|c|}{${\tt brbyyr}$} &
 \multicolumn{1}{|c|}{${\tt bybrry}$} &
 \multicolumn{1}{|c|}{${\tt ryrbby}$} &
 \multicolumn{1}{|c|}{${\tt rbryyb}$} &
 \multicolumn{1}{|c|}{${\tt ybyrrb}$} &
 \multicolumn{1}{|c|}{${\tt yrybbr}$} &
 \multicolumn{1}{c}{} &
 \multicolumn{1}{c}{} \\ \cline{1-6}
\end{tabular}
\caption{Color sequences of the tiles in $T_3$}
\label{tab:colorsequences-three-colors}
\end{table}

Tables~\ref{tab:colorsequences-two-colors}
and~\ref{tab:colorsequences-three-colors} show the color sequences for
the eight tiles in $T_2$ and for the 14 tiles in $T_3$ that are
presented in Figures~\ref{fig:tileset-two-color}
and~\ref{fig:tileset-three-color}, respectively.
Tables~\ref{tab:colorsequences-two-colors-orientations}
and~\ref{tab:colorsequences-three-colors-orientations} give the six
possible orientations for each tile in $T_2$ and in~$T_3$, which
can be described by permuting the color sequences cyclically
and where repetitions of color sequences are omitted.
Regarding the latter, note that some of the tiles in $T_2$ 
(namely, tiles $t_3$, $t_4$, $t_7$, and $t_8$ in
Table~\ref{tab:colorsequences-two-colors-orientations}) have
orientations that yield identical color sequences due to symmetry,
and so repetitions can be omitted.  In contrast,
no such repetitions occur for the 14 tiles in $T_3$ when permuted
cyclically to yield the six possible orientations (see
Table~\ref{tab:colorsequences-three-colors-orientations}).

Note that, for example, tile $t_7$ from $T_2$ (see
Table~\ref{tab:colorsequences-two-colors-orientations}) has the same color
sequence (namely, ${\tt bbbbbb}$) in each of its six orientations.  In
Section~\ref{sec:results}, we will consider the counting versions of
Tantrix\texttrademark\ rotation puzzle problems and will
construct parsimonious reductions.  When counting the solutions of
Tantrix\texttrademark\ rotation puzzles, we will focus on color
sequences only.  That is, whenever some tile (such as $t_7$ from $T_2$)
has distinct orientations
with identical color sequences, we will count this as just one
solution (and disregard such repetitions).  In this sense, our
reduction to be presented in the proof of
Theorem~\ref{thm:two-trp-is-np-complete} will be parsimonious.

\begin{table}[h!]
\centering
\begin{tabular}[h]{|c||c|c|c|c|c|c|}
\hline
Tile     & \multicolumn{6}{|c|}{Orientation} \\ \cline{2-7}
Number   & 1 & 2 & 3 & 4 & 5 & 6 \\
\hline\hline
1 & ${\tt bbrrrr}$ & ${\tt rbbrrr}$ & ${\tt rrbbrr}$ & ${\tt rrrbbr}$ & ${\tt rrrrbb}$ & ${\tt brrrrb}$ \\
\hline
2 & ${\tt rrbbbb}$ & ${\tt brrbbb}$ & ${\tt bbrrbb}$ & ${\tt bbbrrb}$ & ${\tt bbbbrr}$ & ${\tt rbbbbr}$ \\
\hline\hline
3 & ${\tt brrbrr}$ & ${\tt rbrrbr}$ & ${\tt rrbrrb}$ &  &  &  \\
\hline
4 & ${\tt rbbrbb}$ & ${\tt brbbrb}$ & ${\tt bbrbbr}$ &  &  &  \\
\hline\hline
5 & ${\tt rbrrrb}$ & ${\tt brbrrr}$ & ${\tt rbrbrr}$ & ${\tt rrbrbr}$ & ${\tt rrrbrb}$ & ${\tt brrrbr}$ \\
\hline
6 & ${\tt brbbbr}$ & ${\tt rbrbbb}$ & ${\tt brbrbb}$ & ${\tt bbrbrb}$ & ${\tt bbbrbr}$ & ${\tt rbbbrb}$ \\
\hline\hline
7 & ${\tt bbbbbb}$ &  &  &  &  &  \\
\hline
8 & ${\tt rrrrrr}$ &  &  &  &  &  \\
\hline
\end{tabular}
\caption{Color sequences of the tiles in $T_2$ in their six orientations}
\label{tab:colorsequences-two-colors-orientations}
\end{table}

\begin{table}[h!]
\centering
\begin{tabular}[h]{|c||c|c|c|c|c|c|}
\hline
Tile      & \multicolumn{6}{|c|}{Orientation} \\ \cline{2-7}
Number    & 1 & 2 & 3 & 4 & 5 & 6 \\
\hline\hline
 1 & ${\tt yrrbby}$ & ${\tt yyrrbb}$ & ${\tt byyrrb}$ & ${\tt bbyyrr}$ & ${\tt rbbyyr}$ & ${\tt rrbbyy}$ \\
\hline
 2 & ${\tt ryybbr}$ & ${\tt rryybb}$ & ${\tt brryyb}$ & ${\tt bbrryy}$ & ${\tt ybbrry}$ & ${\tt yybbrr}$ \\
\hline\hline
 3 & ${\tt yrrybb}$ & ${\tt byrryb}$ & ${\tt bbyrry}$ & ${\tt ybbyrr}$ & ${\tt rybbyr}$ & ${\tt rrybby}$ \\
\hline
 4 & ${\tt ryyrbb}$ & ${\tt bryyrb}$ & ${\tt bbryyr}$ & ${\tt rbbryy}$ & ${\tt yrbbry}$ & ${\tt yyrbbr}$ \\
\hline
 5 & ${\tt brrbyy}$ & ${\tt ybrrby}$ & ${\tt yybrrb}$ & ${\tt byybrr}$ & ${\tt rbyybr}$ & ${\tt rrbyyb}$ \\
\hline\hline
 6 & ${\tt yrbybr}$ & ${\tt ryrbyb}$ & ${\tt bryrby}$ & ${\tt ybryrb}$ & ${\tt bybryr}$ & ${\tt rbybry}$ \\
\hline
 7 & ${\tt rbyryb}$ & ${\tt brbyry}$ & ${\tt ybrbyr}$ & ${\tt rybrby}$ & ${\tt yrybrb}$ & ${\tt byrybr}$ \\
\hline
 8 & ${\tt brybyr}$ & ${\tt rbryby}$ & ${\tt yrbryb}$ & ${\tt byrbry}$ & ${\tt ybyrbr}$ & ${\tt rybyrb}$ \\
\hline\hline
 9 & ${\tt brbyyr}$ & ${\tt rbrbyy}$ & ${\tt yrbrby}$ & ${\tt yyrbrb}$ & ${\tt byyrbr}$ & ${\tt rbyyrb}$ \\
\hline
10 & ${\tt bybrry}$ & ${\tt ybybrr}$ & ${\tt rybybr}$ & ${\tt rrybyb}$ & ${\tt brryby}$ & ${\tt ybrryb}$ \\
\hline
11 & ${\tt ryrbby}$ & ${\tt yryrbb}$ & ${\tt byryrb}$ & ${\tt bbyryr}$ & ${\tt rbbyry}$ & ${\tt yrbbyr}$ \\
\hline
12 & ${\tt rbryyb}$ & ${\tt brbryy}$ & ${\tt ybrbry}$ & ${\tt yybrbr}$ & ${\tt ryybrb}$ & ${\tt bryybr}$ \\
\hline
13 & ${\tt ybyrrb}$ & ${\tt bybyrr}$ & ${\tt rbybyr}$ & ${\tt rrbyby}$ & ${\tt yrrbyb}$ & ${\tt byrrby}$ \\
\hline
14 & ${\tt yrybbr}$ & ${\tt ryrybb}$ & ${\tt bryryb}$ & ${\tt bbryry}$ & ${\tt ybbryr}$ & ${\tt rybbry}$ \\
\hline
\end{tabular}
\caption{Color sequences of the tiles in $T_3$ in their six orientations}
\label{tab:colorsequences-three-colors-orientations}
\end{table}

\subsubsection{Definition of the Problems}

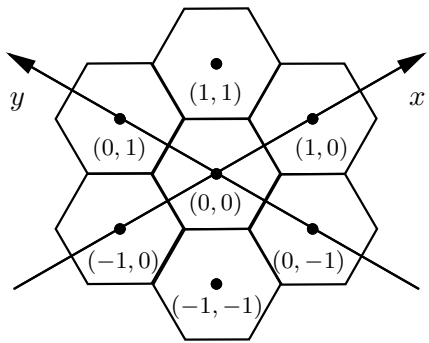
\begin{figure}[h!]
  \centering
\setlength{\unitlength}{0.00053745in}
\begingroup\makeatletter\ifx\SetFigFont\undefined%
\gdef\SetFigFont#1#2#3#4#5{%
  \reset@font\fontsize{#1}{#2pt}%
  \fontfamily{#3}\fontseries{#4}\fontshape{#5}%
  \selectfont}%
\fi\endgroup%
{\renewcommand{\dashlinestretch}{30}
\begin{picture}(4094,3308)(0,-10)
\thicklines
\put(2047,1648){\blacken\ellipse{90}{90}}
\put(2047,1648){\ellipse{90}{90}}
\put(2047,2728){\blacken\ellipse{90}{90}}
\put(2047,2728){\ellipse{90}{90}}
\put(2047,568){\blacken\ellipse{90}{90}}
\put(2047,568){\ellipse{90}{90}}
\put(2992,2188){\blacken\ellipse{90}{90}}
\put(2992,2188){\ellipse{90}{90}}
\put(1102,2188){\blacken\ellipse{90}{90}}
\put(1102,2188){\ellipse{90}{90}}
\put(1102,1108){\blacken\ellipse{90}{90}}
\put(1102,1108){\ellipse{90}{90}}
\put(2992,1108){\blacken\ellipse{90}{90}}
\put(2992,1108){\ellipse{90}{90}}
\path(2362,1108)(2672,1651)(2357,2191)
	(1732,2188)(1422,1645)(1737,1105)(2362,1108)
\path(3307,1648)(3617,2191)(3302,2731)
	(2677,2728)(2367,2185)(2682,1645)(3307,1648)
\path(3312,565)(3622,1108)(3307,1648)
	(2682,1645)(2372,1102)(2687,562)(3312,565)
\path(2367,25)(2677,568)(2362,1108)
	(1737,1105)(1427,562)(1742,22)(2367,25)
\path(1412,571)(1722,1114)(1407,1654)
	(782,1651)(472,1108)(787,568)(1412,571)
\path(1412,1651)(1722,2194)(1407,2734)
	(782,2731)(472,2188)(787,1648)(1412,1651)
\path(67,523)(4072,2818)
\path(67,523)(4072,2818)
\blacken\path(3893.597,2646.616)(4072.000,2818.000)(3833.934,2750.733)(3893.597,2646.616)
\path(4027,523)(22,2818)
\path(4027,523)(22,2818)
\blacken\path(260.066,2750.733)(22.000,2818.000)(200.403,2646.616)(260.066,2750.733)
\path(2362,2188)(2672,2731)(2357,3271)
	(1732,3268)(1422,2725)(1737,2185)(2362,2188)
\put(3937,2323){\makebox(0,0)[lb]{\smash{{\SetFigFont{11}{7.2}{\rmdefault}{\mddefault}{\updefault}$x$}}}}
\put(22,2323){\makebox(0,0)[lb]{\smash{{\SetFigFont{11}{7.2}{\rmdefault}{\mddefault}{\updefault}$y$}}}}
\put(1772,2368){\makebox(0,0)[lb]{\smash{{\SetFigFont{9.5}{7.2}{\rmdefault}{\mddefault}{\updefault}$(1,1)$}}}}
\put(1772,1288){\makebox(0,0)[lb]{\smash{{\SetFigFont{9.5}{7.2}{\rmdefault}{\mddefault}{\updefault}$(0,0)$}}}}
\put(1602,288){\makebox(0,0)[lb]{\smash{{\SetFigFont{9.5}{7.2}{\rmdefault}{\mddefault}{\updefault}$(-1,-1)$}}}}
\put(2812,1828){\makebox(0,0)[lb]{\smash{{\SetFigFont{9.5}{7.2}{\rmdefault}{\mddefault}{\updefault}$(1,0)$}}}}
\put(832,1828){\makebox(0,0)[lb]{\smash{{\SetFigFont{9.5}{7.2}{\rmdefault}{\mddefault}{\updefault}$(0,1)$}}}}
\put(777,728){\makebox(0,0)[lb]{\smash{{\SetFigFont{9.5}{7.2}{\rmdefault}{\mddefault}{\updefault}$(-1,0)$}}}}
\put(2592,728){\makebox(0,0)[lb]{\smash{{\SetFigFont{9.5}{7.2}{\rmdefault}{\mddefault}{\updefault}$(0,-1)$}}}}
\end{picture}
}
  \caption{A two-dimensional hexagonal coordinate system}
  \label{fig:coordinate-system}
\end{figure}

We now recall some useful notation that Holzer and
Holzer~\cite{hol-hol:j:tantrix} introduced in order to formalize
problems related to the Tantrix\texttrademark\ rotation puzzle.  The
instances of such problems are Tantrix\texttrademark\ tiles firmly
arranged in the plane.  To represent their positions, we use a
two-dimensional hexagonal coordinate system shown in
Figure~\ref{fig:coordinate-system}.  Let $T \in \{T_1, T_2, T_3,
T_4\}$ be some tile set as defined above.  Let $\mathcal{A} :
\mathbb{Z}^2 \rightarrow T$ be a function mapping points in
$\mathbb{Z}^2$ to tiles in~$T$, i.e., $\mathcal{A}(x)$ is the type of
the tile located at position~$x$.  Note that $\mathcal{A}$ is a
partial function; throughout this paper (except in
Theorem~\ref{thm:inf-3-trp-and-inf-2-trp-are-undecidable} and its
proof), we restrict our problem instances to finitely many
given tiles, and the regions of $\mathbb{Z}^2$ they cover may have
holes (which is a difference to the original Tantrix\texttrademark\
game).

Define $\shape(\mathcal{A})$ to be the set of points $x \in
\mathbb{Z}^2$ for which $\mathcal{A}(x)$ is defined.
For any two distinct points $x = (a,b)$ and $y = (c,d)$ in
$\integers^2$, $x$ and $y$ are neighbors if and only if $(a = c$ and
$|b-d| = 1)$ or $(|a-c| = 1$ and $b = d)$ or $(a-c = 1$ and $b-d = 1)$
or $(a-c = -1$ and $b-d = -1)$.  For any two points $x$ and $y$ in
$\shape(\mathcal{A})$, $\mathcal{A}(x)$ and $\mathcal{A}(y)$ are said
to be neighbors exactly if $x$ and $y$ are neighbors.

We now define the Tantrix\texttrademark\ rotation puzzle problems we
are interested in, where the parameter $k$ is chosen from $\{1, 2, 3,
4\}$:
\begin{desctight}
\item[Name:] $k$-Color Tantrix\texttrademark\ Rotation Puzzle
($\ktrp{k}$, for short).

\item[Instance:] A finite shape function $\mathcal{A} : \mathbb{Z}^2
\rightarrow T_k$, appropriately encoded as a string in~$\sigmastar$.

\item[Question:] Is there a solution to the rotation puzzle defined by
$\mathcal{A}$, i.e., does there exist a rotation of the given tiles in
$\shape(\mathcal{A})$ such that the colors of the lines of any two
adjacent tiles match at their joint edge?
\end{desctight}

Clearly, $\ktrp{1}$ can be solved trivially, so $\ktrp{1}$ is in~$\p$.
On the other hand, Holzer and Holzer~\cite{hol-hol:j:tantrix} showed
that $\ktrp{4}$ is $\np$-complete and that the infinite variant of
$\ktrp{4}$ is undecidable.  Baumeister and
Rothe~\cite{bau-rot:c:tantrix}
investigated the
counting and the unique variant of $\ktrp{4}$ and, in
particular, provided a parsimonious
reduction from
$\sat$ to $\ktrp{4}$.  In this
paper, we
study the three-color and two-color versions of this
problem, $\ktrp{3}$ and $\ktrp{2}$,
and their counting, unique, another-solution, and infinite variants.

\begin{definition}
\label{def:sharpktrp-uniquektrp}
\begin{enumerate}
\item A \emph{solution to a $\ktrp{k}$ instance $\mathcal{A}$}
specifies an orientation of each tile in $\shape(\mathcal{A})$ such
that the colors of the lines of any two adjacent tiles match at their
joint edge.  Let $\solutionktrp{k}{\mathcal{A}}$ denote the set of
solutions of~$\mathcal{A}$.

\item Define the \emph{counting version} of $\ktrp{k}$ to be the
function $\sharpktrp{k}$ mapping from $\sigmastar$ to $\nats$
such that
$\sharpktrp{k}(\mathcal{A}) = \| \solutionktrp{k}{\mathcal{A}} \|$.

\item Define the \emph{unique version} of $\ktrp{k}$ as
$\uktrp{k} = \{\mathcal{A} \condition \sharpktrp{k}(\mathcal{A}) = 1 \}$.

\item Define the \emph{another-solution problem associated with $\ktrp{k}$}
as
\[
\asktrp{k} = \{(\mathcal{A},y_1, \ldots , y_n) \condition
y_1, \ldots , y_n \in \solutionktrp{k}{\mathcal{A}} \mbox{ and }
\|\solutionktrp{k}{\mathcal{A}} \| > n\}.
\]
\end{enumerate}
\end{definition}

The above problems are defined for the case of finite problem
instances.  The infinite Tantrix\texttrademark\ rotation puzzle
problem with $k$ colors ($\infktrp{k}$, for short) is defined exactly
as $\ktrp{k}$, the only difference being that the shape function
$\mathcal{A}$
is not required to be finite
and is represented by the encoding of a Turing machine computing
$\mathcal{A} : \mathbb{Z}^2 \rightarrow T_k$.

\section{Results}
\label{sec:results}

\subsection{Parsimonious Reduction from SAT to 3-TRP}
\label{sec:sharpsat-parsimoniously-reduces-to-sharp3trp}

Theorem~\ref{thm:3trp-npc} below is the main result of this section.
Notwithstanding that our proof
follows the
general approach of
Holzer and Holzer~\cite{hol-hol:j:tantrix},
our specific construction and our proof of correctness will
differ substantially from theirs. 
We will provide a parsimonious reduction from
$\sat$ to $\ktrp{3}$.  Let $\circuitandnotsat$ denote the
problem of deciding, given a boolean circuit $c$ with AND and NOT
gates only, whether or not there is a satisfying truth assignment to the
input variables of~$c$.  The $\np$-completeness of $\circuitandnotsat$
was shown by Cook~\cite{coo:c:theorem-proving}.  The following lemma
(stated, e.g., in~\cite{bau-rot:c:tantrix})
is straightforward.

\begin{lemma}
\label{lem:sat-parsimoniously-reduces-to-circuitsat-and-not}
$\sat$ parsimoniously reduces to~$\circuitandnotsat$.
\end{lemma}

\begin{theorem}
\label{thm:3trp-npc}
$\sat$ parsimoniously reduces to $\threetrp$.
\end{theorem}

\begin{proofs}
By Lemma~\ref{lem:sat-parsimoniously-reduces-to-circuitsat-and-not},
it is enough to show that $\circuitandnotsat$ parsimoniously
reduces to $\ktrp{3}$.  The resulting $\ktrp{3}$ instance
simulates a boolean circuit with AND and NOT gates such that
the number of solutions of the rotation puzzle equals the
number of satisfying
truth assignments to the variables of the circuit.  

\paragraph{General remarks on our proof approach:}
The rotation puzzle to be constructed from a given circuit consists of
different subpuzzles each using only three colors.  The color
\emph{green} was employed by Holzer and
Holzer~\cite{hol-hol:j:tantrix} only to exclude certain rotations, so
we choose to eliminate this color in our three-color rotation puzzle.
Thus, letting $C_3$
contain the colors \emph{blue},
\emph{red}, and \emph{yellow}, we have the tile set $T_3 = \{t_1, t_2,
\ldots, t_{14}\}$, where the enumeration of tiles corresponds to
Figure~\ref{fig:tileset-three-color}.
Furthermore, our construction will be parsimonious, i.e., there will be
a one-to-one correspondence between the solutions of the given
$\circuitandnotsat$ instance and the solutions of the resulting
rotation puzzle instance.
Note that part of our work is already done, since some subpuzzles
constructed in~\cite{bau-rot:c:tantrix} use only three
colors and they each have unique solutions.  However,
the remaining subpuzzles have to be either modified substantially or
to be constructed completely differently, and the arguments of why our
modified construction is correct differs
considerably from previous
work~\cite{hol-hol:j:tantrix,bau-rot:c:tantrix}.

Since it is not so easy to exclude undesired rotations without having
the color \emph{green} available, let us first analyze the 14 tiles
in~$T_3$.  For $u,v \in C_3$ and for each tile $t_i$ in~$T_3$, where
$1 \leq i \leq 14$, Table~\ref{tab:subcolorsequences-three-colors}
shows which substrings of the form $uv$ occur in the color sequence of
$t_i$ (as indicated by an \x\ entry in row $uv$ and column~$i$).  In
the remainder of this proof, when showing that our construction is
correct, our arguments will often be based on which substrings do or
do not occur in the color sequences of certain tiles from~$T_3$, and
Table~\ref{tab:subcolorsequences-three-colors} may then be looked up
for convenience.

\begin{table}[h!t]
\centering
\begin{tabular}[c]{|l||c|c||c|c|c||c|c|c||c|c|c|c|c|c|}

\hline
 & \multicolumn{2}{|c||}{\emph{Rond}} & 
\multicolumn{3}{|c||}{\emph{Brid}} & 
\multicolumn{3}{|c||}{\emph{Chin}} & 
\multicolumn{6}{|c|}{\emph{Sint}}  \\
\hline
$uv$  &  1 &  2 &  3 &  4 &  5 &  6 &  7 &  8 &  9 & 10 & 11 & 12 & 13 & 14 \\
 \hline\hline
${\tt bb}$    & \x & \x & \x & \x &    &    &    &    &    &    & \x &    &    & \x \\
 \hline
${\tt rr}$    & \x & \x & \x &    & \x &    &    &    &    & \x &    &    & \x &    \\
 \hline
${\tt yy }$   & \x & \x &    & \x & \x &    &    &    & \x &    &    & \x &    &    \\
 \hline\hline
${\tt br }$   &    & \x &    & \x & \x & \x & \x & \x & \x & \x &    & \x &    & \x \\
 \hline
${\tt rb }$   & \x &    &    & \x & \x & \x & \x & \x & \x &    & \x & \x & \x &    \\
 \hline
${\tt by }$   & \x &    & \x &    & \x & \x & \x & \x & \x & \x & \x &    & \x &    \\
 \hline
${\tt yb }$   &    & \x & \x &    & \x & \x & \x & \x &    & \x &    & \x & \x & \x \\
 \hline
${\tt ry}$    &    & \x & \x & \x &    & \x & \x & \x &    & \x & \x & \x &    & \x \\
 \hline
${\tt yr}$    & \x &    & \x & \x &    & \x & \x & \x & \x &    & \x &    & \x & \x \\
 \hline
\end{tabular}
\caption{Substrings $uv$ that occur in the color sequences of the
tiles in $T_3$}
\label{tab:subcolorsequences-three-colors}
\end{table}












Holzer and
Holzer~\cite{hol-hol:j:tantrix} consider a boolean circuit $c$ on
input variables $x_1, x_2, \ldots , x_n$ as a sequence $(\alpha_1,
\alpha_2, \ldots ,\alpha_m)$ of computation steps (or
``instructions''), and we adopt this approach here.  For the $i$th
instruction, $\alpha_i$, we have $\alpha_i = x_i$ if $1 \leq i \leq
n$, and if $n+1 \leq i \leq m$ then we have 
either $\alpha_i = \mbox{NOT}(j)$ or $\alpha_i = \mbox{AND}(j,k)$,
where $j\leq k < i$.  Circuits
are evaluated in the standard way.
We will represent the truth value \emph{true} by the color \emph{blue}
and the truth value \emph{false} by the color \emph{red}
in our rotation puzzle.

A technical difficulty in the construction results
from the wire crossings that circuits can have.  To construct rotation
puzzles from \emph{planar} circuits, Holzer and Holzer use McColl's
planar ``cross-over'' circuit with AND and NOT gates to simulate such
wire crossings~\cite{mcc:j:planar-crossovers}, and in particular they
employ Goldschlager's log-space transformation from general to planar
circuits~\cite{gol:j:monotone-planar-circuits}.  For the details of
this transformation, we refer to Holzer and Holzer's
work~\cite{hol-hol:j:tantrix}.

We use a different approach to overcome the difficulty caused by wire
crossings.  Our construction will employ a new subpuzzle for this purpose.
Holzer and Holzer's circuit
construction uses several cross-over
circuits, and each of them consists of twelve AND and nine NOT gates, 
and in addition it increases
the number of instruction steps by 14. We will avoid this blow-up
by using the
CROSS subpuzzle, which achieves a direct crossing of two
adjacent wires in our Tantrix\texttrademark\ puzzle and thus is much more
efficient.

For the sake of comparison, we also present the original subpuzzles from 
Holzer and Holzer's construction (\cite{hol-hol:j:tantrix}) in this section, 
with the following conventions:
Tiles having more than one possible orientation as well as tiles 
containing \emph{green} lines will always have a grey instead of a 
black
edging, and modified or inserted tiles
in our new subpuzzles will always be highlighted by having a grey background.
This will illustrate the differences between our new and the
previously known original subpuzzles.

\paragraph{Wire subpuzzles:}
Wires of the circuit are simulated by the subpuzzles WIRE, MOVE, and COPY.

A vertical wire is represented by a WIRE subpuzzle, which is shown in
Figure~\ref{fig:wire-3trp}.  The original WIRE subpuzzle
from~\cite{hol-hol:j:tantrix} 
(see Figure~\ref{fig:wire-4trp})
does not contain \emph{green} but it does not have a
unique solution, while the WIRE subpuzzle
from~\cite{bau-rot:c:tantrix}, which is not displayed here, ensures the
uniqueness of the solution but is using a tile with a \emph{green} line.  
In the original WIRE subpuzzle, both tiles, $a$ and $b$, have two possible 
orientations for each input color.
Inserting two new tiles
at positions $x$ and $y$ (see Figure~\ref{fig:wire-3trp})
makes the solution unique.  If the input color is \emph{blue}, tile $x$ must
contain one of the following color-sequence substrings
for the edges joint with tiles
$b$ and $a$: ${\tt ry}$, ${\tt rr}$, ${\tt yy}$, or ${\tt yr}$.  If the 
input color is \emph{red}, $x$ must contain one of these substrings:
${\tt bb}$, 
${\tt yb}$, ${\tt yy}$, or ${\tt by}$. Tile $t_{12}$ satisfies the conditions 
${\tt yy}$ and ${\tt ry}$ for the input color \emph{blue}, and the conditions
${\tt yb}$ and ${\tt yy}$ for the input color \emph{red}.

The solution must now be fixed with tile~$y$.
The possible color-sequence substrings of $y$ at the edges joint with $a$ and 
$b$ are ${\tt rr}$ and ${\tt ry}$ for the input color 
\emph{blue},
and
${\tt yb}$ and ${\tt bb}$ for the input
color \emph{red}. Tile $t_{13}$ has exactly one of these sequences for each
input color.  Thus, the solution for this subpuzzle contains only three
colors and is unique.

\begin{figure}[h!]
  \centering
  \subfigure[In: \emph{true}]{
    \label{fig:wire-4trp-t}
    \quad
    \quad
    \includegraphics[height=2.8cm]{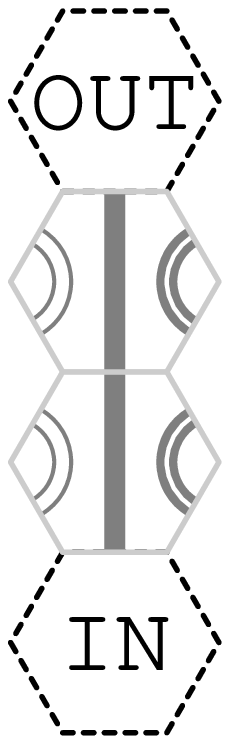}
    \quad
    \quad
  }
  \subfigure[In: \emph{false}]{
    \label{fig:wire-4trp-f}
    \quad
    \quad
    \includegraphics[height=2.8cm]{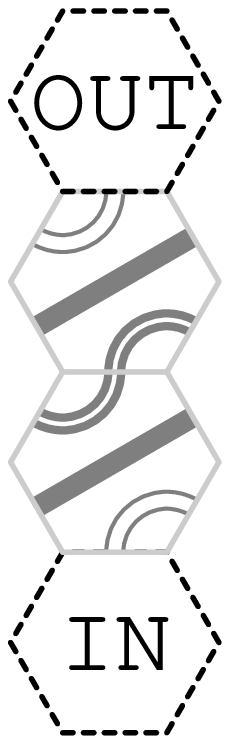}
    \quad
    \quad
  }
  \subfigure[Scheme]{
    \label{fig:wire-4trp-s}
    \quad
    \quad
    \includegraphics[height=2.8cm]{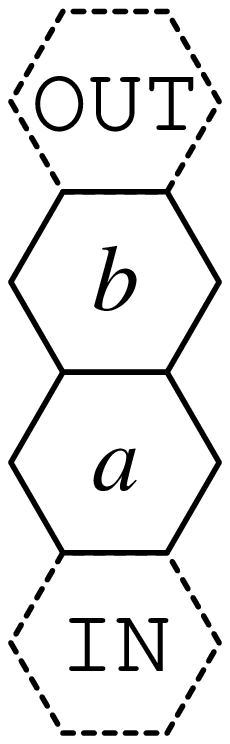}
    \quad
    \quad
  }
  \caption{Original WIRE subpuzzle, see~\cite{hol-hol:j:tantrix}}
  \label{fig:wire-4trp}
\end{figure}

\begin{figure}[h!]
  \centering
  \subfigure[In: \emph{true}]{
    \label{fig:wire-3trp-t}
    \quad 
    \includegraphics[height=2.8cm]{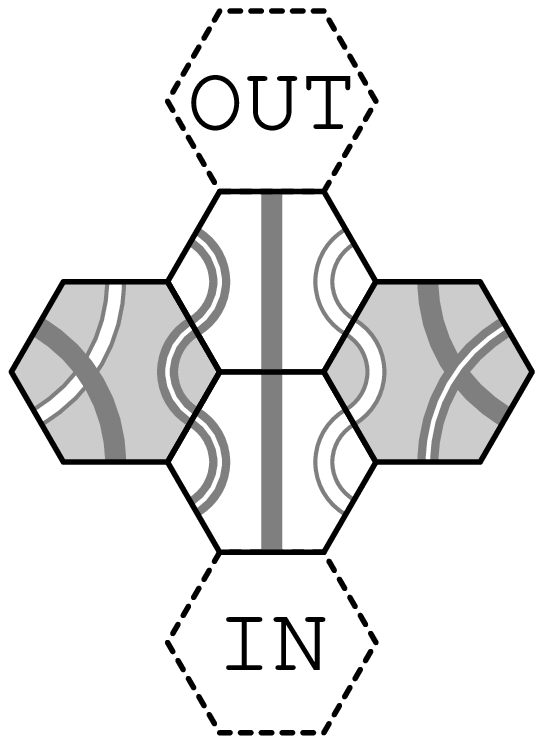}
    \quad 
  }
  \subfigure[In: \emph{false}]{
    \label{fig:wire-3trp-f}
    \quad 
    \includegraphics[height=2.8cm]{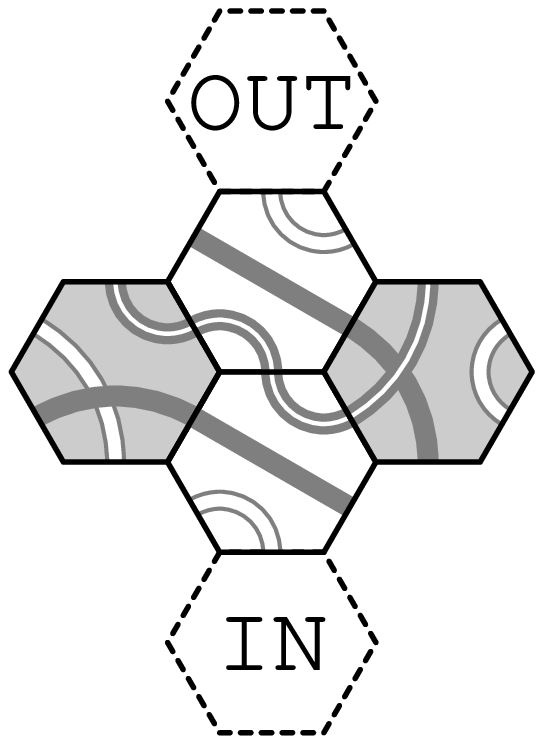}
    \quad 
  }
  \subfigure[Scheme]{
    \label{fig:wire-3trp-s}
    \quad 
    \includegraphics[height=2.8cm]{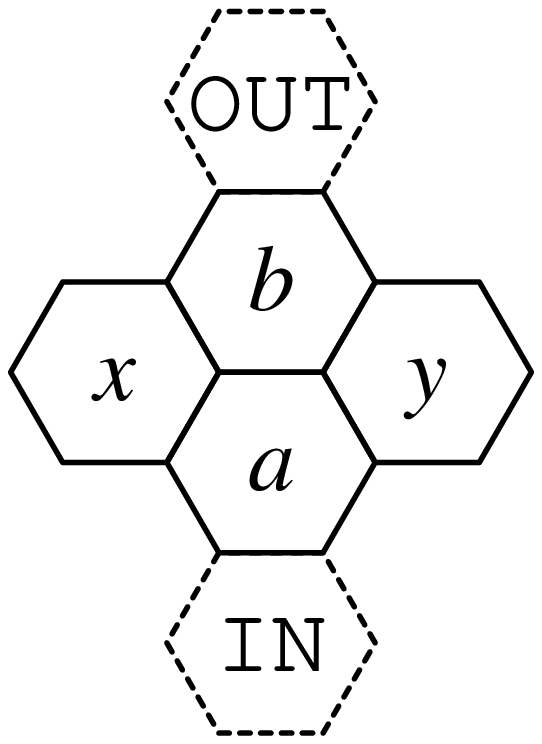}
    \quad 
  } 
  \caption{Three-color WIRE subpuzzle}
  \label{fig:wire-3trp}
\end{figure}

The MOVE subpuzzle is needed to move a wire by two positions to the left
or to the right.  
The original MOVE subpuzzle from~\cite{hol-hol:j:tantrix}
contains only
three colors but has several solutions.  
One solution for each input color is shown in Figure~\ref{fig:move-4trp}, 
where the tiles with a grey edging have more than one possible orientation.
However, the modified subpuzzle
from~\cite{bau-rot:c:tantrix}, which is presented in
Figure~\ref{fig:move-3trp},
contains also only three colors but has a
unique solution.

\begin{figure}[h!]
  \centering
  \subfigure[In: \emph{true}]{
    \label{fig:move-4trp-t}
    \quad
    \includegraphics[height=4.2cm]{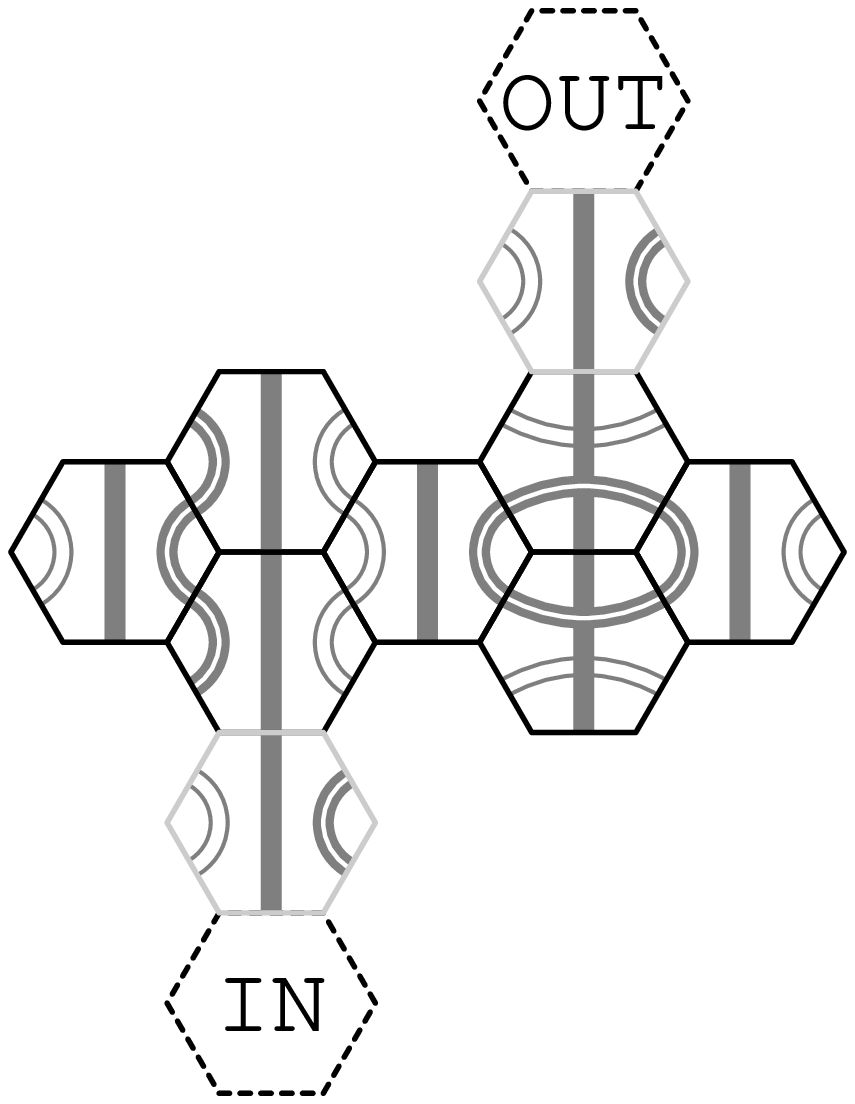}
    \quad
  }
    \subfigure[In: \emph{false}]{
    \label{fig:move-4trp-f}
    \quad
    \includegraphics[height=4.2cm]{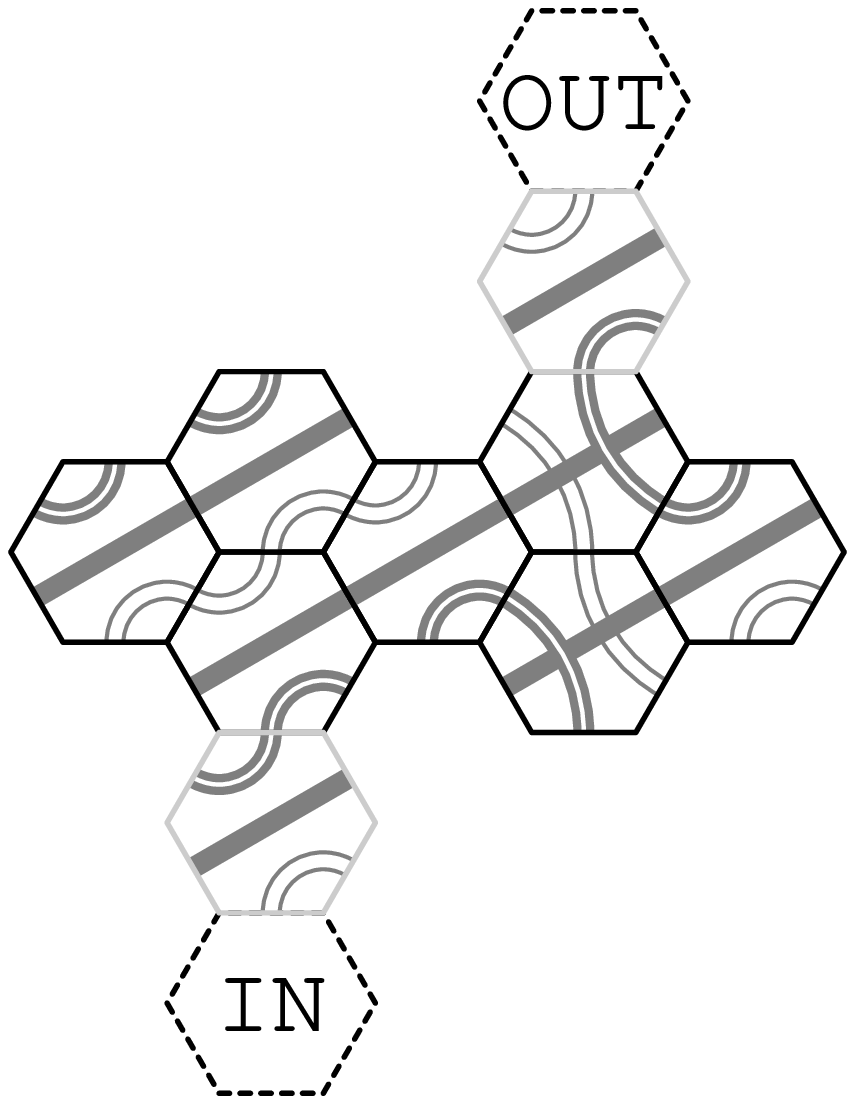}
    \quad
  }
  \caption{Original MOVE subpuzzle, see~\cite{hol-hol:j:tantrix}}
  \label{fig:move-4trp}
\end{figure}

\begin{figure}[h!]
  \centering
  \subfigure[In: \emph{true}]{
    \label{fig:move-3trp-t}
    \quad 
    \includegraphics[height=4.2cm]{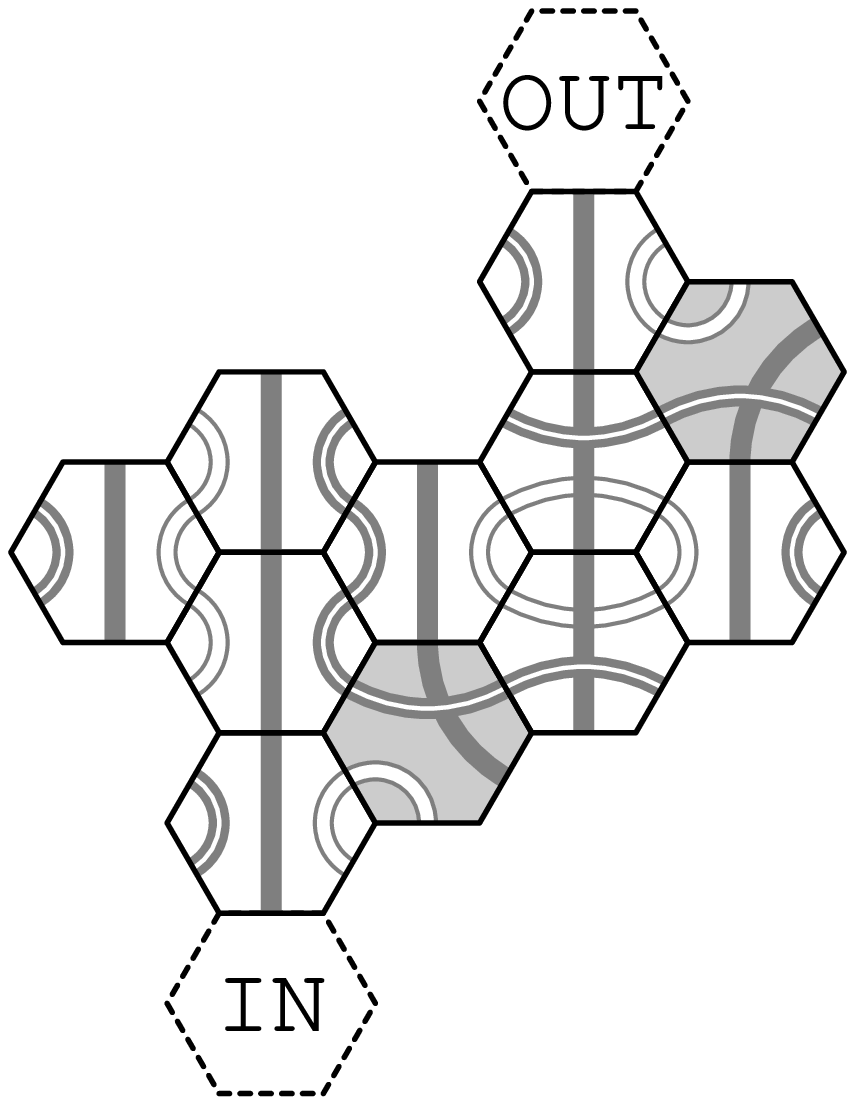}
    \quad 
  }
  \subfigure[In: \emph{false}]{
    \label{fig:move-3trp-f}
    \quad 
    \includegraphics[height=4.2cm]{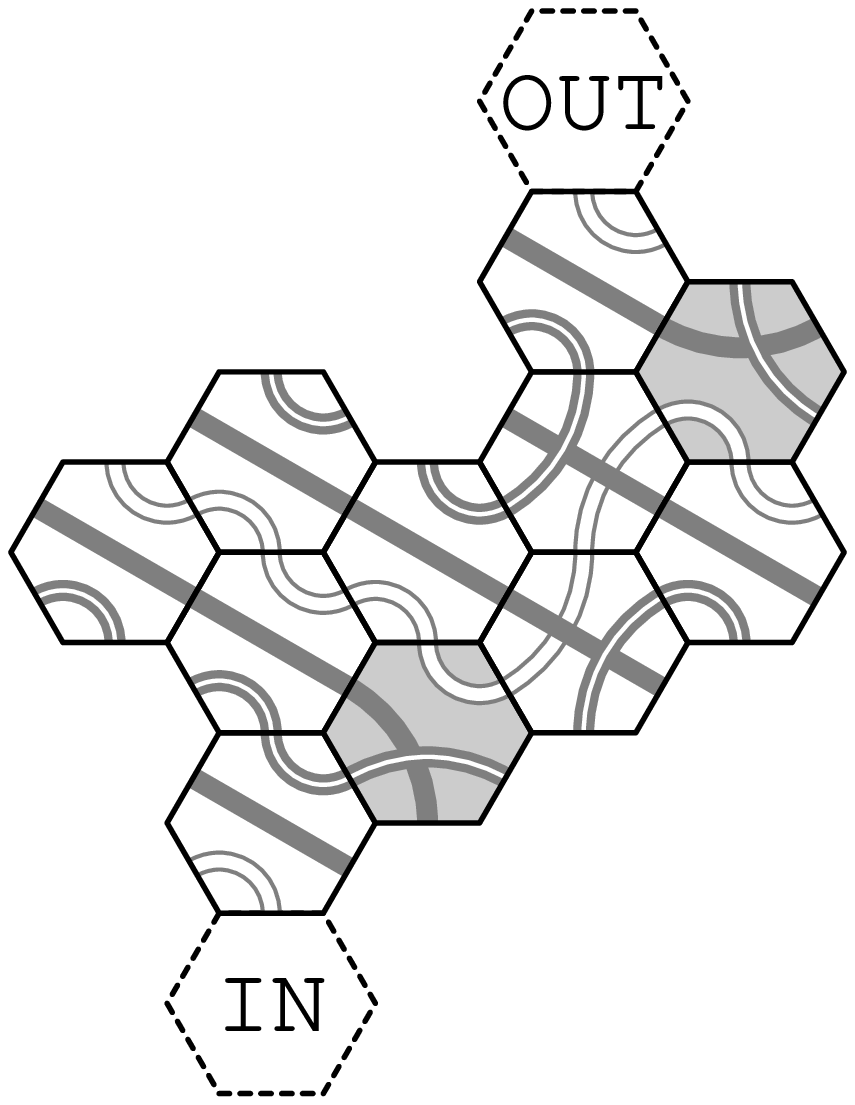}
    \quad 
  }
  \caption{Three-color MOVE subpuzzle, see~\cite{bau-rot:c:tantrix}}
  \label{fig:move-3trp}
\end{figure}

The COPY subpuzzle is used to ``split'' a wire into two copies.  By the
same arguments as above we can take the modified COPY subpuzzle
from~\cite{bau-rot:c:tantrix}, which is presented in
Figure~\ref{fig:copy-3trp}.  Figure~\ref{fig:copy-4trp} shows
the original COPY subpuzzle from~\cite{hol-hol:j:tantrix}.

\begin{figure}[h!]
  \centering
  \subfigure[In: \emph{true}]{
    \label{fig:copy-4trp-t}
    \quad
    \includegraphics[height=4.2cm]{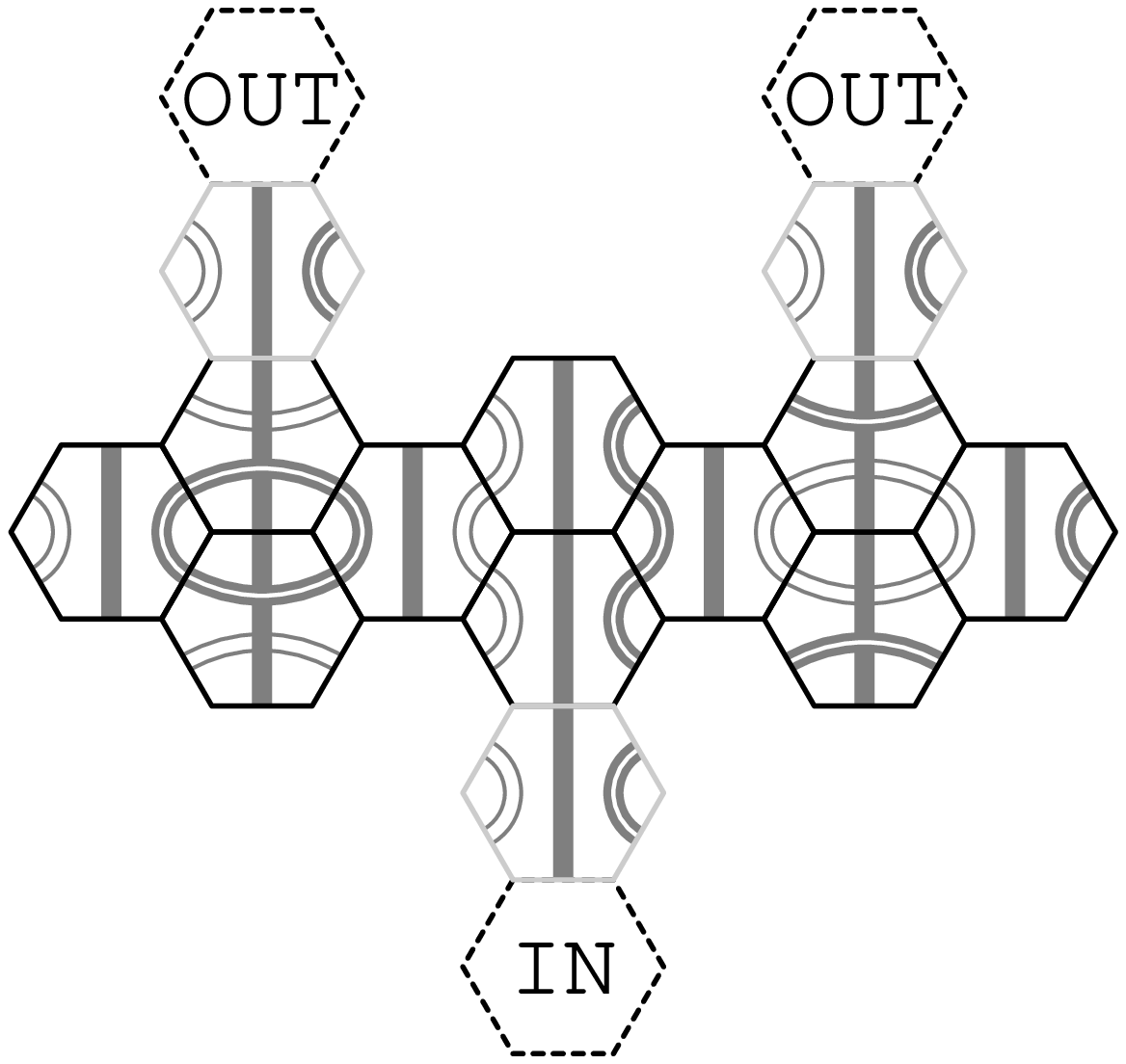}
    \quad
  }
    \subfigure[In: \emph{false}]{
    \label{fig:copy-4trp-f}
    \quad
    \includegraphics[height=4.2cm]{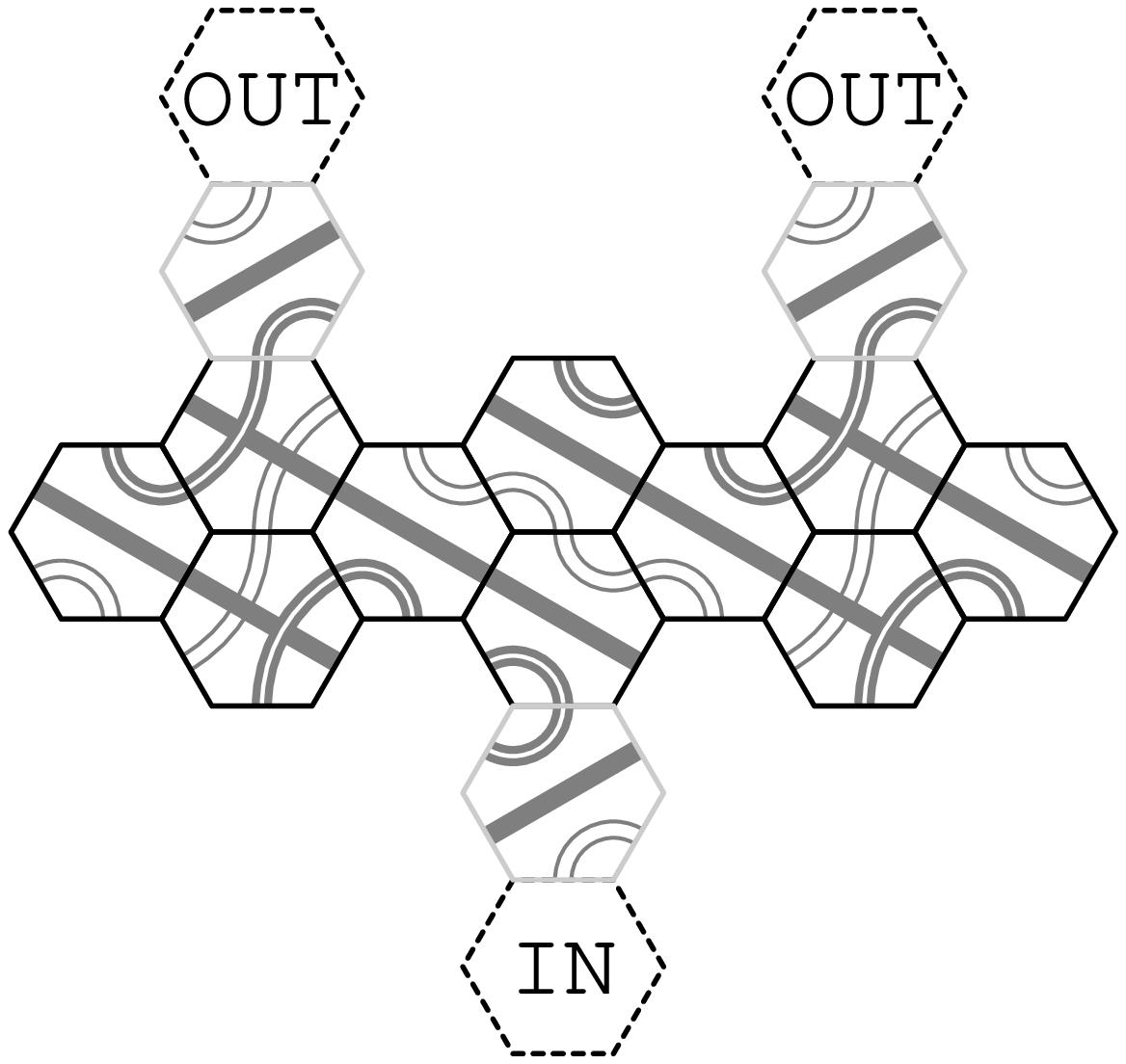}
    \quad
  }
  \caption{Original COPY subpuzzle, see~\cite{hol-hol:j:tantrix}}
  \label{fig:copy-4trp}
\end{figure}

\begin{figure}[h!]
  \centering
  \subfigure[In: \emph{true}]{
    \label{fig:copy-3trp-t}
    \quad 
    \includegraphics[height=4.2cm]{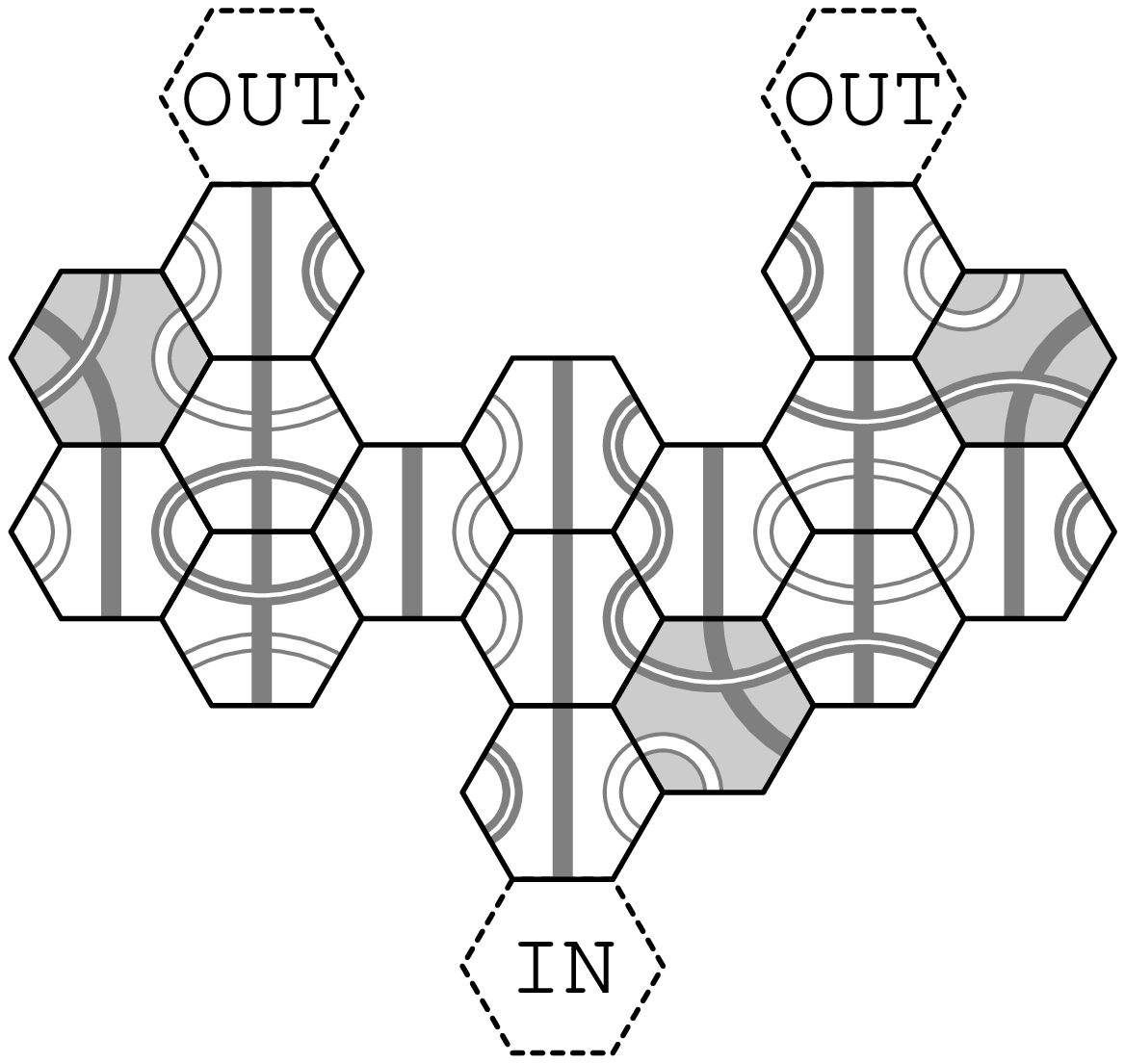}
    \quad 
  }
  \subfigure[In: \emph{false}]{
    \label{fig:copy-3trp-f}
    \quad 
    \includegraphics[height=4.2cm]{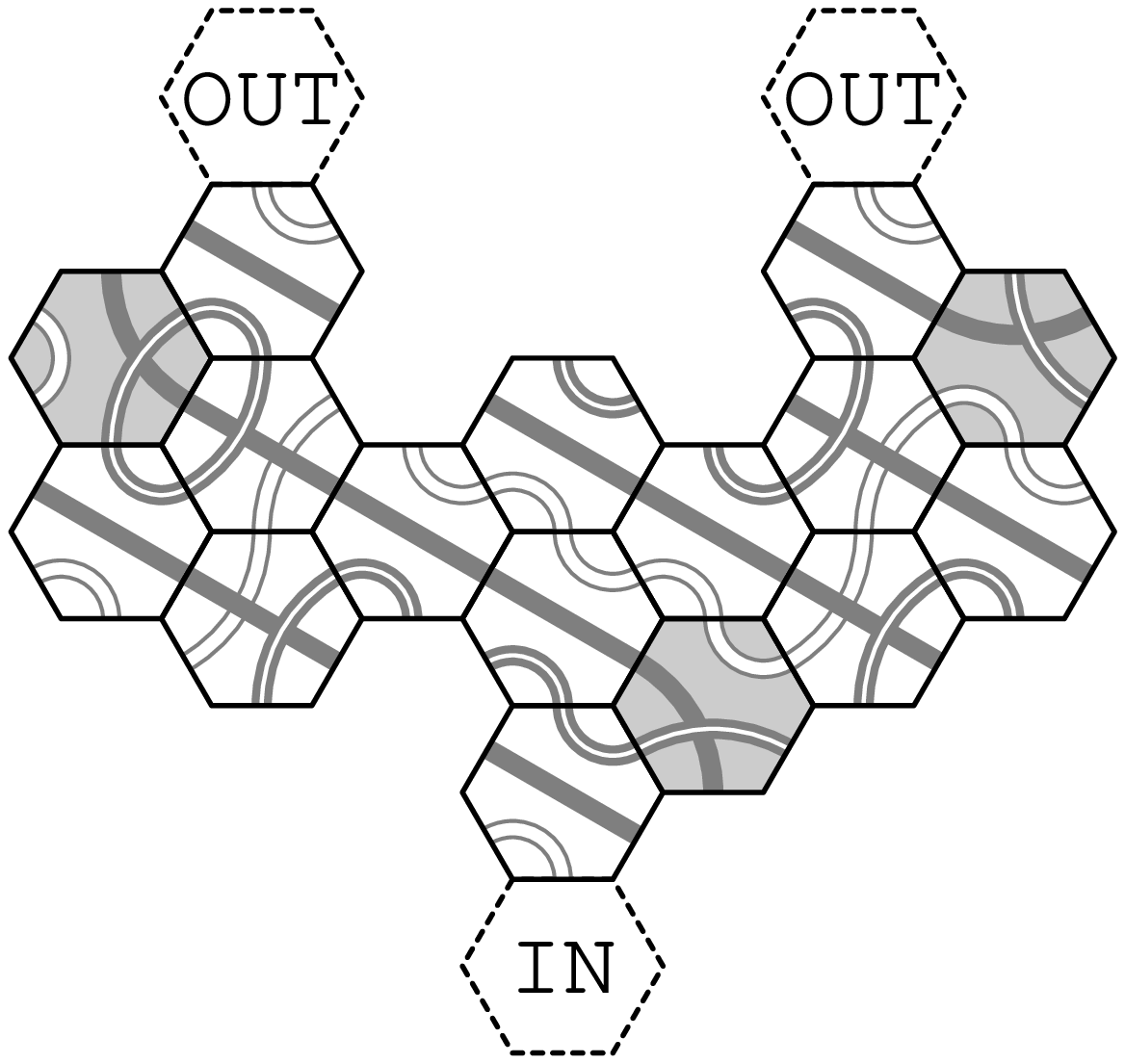}
    \quad 
  }
  \caption{Three-color COPY subpuzzle, see~\cite{bau-rot:c:tantrix}}
  \label{fig:copy-3trp}
\end{figure}

The last subpuzzle needed to simulate the wires of the circuit is our
new CROSS subpuzzle shown in Figure~\ref{fig:cross-3trp}. This
subpuzzle has two inputs and two outputs, and it ensures that
the input colors will
be swapped at the outputs.  This subpuzzle uses only three colors and
has unique solutions for each combination of input colors.

\begin{figure}[h!]
  \centering
  \subfigure[In: \emph{true}, \emph{true}]{
    \label{fig:cross-3trp-tt}
    \includegraphics[height=7cm]{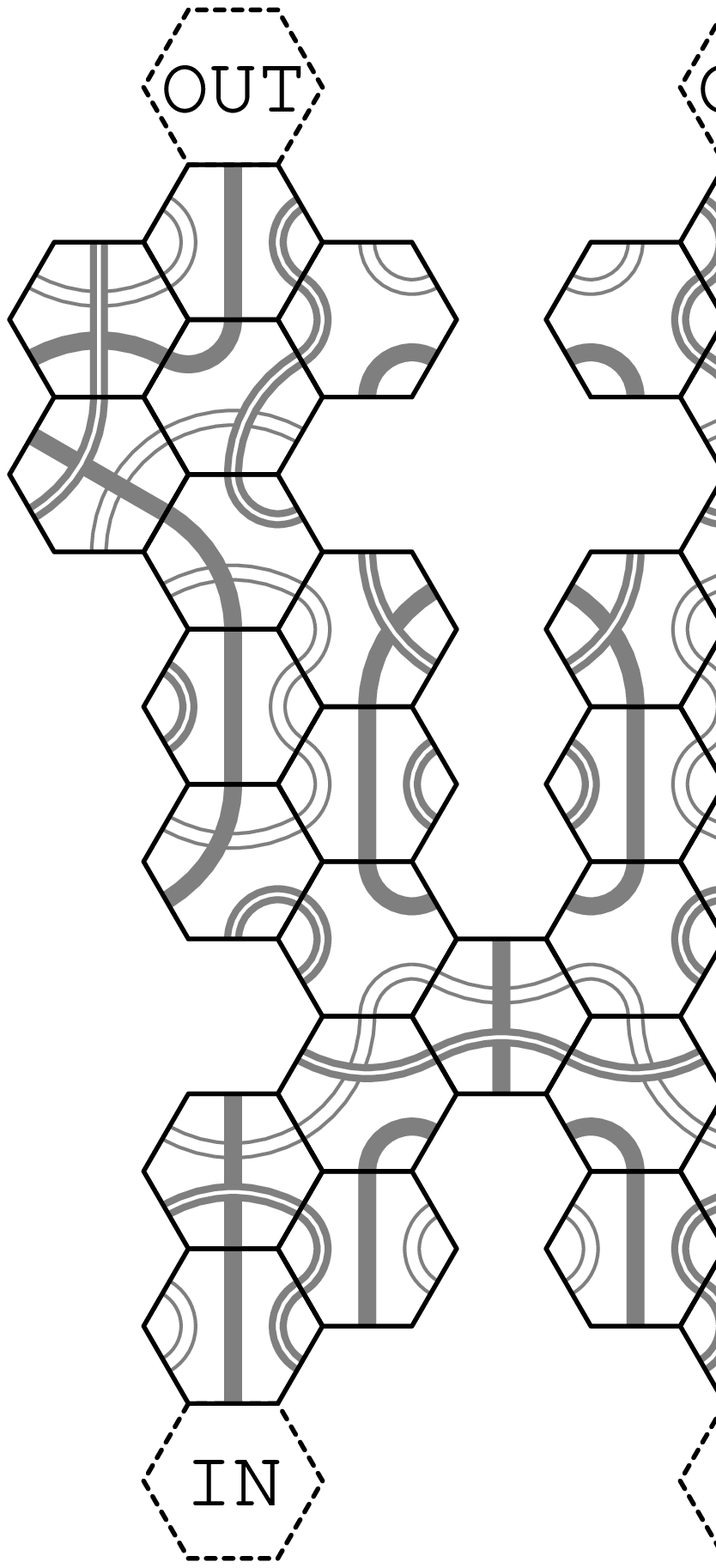}
  }
  \subfigure[In: \emph{true}, \emph{false}]{
    \label{fig:cross-3trp-tf}
    \includegraphics[height=7cm]{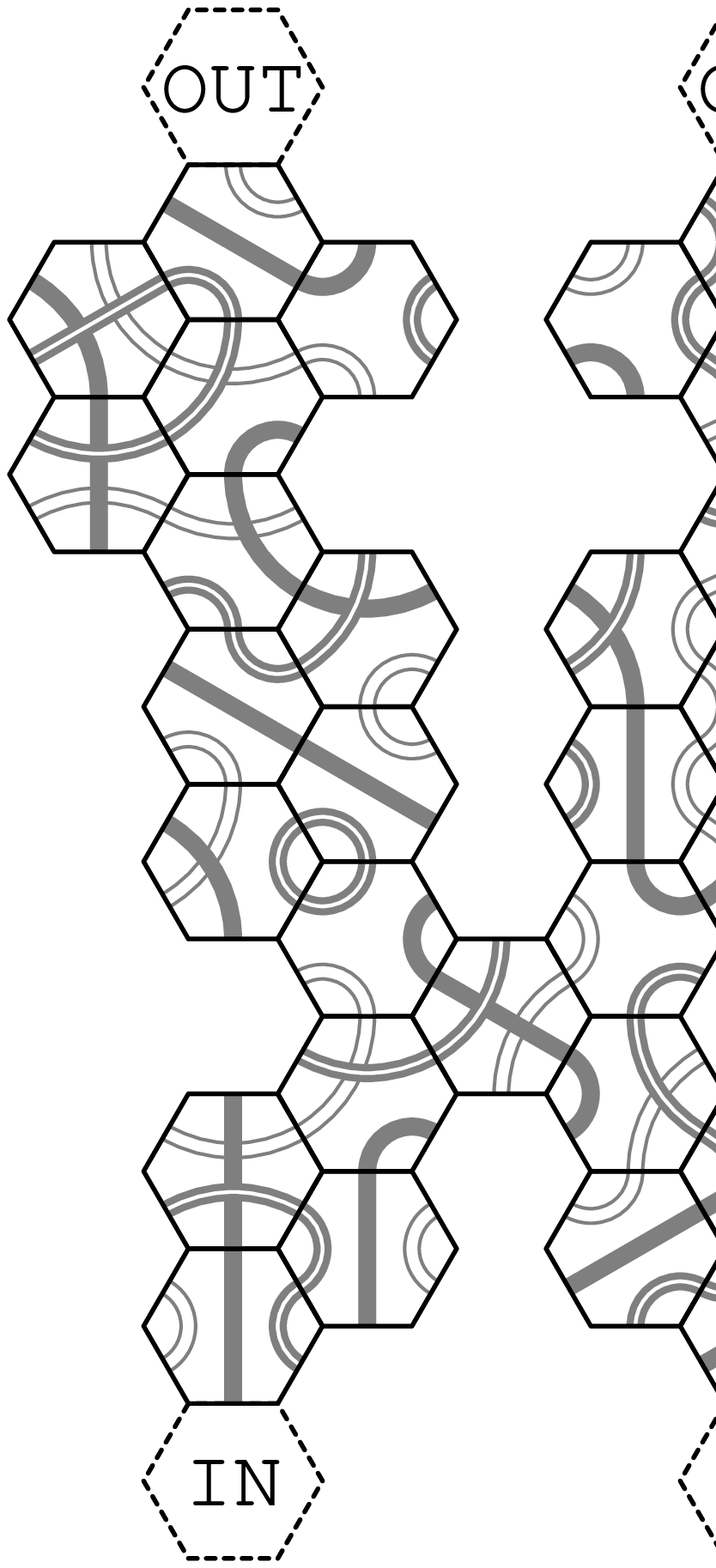}
  }
  \subfigure[In: \emph{false}, \emph{true}]{
    \label{fig:cross-3trp-ft}
    \includegraphics[height=7cm]{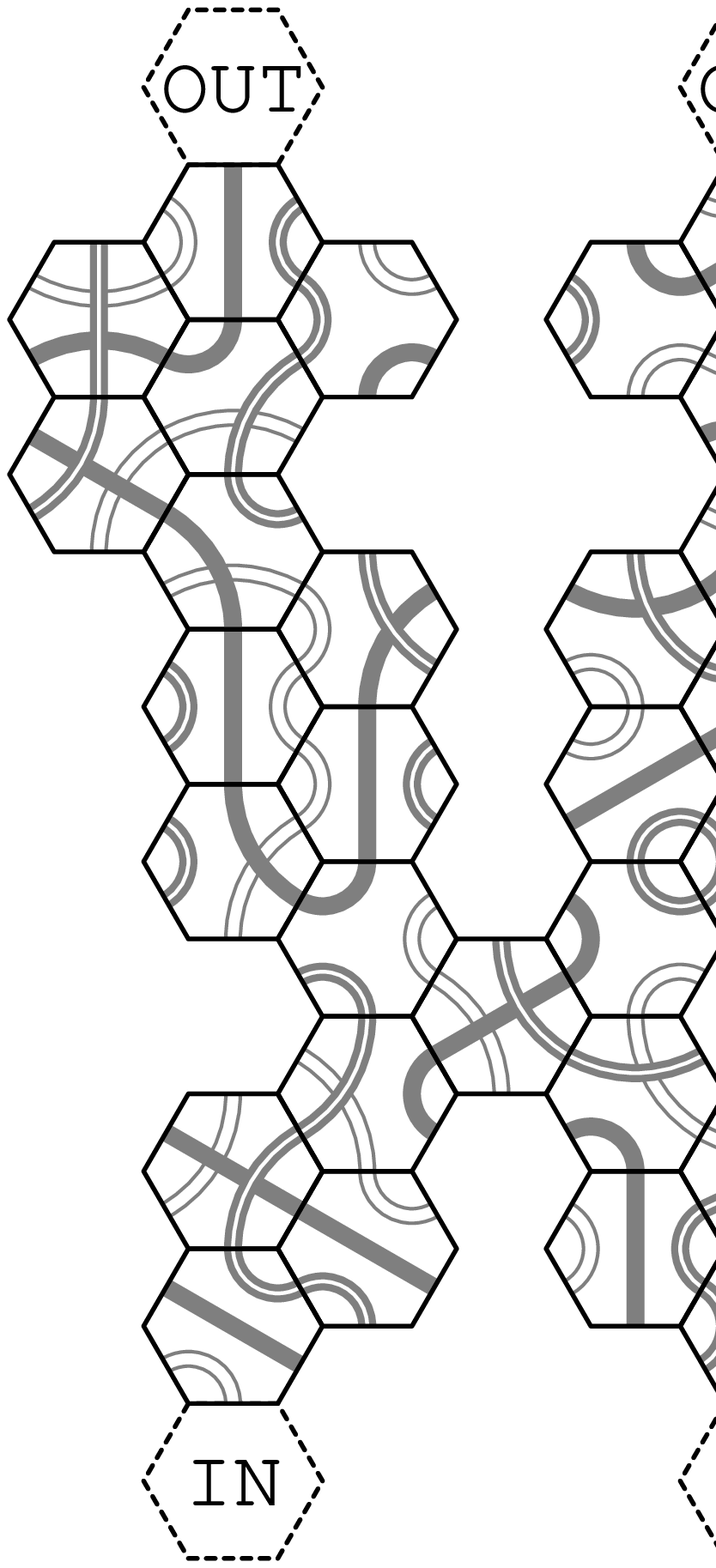}
  }
  \subfigure[In: \emph{false}, \emph{false}]{
    \label{fig:cross-3trp-ff}
    \quad
    \includegraphics[height=7cm]{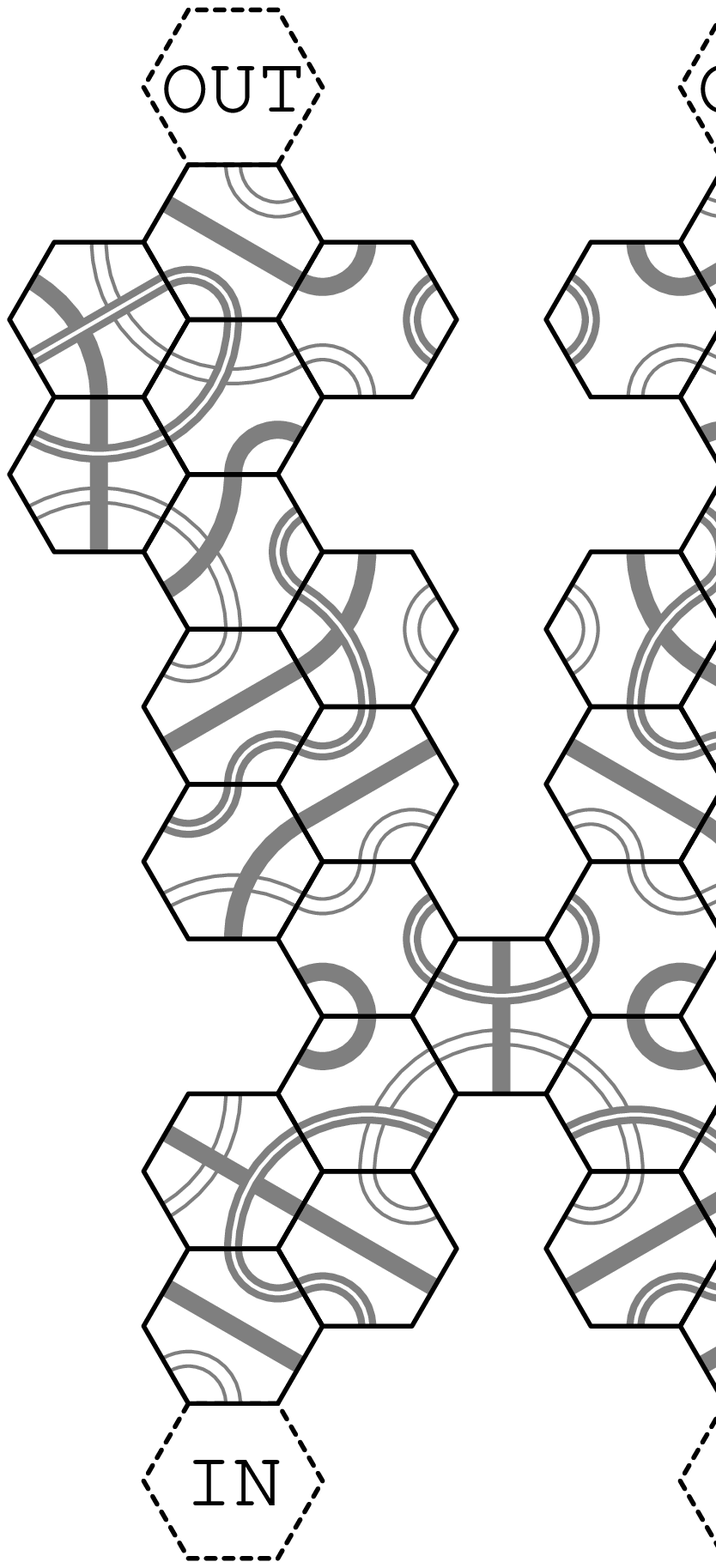}
    \quad
  }
  \subfigure[Scheme]{
    \label{fig:cross-3trp-s}
    \quad 
    \psfrag{l1}{$l_1$}
    \psfrag{m1}{$m_1$}
    \psfrag{n1}{$n_1$}
    \psfrag{o1}[cl][cl]{$o_1$}
    \psfrag{p1}[cl][cl]{$p_1$}
    \psfrag{q1}[cl][cl]{$q_1$}
    \psfrag{r1}{$r_1$}
    \psfrag{s1}{$s_1$}
    \psfrag{t1}{$t_1$}
    \psfrag{u1}{$u_1$}
    \psfrag{l2}{$l_2$}
    \psfrag{m2}{$m_2$}
    \psfrag{n2}{$n_2$}
    \psfrag{o2}[cl][cl]{$o_2$}
    \psfrag{p2}[cl][cl]{$p_2$}
    \psfrag{q2}[cl][cl]{$q_2$}
    \psfrag{r2}{$r_2$}
    \psfrag{s2}{$s_2$}
    \psfrag{t2}{$t_2$}
    \psfrag{u2}{$u_2$} 
    \includegraphics[height=7cm]{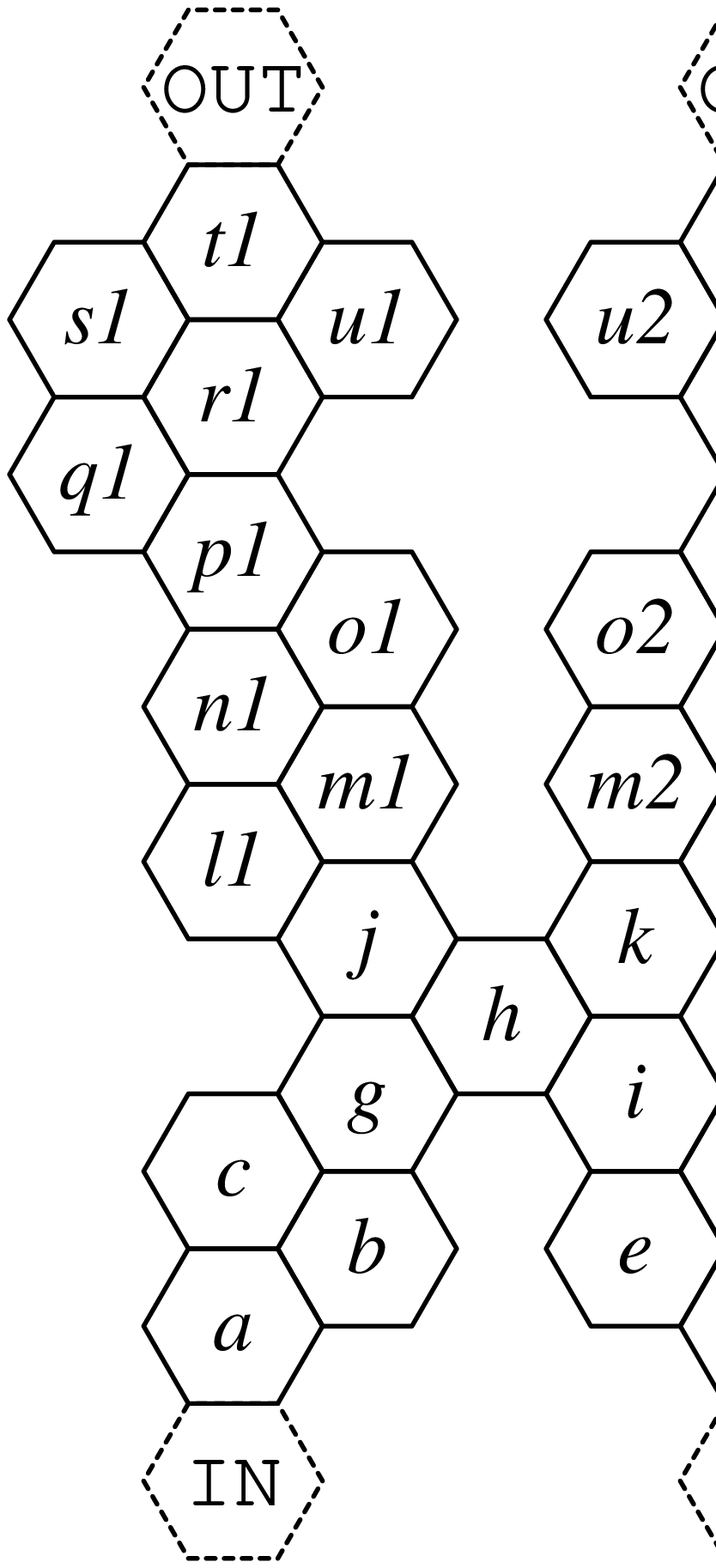}
    \quad 
  }
  \caption{CROSS subpuzzle}
  \label{fig:cross-3trp}
\end{figure}

The CROSS subpuzzle can be subdivided into three distinct parts: the
lower part consisting of tiles $a$ through~$k$, the upper left part
consisting of tiles $l_1$ through $u_1$,
and the upper right part consisting of tiles $l_2$ through~$u_2$.

Let us first consider the upper left part.  Consider
the three possible colors that can occur at the edge of
tile $j$ joint with tile~$m_1$.
\begin{description}
\item[{\bf Case~1:}]
Assume that the joint edge of these two tiles is \emph{blue}. One
possible orientation for tile $m_1$ has \emph{yellow} at the edge
joint with tile $l_1$. This leaves two possible orientations for tile
$l_1$. The first one has \emph{red} at the edge joint with tile $n_1$,
but $n_1$ does not contain the color sequence ${\tt yr}$. The second
possible orientation has \emph{yellow} at the edge joint with tile
$n_1$, but this leads to \emph{blue} at the edges of tiles $m_1$ and
$n_1$ with tile $o_1$. Since $o_1$ does not contain the color sequence
${\tt bb}$ this is not possible either.  The orientation of tile $m_1$
is now fixed with \emph{red} at the edge joint with tile $l_1$.

There
are two orientations of tile $l_1$, but they both have \emph{blue} at
the edge joint with tile $n_1$.  In the analysis of the lower part we
will see that both solutions are needed. The first one has
\emph{yellow} at the edge joint with tile $j$ and the second one has
\emph{blue} at this edge.  The orientation of tile $n_1$ is fixed with
\emph{red} and \emph{blue} at the edges joint with tiles $m_1$ and
$l_1$.  Tile $o_1$ has a fixed orientation due to the color-sequence substring
${\tt br}$ at the edges joint with tiles $m_1$ and $n_1$.  For tile
$p_1$ there are two orientations left, because this tile contains the
color-sequence substring ${\tt rb}$ for the edges joint with tiles $o_1$ and
$n_1$, twice.  The first one has \emph{red} at the edge joint with
tile $r_1$ and \emph{yellow} at the edge joint with tile $q_1$.  Thus,
it is not posibble that tile $r_1$ has \emph{yellow} at the edge joint
with tile $q_1$, since $q_1$ does not contain the color-sequence substring
${\tt yy}$.  Neither is it possible that $r_1$ has \emph{blue} at
the edge joint with tile $q_1$, because this leads to the
color-sequence substring ${\tt yr}$ at the edges of tiles $r_1$ and $s_1$ with
tile $t_1$.  So the orientation of tile $p_1$ is fixed with
\emph{blue} at the edge joint with tile $q_1$ and \emph{yellow} at the
edge joint with tile $r_1$.  Tile $r_1$ forces the edge joint with
tile $q_1$ to be \emph{red}, and since $s_1$ does not contain the
color sequence ${\tt yy}$, the orientation of tiles $r_1$ and $s_1$ is
fixed with \emph{blue} at their joint edge. This immediately fixes the
orientation of all other tiles, and the output color at the left output
tile will be \emph{blue}.

\item[{\bf Case~2:}]
Now we assume that the joint edge of tiles $j$ and $m_1$ is
\emph{red}. There are two possible orientations for tile $m_1$. The
first one has \emph{red} at the edge joint with tile $l_1$ and
\emph{blue} at the edge joint with tile $n_1$. This is not possible
because then the joint edge of tiles $l_1$ and $n_1$ would have to be
\emph{blue}, but tile $n_1$ does not contain the color-sequence substring ${\tt
bb}$. So the orientation of tile $m_1$ is fixed with \emph{blue} at
the edge joint with tile $l_1$ and \emph{yellow} at the edges joint
with tiles $n_1$ and $o_1$. Since $n_1$ does not contain the
color-sequence substring ${\tt yr}$, the orientation of tiles $l_1$ and $n_1$ is
fixed with \emph{yellow} at their joint edge. The joint edge of tiles
$o_1$ and $p_1$ cannot be \emph{red}, since $p_1$ does not contain the
color-sequence substring ${\tt rr}$ for the edges joint with tiles $o_1$ and
$n_1$, so the joint edge of tiles $o_1$ and $p_1$ is \emph{yellow},
and their orientation is fixed.

Now, there are two possible
orientations for tile $r_1$. The first one with \emph{yellow} at the
edge joint with tile $s_1$ is not possible, since this would lead to the
color-sequence substring ${\tt yb}$ for tile $u_1$ at the edges joint with
tiles $r_1$ and $t_1$. So we fix the orientation of tile $r_1$ with
\emph{yellow} at the edge joint with tile $q_1$. This also fixes the
orientation of tile $q_1$ with \emph{blue} at the edge joint with tile
$s_1$. The edges of tile $t_1$ joint with tiles $r_1$ and $s_1$ are
both \emph{yellow}, and the orientation of all other tiles is
fixed. The output of the subpuzzle's left output tile will thus 
be \emph{red}.

\item[{\bf Case~3:}]
The last possible color for the joint edge of tiles $j$ and
$m_1$ is \emph{yellow}. We first assume that the edge of tile $m_1$
joint with tile $l_1$ is \emph{blue}.

There are two possible
orientations for tile $l_1$.  The first one has \emph{yellow} at the
edge joint with tile $n_1$ and thus is not possible, since $n_1$ does not
contain the color-sequence substring ${\tt ry}$. The second one has \emph{red}
at the edge joint with tile $n_1$. Since the edge of tile $m_1$ joint
with tile $o_1$ is \emph{red}, this is not possible either, because
$o_1$ does not contain the color-sequence substring ${\tt rb}$. So the
orientation of tile $m_1$ is fixed with \emph{yellow} at the edge
joint with tile $l_1$. And since tile $j$ does not contain the 
color-sequence substring
${\tt by}$, the orientation of tile $l_1$ is fixed as well

The given colors at the edges of tiles $l_1$ and $m_1$ immediately fix
the orientation of tiles $n_1$ and $o_1$ with \emph{blue} and
\emph{yellow} at the edges joint with tile $p_1$, which contains the
color-sequence substring ${\tt by}$ only once and so has a fixed
orientation as well. Now we have the same situation as in the previous
case,
since the joint edge of tile $p_1$ with $r_1$ is \emph{blue} and the
joint edge of $p_1$ with tile $q_1$ is \emph{red}. As to color \emph{red} at
the joint edge of tiles $j$ and $m_1$ this case will also result in a
unique solution with the output color \emph{red} at the left output tile.
\end{description}

Due to symmetry
the upper right part can be handled analogously with the upper left part.
All \emph{Brid} and \emph{Chin} tiles are the same, and the \emph{Rond} is
replaced by the other \emph{Rond}, and the \emph{Sint} tiles are replaced
by the respective other \emph{Sint} tiles
having a small arc of the same color. So we
obtain a symmetrical subpuzzle and similar arguments as for the upper
left part apply.

We now analyze the lower part of this subpuzzle. We first consider
tiles $a$, $b$, and $c$. If the left input is \emph{blue} then there is
only one possible solution to these tiles. Obviously tiles $a$ and $c$
must have a vertical \emph{blue} line, and since tile $g$ does not
contain the color-sequence substring ${\tt by}$, the orientation of these three
tiles is fixed with \emph{yellow} at the edges of tiles $b$ joint with
tiles $c$ and $a$. The orientation of tile $g$ is fixed as well, since
it contains the color-sequence substring ${\tt br}$ only once.  If the input to
this part is \emph{red}, we have a fixed orientation with the
color-sequence substring ${\tt ry}$ for the edges joint with tile $g$ by similar
arguments. Note that tile $g$ has two possible solutions left.  Since
tiles $d$, $e$, and $f$ are the same as tiles $a$, $b$, and $c$, and
tile $i$ is a mirrored tile $g$, the same arguments hold for the right
input.  To analyze the whole lower part, we will distinguish the 
following four possible pairs of input colors:
\begin{itemize}

\item First we assume that both input colors are \emph{blue} (see
Figure~\ref{fig:cross-3trp-tt}). We have seen that the orientation of
tiles $g$ and $i$ is fixed with \emph{yellow} at their edges joint
with tile $h$, and \emph{red} at their edges joint with tiles $j$ and
$k$, respectively. The orientation of tile $h$ is fixed with
\emph{red} at the edges joint with tiles $j$ and $k$, and so they are
fixed with the color-sequence substring ${\tt by}$ for the edges joint with
tiles $l_1$ and $m_1$ and with the color-sequence substring ${\tt yb}$ for the
edges joint with tiles $m_2$ and $l_2$. In the analysis of the upper
part we have seen, that both output colors will be \emph{blue} in this case,
as desired.

\item Now, let the right input color be \emph{blue} and let the left
input color be \emph{red} (see Figure~\ref{fig:cross-3trp-ft}). The
two possible colors for tile $g$ joint with tile $h$ are \emph{blue}
and \emph{red}. The color for the joint edge of tiles $i$ and $h$ is
\emph{yellow}, and since $h$ contains the color-sequence substring ${\tt yxb}$
but not ${\tt yxr}$, where ${\tt x}$ stands for an arbitrary color
(chosen among \emph{blue}, \emph{red}, and \emph{yellow}),
the orientation of tiles $g$ and $h$ is fixed. This also fixes the
orientation of tiles $j$ and $k$. Tile $j$ has \emph{blue} at the
edges joint with tiles $l_1$ and $m_1$, and (as we have seen in the
analysis of the upper part) the left output color will be \emph{blue}, 
just like
the right input color. The edges of tile $k$ joint with tiles $m_2$ and
$l_2$ are \emph{yellow}, and so the right output color
will be \emph{red}, as desired.

\item The case of \emph{blue} being the left input color and
\emph{red} being the right input color (see
Figure~\ref{fig:cross-3trp-tf}) is similar to the second case. The
output colors will again be the exchanged input colors, as desired.

\item The last case is that both input colors are \emph{red} (see
Figure~\ref{fig:cross-3trp-ff}). We have seen that the two possible
colors for tiles $g$ and $i$ joint with tile $h$ are \emph{blue} and
\emph{red}. Obviously, they cannot both be \emph{blue}. If the joint
edge of tiles $g$ and $h$ is \emph{blue}, the joint edges of tiles $g$
and $h$ with $j$ are both \emph{yellow}. This is not possible, because
the combination of \emph{blue} at the joint edge of tiles $j$ and
$l_1$ and \emph{red} at the joint edge of tiles $j$ and $m_1$ is not
possible. The case of \emph{blue} at the edge of tile $i$ joint with
tile $h$ is not possible due to similar arguments for tile $k$ and the
upper right part. So the edges of tiles $g$ and $i$ joint with tile
$h$ must both be \emph{red}. This leads to \emph{red} at the edges of
tile $j$ joint with the upper left part, and tile $k$ joint with the
upper right part. We have already seen that this combination leads to
both output colors being \emph{red}, as desired.
\end{itemize}

So we have unique solutions with the desired effect of exchanging the
input colors at the output tiles for all four possible combinations of
input colors for the CROSS subpuzzle.

\paragraph{Gate subpuzzles:}
The boolean gates AND and NOT are represented by the
AND and NOT subpuzzles.  Both the original four-color NOT subpuzzle
from~\cite{hol-hol:j:tantrix} (see Figure~\ref{fig:not-4trp})
and the modified four-color NOT subpuzzle
from~\cite{bau-rot:c:tantrix}, which is not displayed here,
use tiles with \emph{green} lines to exclude certain rotations.
Our three-color NOT subpuzzle is shown in
Figure~\ref{fig:not-3trp}.  Tiles $a$, $b$, $c$, and $d$
from the original NOT subpuzzle shown in Figure~\ref{fig:not-4trp}
remain unchanged.
Tiles $e$, $f$, and $g$ in this original NOT subpuzzle ensure that 
the output color will be correct, since the joint edge of $e$  
and $b$ is always \emph{red}.
So for our new NOT subpuzzle in
Figure~\ref{fig:not-3trp}, we have to show that the edge between 
tiles $x$ and $b$ is always \emph{red}, and that we have unique solutions 
for both input colors.

First, let the input color be \emph{blue} and suppose for a
contradiction that the joint edge of tiles 
$b$ and $x$ were \emph{blue}.  Then the joint edge of tiles $b$ and
$c$ would be \emph{yellow}.  Since $x$ is a tile of type $t_{13}$ and
so does not contain the color-sequence substring 
substring ${\tt bb}$, the edge between tiles $c$ and $x$
must be \emph{yellow}.
But then the edges of tile $w$ joint with tiles $c$ and $x$ must 
both be \emph{blue}.  This is not possible, however, because $w$
(which is of type~$t_{10}$) does not 
contain the color-sequence substring substring ${\tt bb}$.
So if the input color is \emph{blue}, the orientation of tile $b$ is 
fixed with \emph{yellow} at the edge of 
$b$ joint with tile $y$, and with \emph{red} 
at the edges of 
$b$ joint with tiles $c$ and~$x$. This already ensures that 
the output color will be \emph{red}, because tiles $c$ and $d$ behave 
like a WIRE subpuzzle. Tile $x$ does not contain the color-sequence substring
${\tt br}$, so the orientation of tile $c$ is also fixed with \emph{blue} 
at the joint edge of tiles $c$ and~$w$.
As a consequence, the joint edge of tiles $w$
and $d$ is \emph{yellow}, and due to the fact that the 
joint edge of tiles $w$ and $x$ is also \emph{yellow}, the 
orientation of $w$ and $d$ is fixed as well.  Regarding tile~$a$, 
the edge joint with tile $y$ can be \emph{yellow} or \emph{red}, but tile 
$x$ has \emph{blue} at the edge joint with tile~$y$, so the joint edge of
tiles $y$ and $a$ is \emph{yellow}, and the orientation of all 
tiles is fixed for the input color \emph{blue}.
The case of \emph{red} being the input color can be handled analogously.

\begin{figure}[h]
  \centering
  \subfigure[In: \emph{true}]{
    \label{fig:not-4trp-t}
    \quad
    \includegraphics[height=4.2cm]{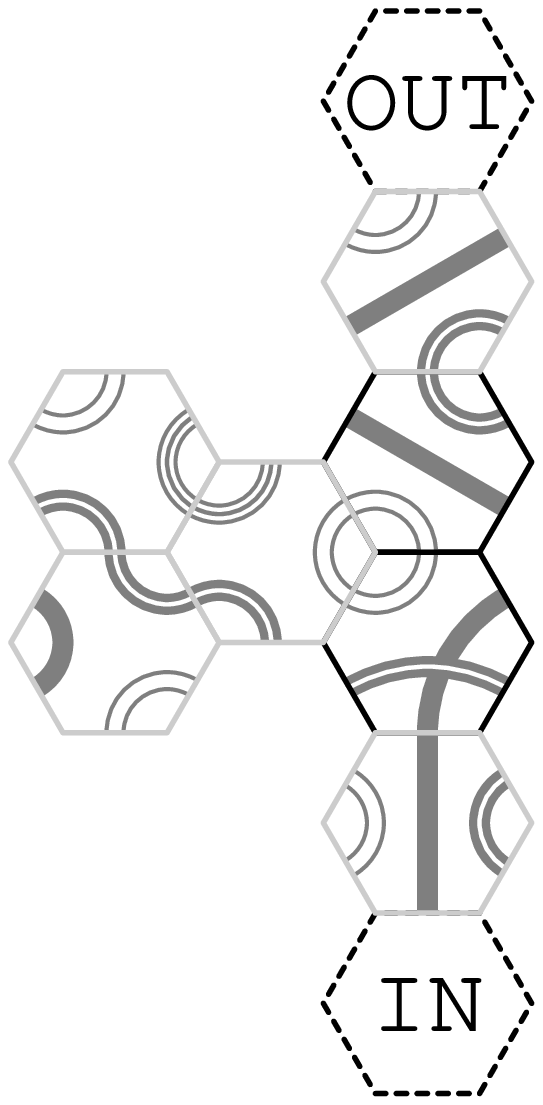}
    \quad
  }
    \subfigure[In: \emph{false}]{
    \label{fig:not-4trp-f}
    \quad
    \includegraphics[height=4.2cm]{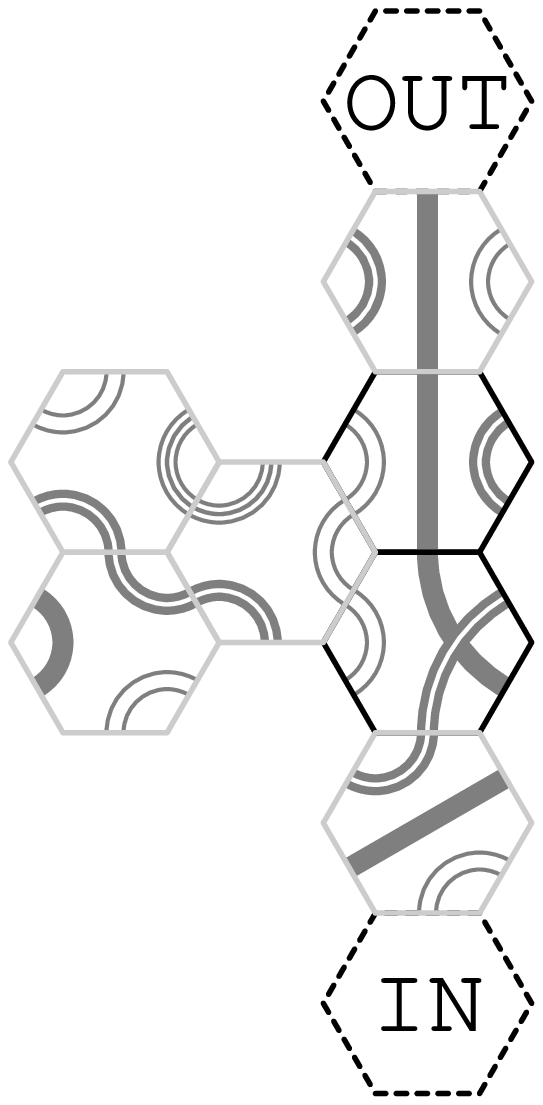}
    \quad
  }
    \subfigure[Scheme]{
    \label{fig:not-4trp-s}
    \quad
    \includegraphics[height=4.2cm]{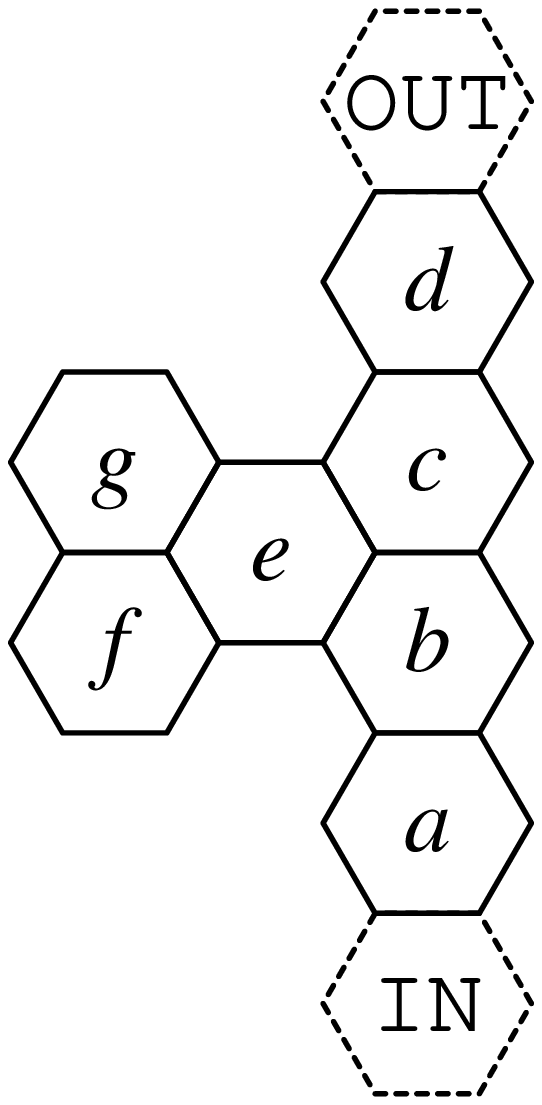}
    \quad
  }
  \caption{Original NOT subpuzzle, see~\cite{hol-hol:j:tantrix}}
  \label{fig:not-4trp}
\end{figure}

\begin{figure}[h!]
  \centering
  \subfigure[In: \emph{true}]{
    \label{fig:not-3trp-t}
    \quad 
    \includegraphics[height=4.2cm]{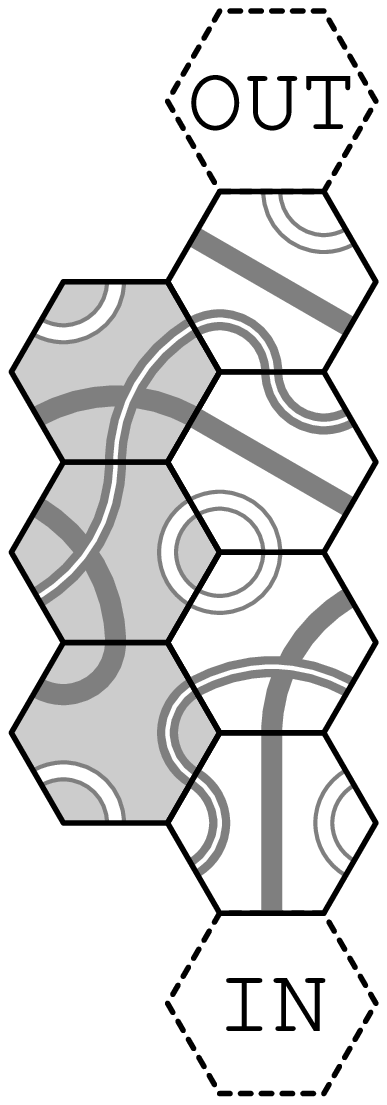}
    \quad 
  }
  \subfigure[In: \emph{false}]{
    \label{fig:not-3trp-f}
    \quad 
    \includegraphics[height=4.2cm]{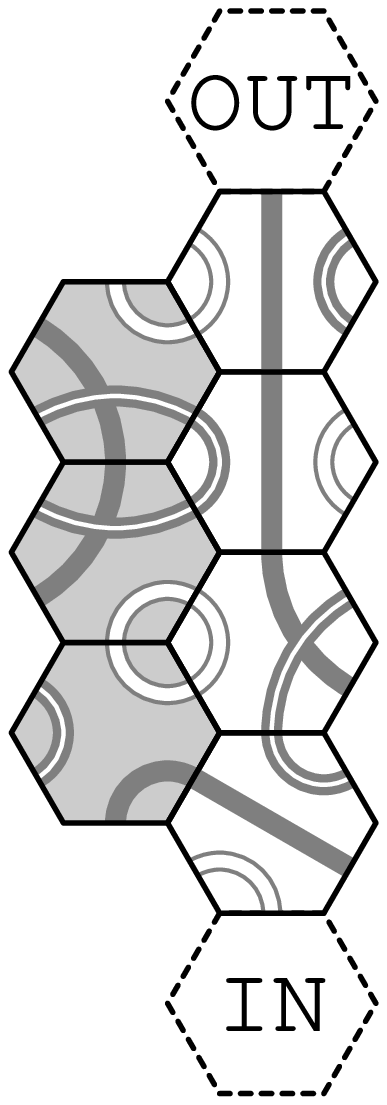}
    \quad 
  }
  \subfigure[Scheme]{
    \label{fig:not-3trp-s}
    \quad 
    \includegraphics[height=4.2cm]{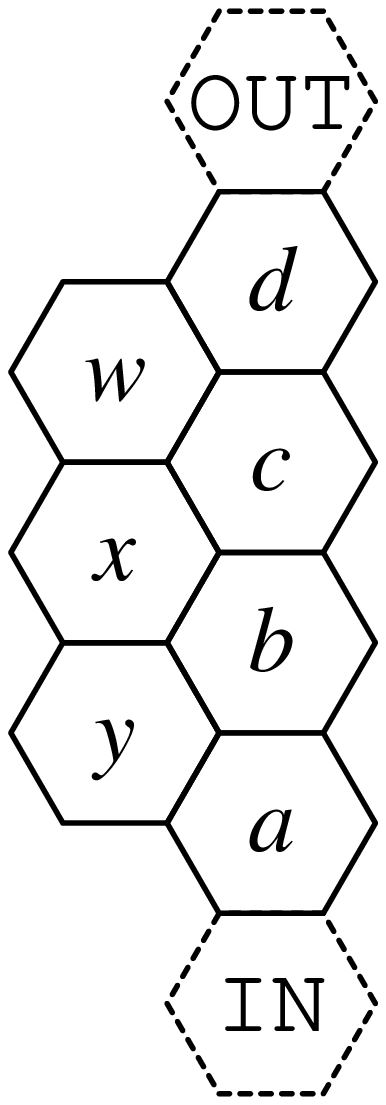}
    \quad 
  }
  \caption{Three-color NOT subpuzzle}
  \label{fig:not-3trp}
\end{figure}

The most complicated figure (besides the CROSS)
is the AND subpuzzle.  The original
four-color version from~\cite{hol-hol:j:tantrix}
(see Figure~\ref{fig:and-4trp}) 
uses four tiles with \emph{green} lines
and the modified four-color AND subpuzzle
from~\cite{bau-rot:c:tantrix}, which is not displayed here, uses
seven tiles with \emph{green} lines.  Figure~\ref{fig:and-3trp} shows 
our new AND subpuzzle using only three colors and having
unique solutions for all four possible combinations of input colors.
To analyze this subpuzzle, we subdivide it into a lower and an upper
part.  The lower part ends with tile $c$ and has four possible
solutions (one for each combination of input colors),
while the upper part, which begins with tile~$j$, has only
two possible solutions (one for each possible output color).
The lower part can again be subdivided into three different parts.

\begin{figure}
  \centering
  \subfigure[In: \emph{true, true}]{
    \label{fig:and-4trp-tt}
    \quad
    \includegraphics[height=4.8cm]{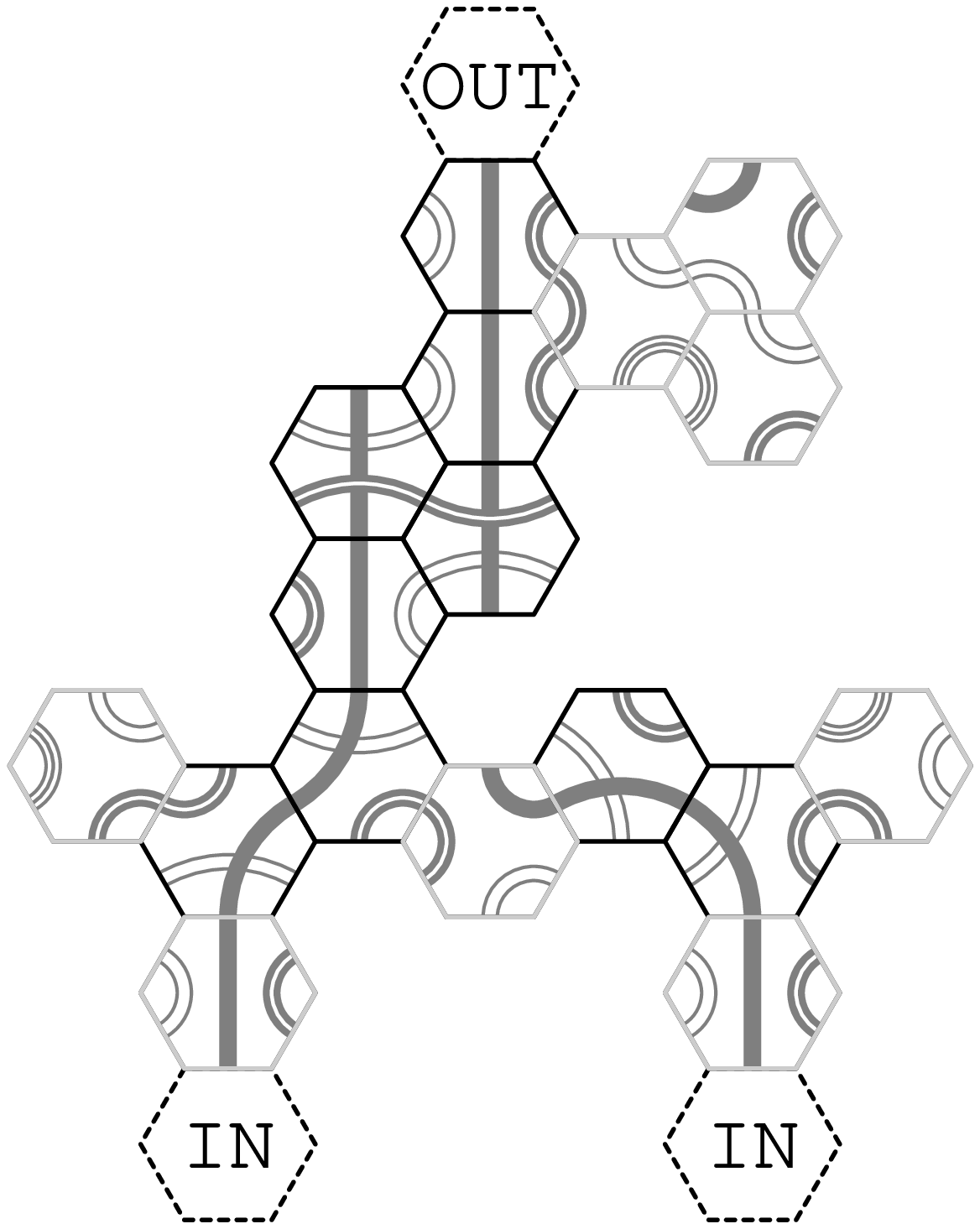}
    \quad
  }
  \subfigure[In: \emph{true, false}]{
    \label{fig:and-4trp-tf}
    \quad
    \includegraphics[height=4.8cm]{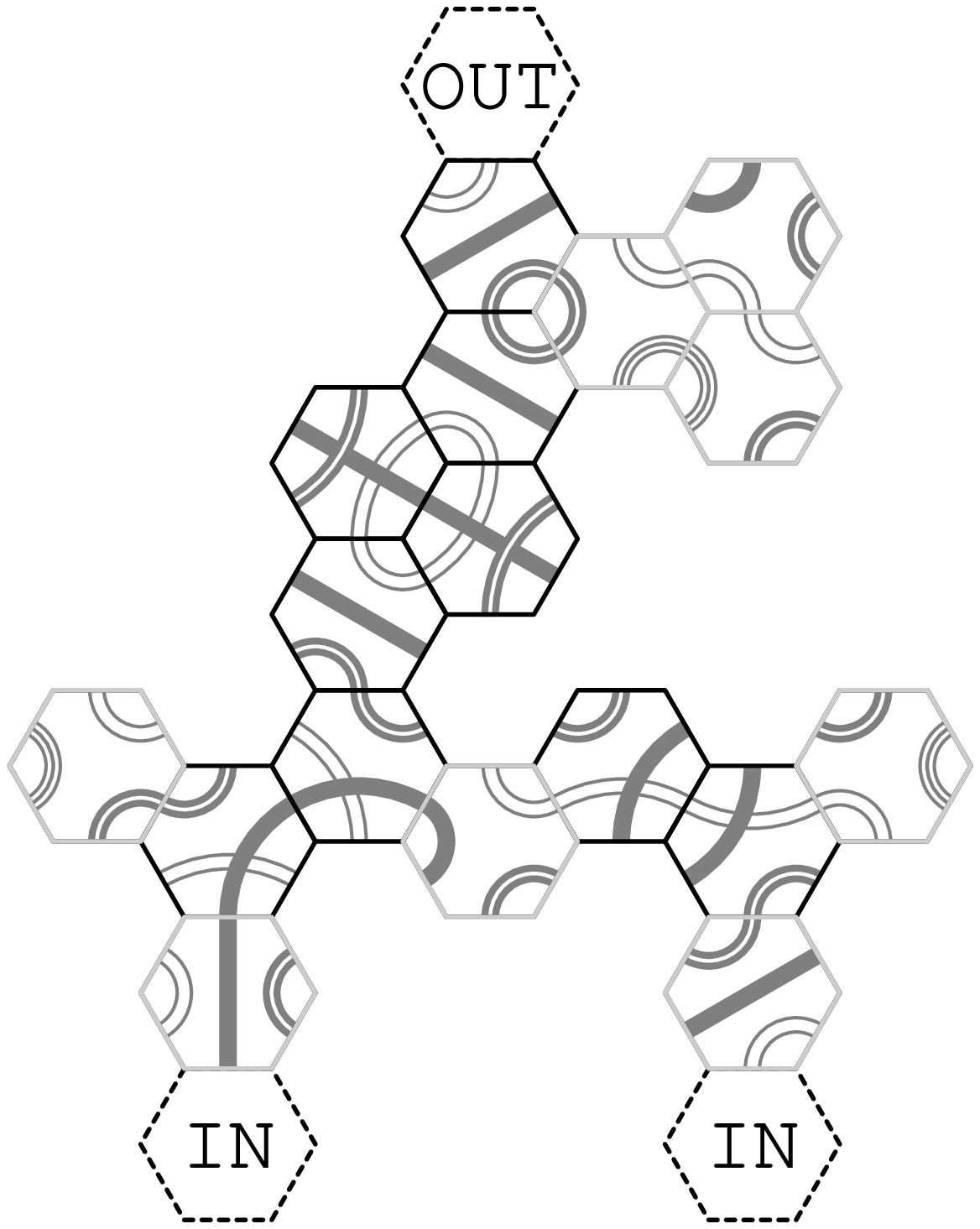}
    \quad
  }
  \subfigure[In: \emph{false, true}]{
    \label{fig:and-4trp-ft}
    \quad
    \includegraphics[height=4.8cm]{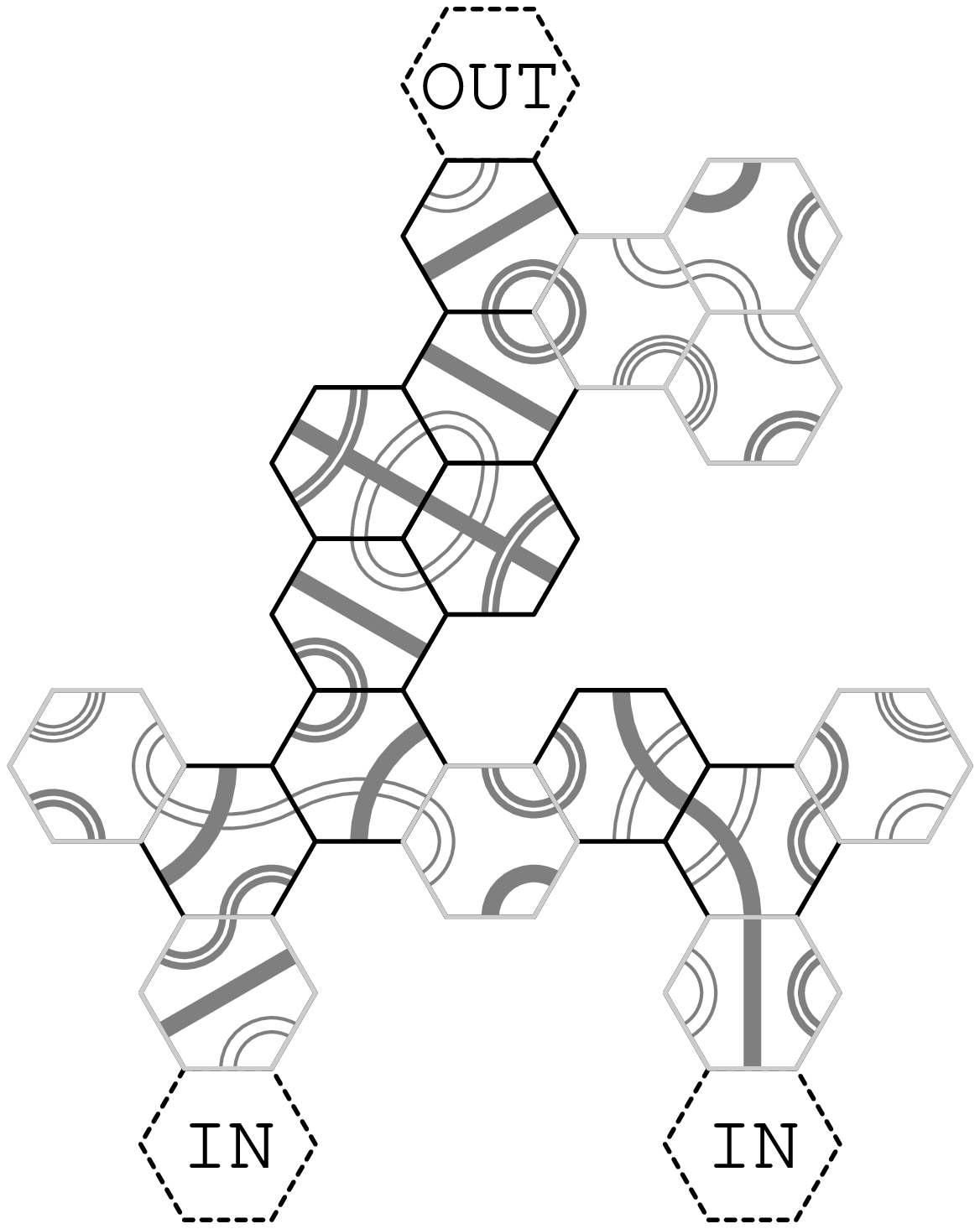}
    \quad
  }
  \subfigure[In: \emph{false, false}]{
    \label{fig:and-4trp-ff}
    \quad
    \includegraphics[height=4.8cm]{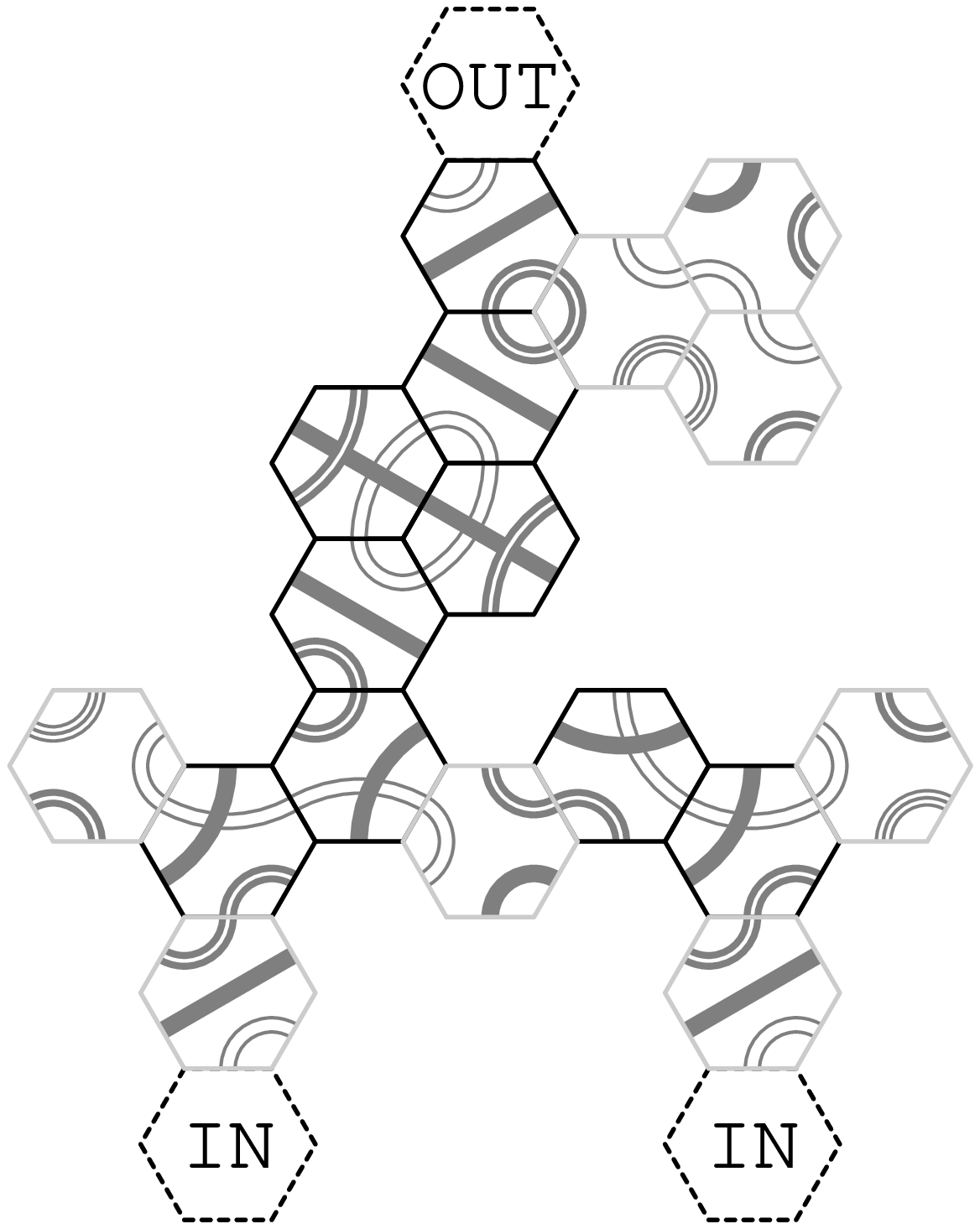}
    \quad
  }
  \subfigure[Scheme]{
    \label{fig:and-4trp-s}
    \quad
    \includegraphics[height=4.8cm]{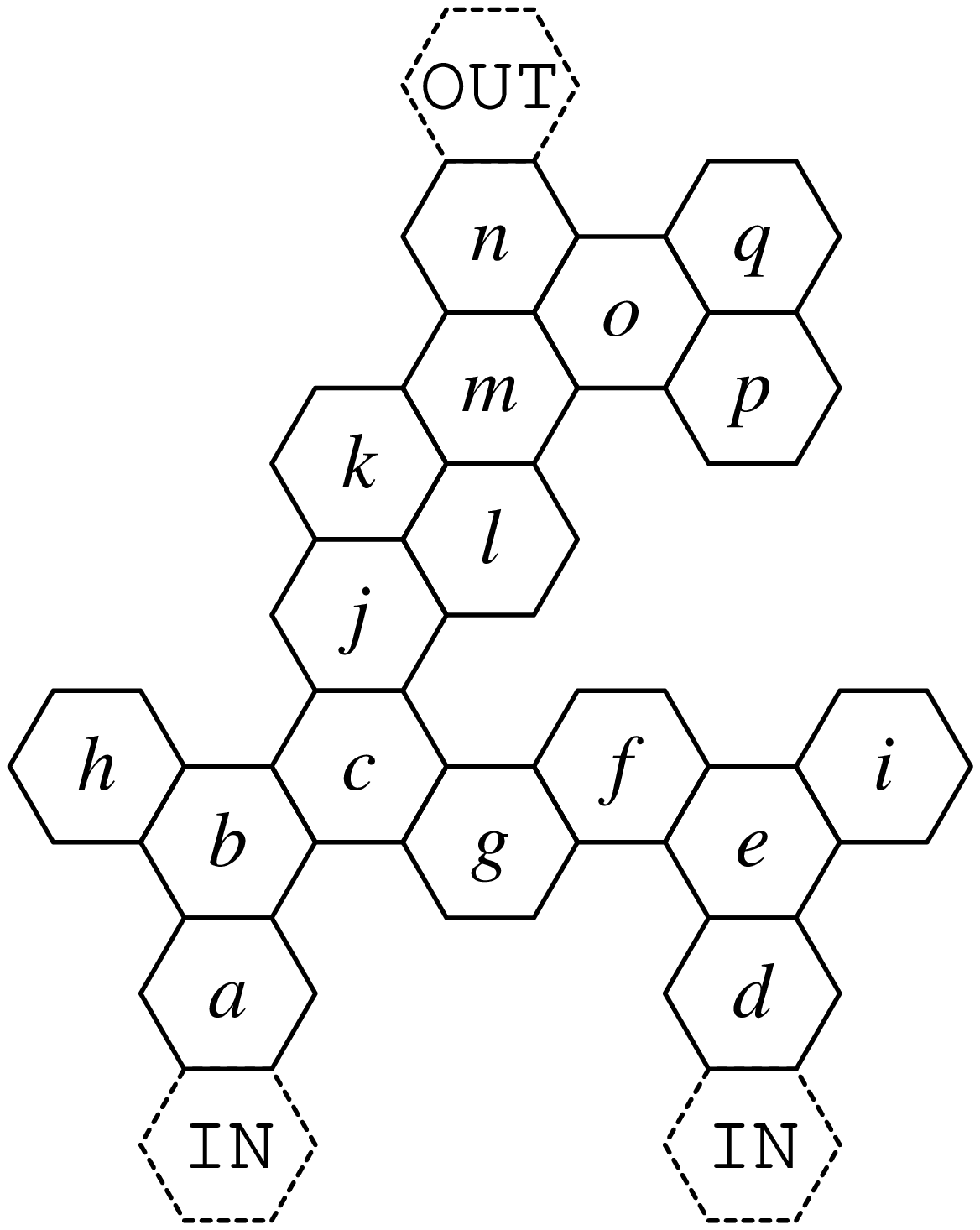}
    \quad
  }
  \caption{Original AND subpuzzle, see~\cite{hol-hol:j:tantrix}}
  \label{fig:and-4trp}
\end{figure}

\begin{figure}[h!]
  \centering
  \subfigure[In: \emph{true, true}]{
    \label{fig:and-3trp-tt}
    \quad 
    \includegraphics[height=4.8cm]{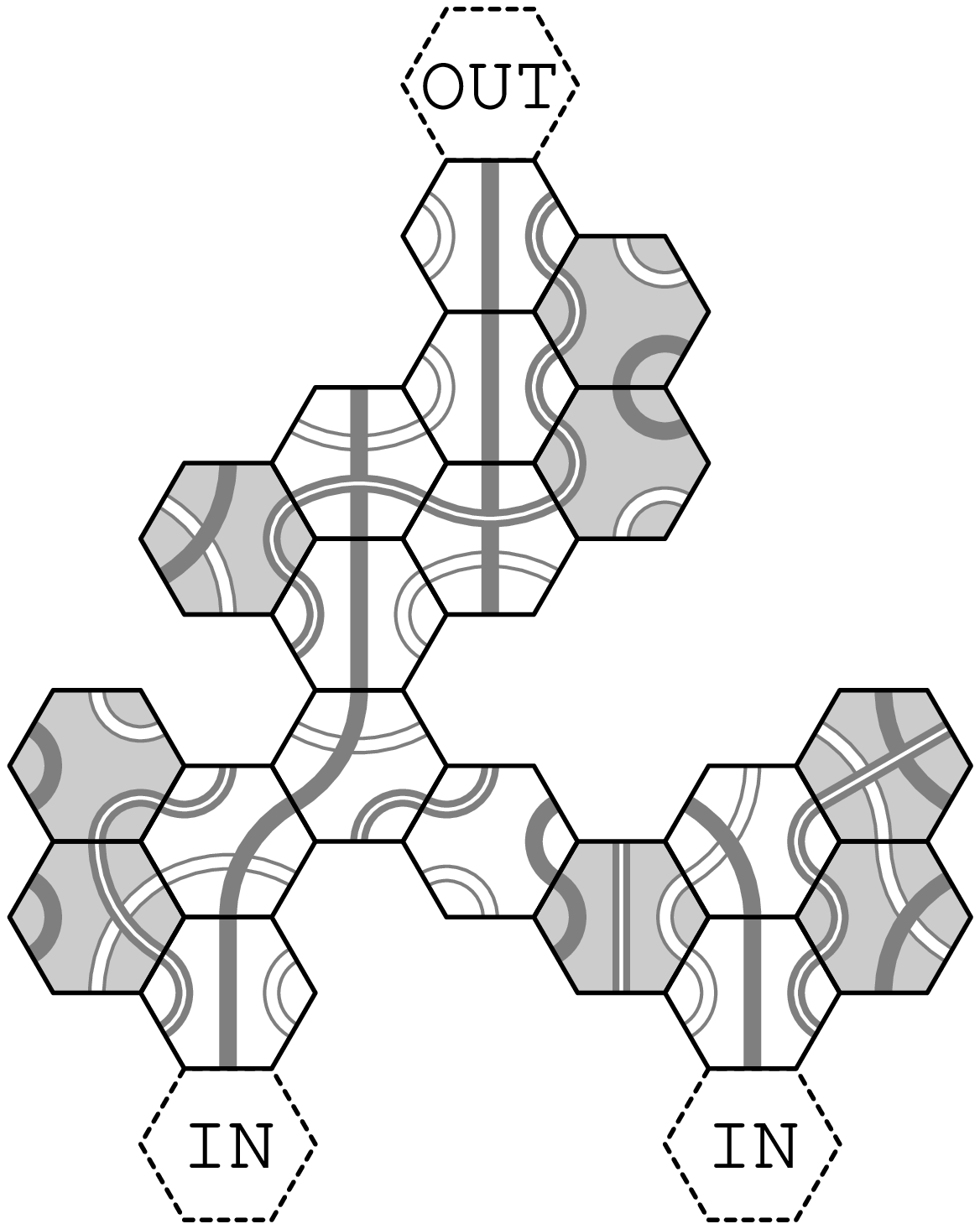}
    \quad 
  }
  \subfigure[In: \emph{true, false}]{
    \label{fig:and-3trp-tf}
    \quad 
    \includegraphics[height=4.8cm]{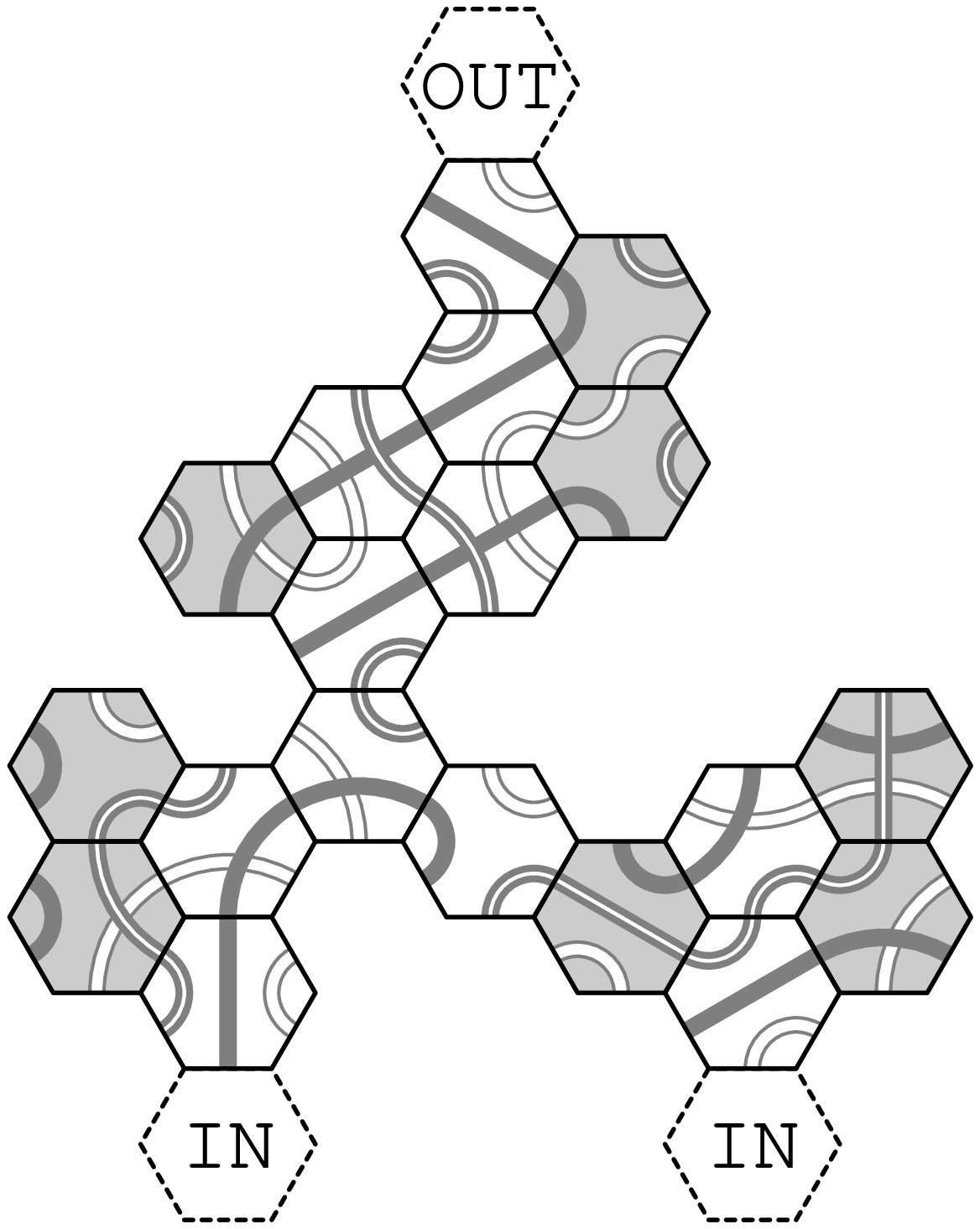}
    \quad 
  }
  \subfigure[In: \emph{false, true}]{
    \label{fig:and-3trp-ft}
    \quad 
    \includegraphics[height=4.8cm]{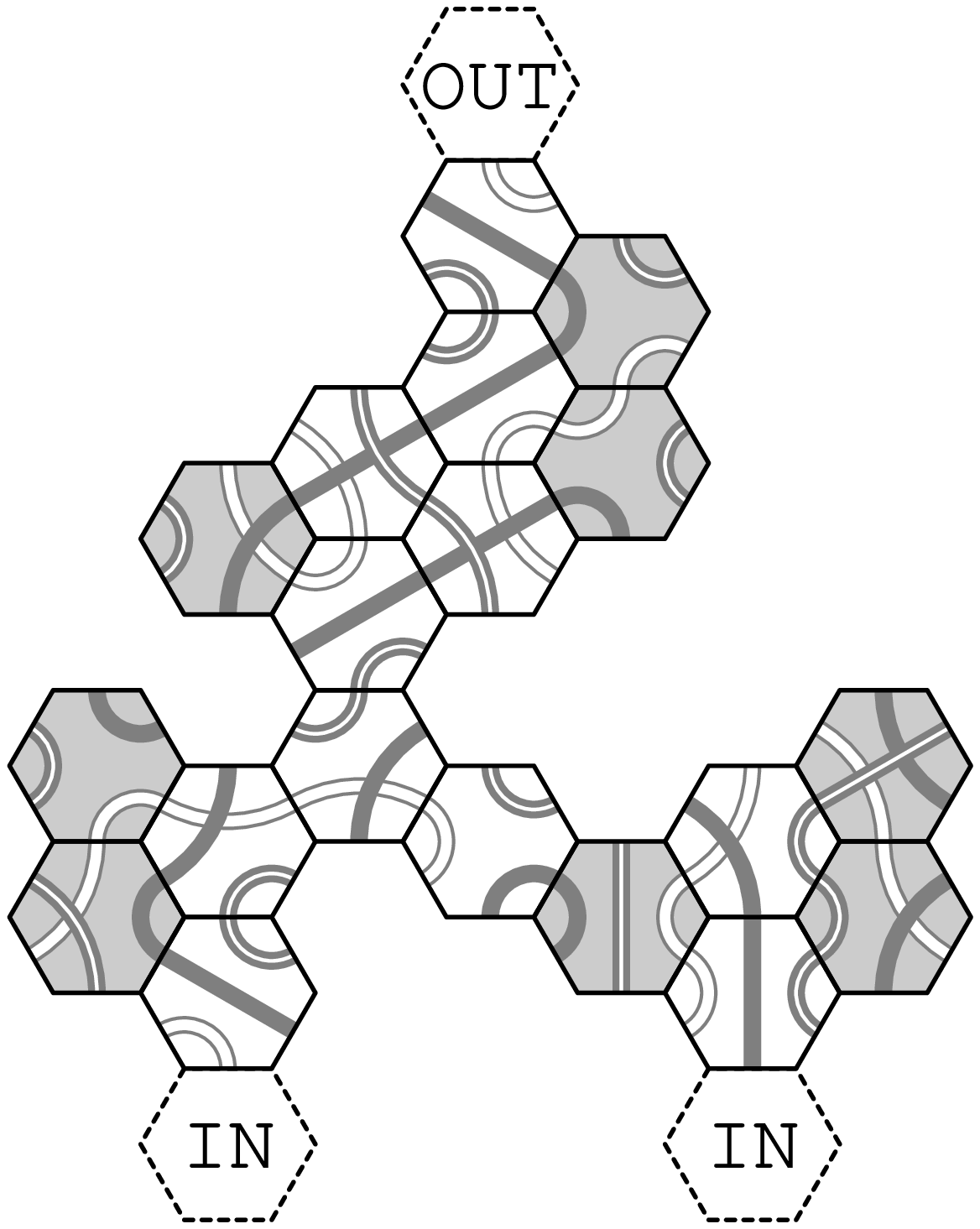}
    \quad 
  }
  \subfigure[In: \emph{false, false}]{
    \label{fig:and-3trp-ff}
    \quad 
    \includegraphics[height=4.8cm]{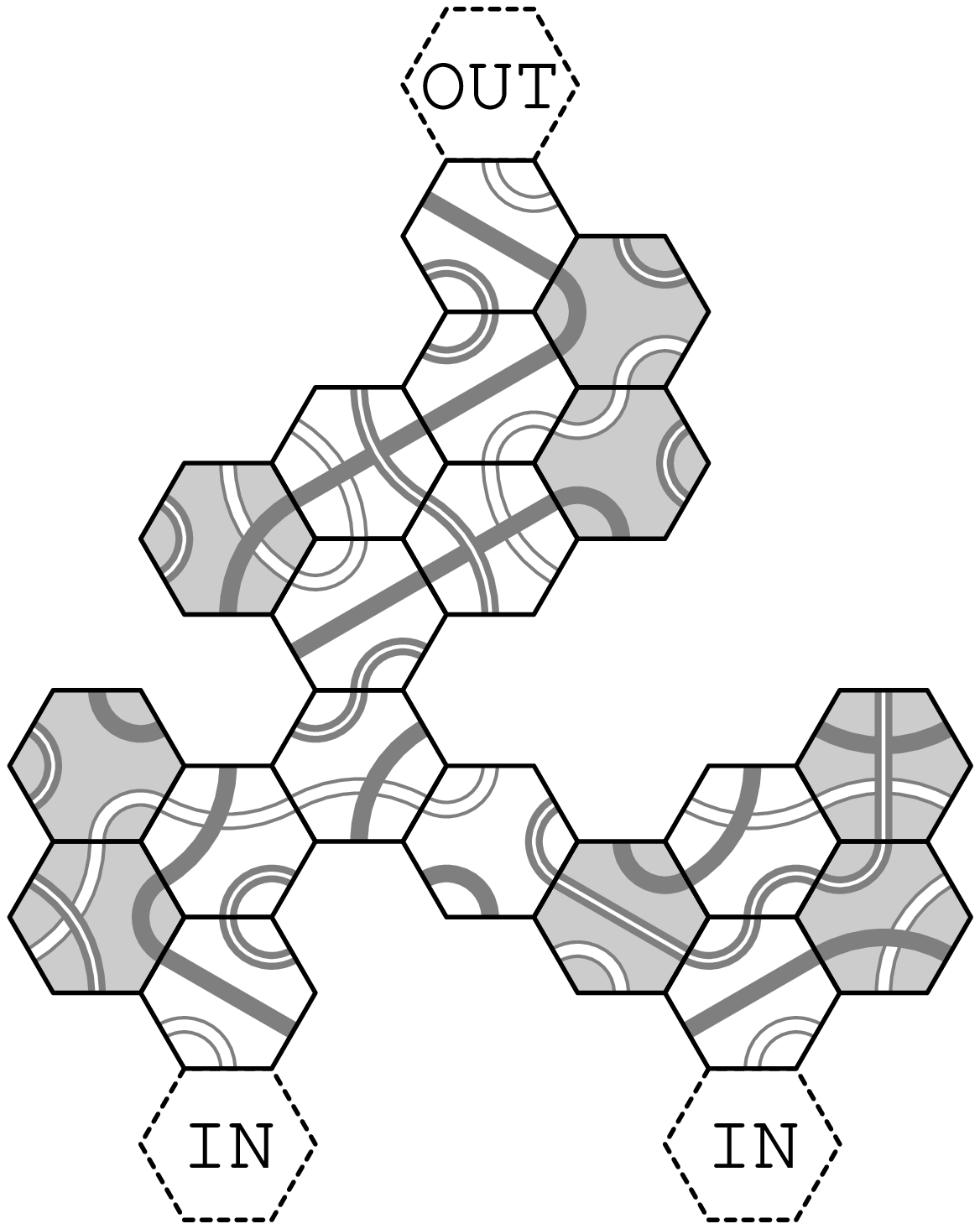}
    \quad 
  }
  \subfigure[Scheme]{
    \label{fig:and-3trp-s}
    \quad 
    \includegraphics[height=4.8cm]{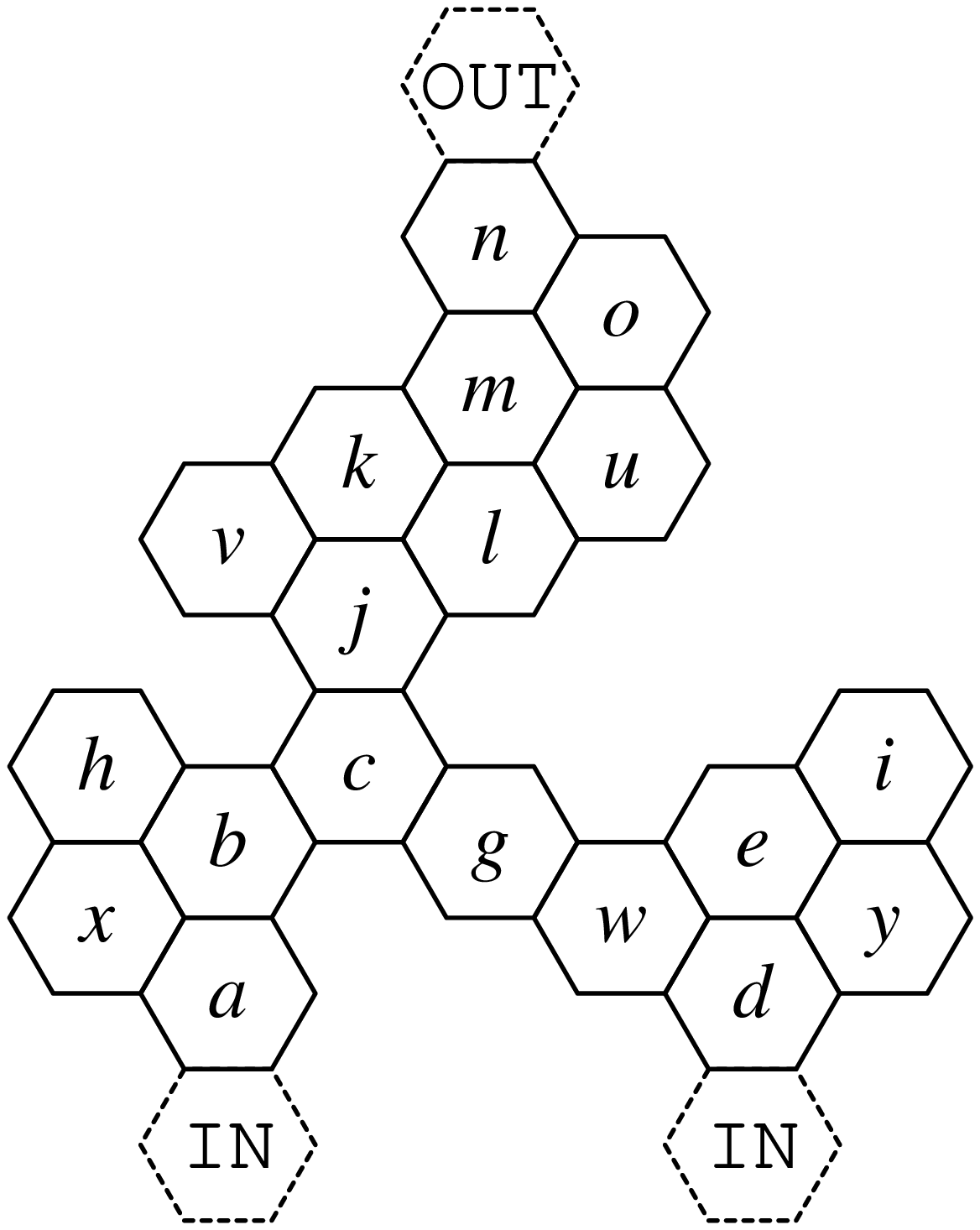}
    \quad 
  }
  \caption{Three-color AND subpuzzle}
  \label{fig:and-3trp}
\end{figure}

The lower left part contains the tiles $a$, $b$, $x$, 
and~$h$. If the input color to this part is \emph{blue} (see 
Figures~\ref{fig:and-3trp-tt} and~\ref{fig:and-3trp-tf}),
the joint edge of tiles $b$ and $x$ is always \emph{red}, and since tile $x$
(which is of type~$t_{11}$) does not contain the 
color-sequence substring ${\tt rr}$,
the orientation of tiles $a$ and $x$ is fixed. 
The orientation of tiles $b$ and $h$ is also fixed, since $h$
(which is of type~$t_2$) does not contain the color-sequence
substring ${\tt by}$ but the color-sequence substring ${\tt yy}$ 
for the edges joint with tiles $b$ and $x$. By similar arguments we 
obtain a unique solution for these tiles if the left input color is \emph{red} 
(see Figures~\ref{fig:and-3trp-ft} and~\ref{fig:and-3trp-ff}).
The connecting edge to the rest of the subpuzzle is the joint edge
between tiles $b$ and~$c$, and tile $b$ will have the same color
at this edge as the left input color.

Tiles $d$, $e$,
$i$, $w$, and~$y$ form the lower right part.
If the input color to this part is \emph{blue} (see
Figures~\ref{fig:and-3trp-tt} and~\ref{fig:and-3trp-ft}),
the joint edge of tiles $d$ and $y$ must be \emph{yellow}, 
since tile $y$ (which is of type~$t_9$)
does not contain the color-sequence substrings
${\tt rr}$ nor ${\tt ry}$ 
for the edges joint with tiles $d$ and~$e$.  Thus the joint edge of tiles $y$  
and $e$ must be \emph{yellow}, since $i$ (which is of type~$t_6$)
does not contain the color-sequence substring ${\tt bb}$
for the edges joint with tiles $y$ and~$e$. This implies that the 
tiles $i$ and $w$ also have a fixed orientation. If the input color to the
lower right part is \emph{red} (see Figures~\ref{fig:and-3trp-tf}
and~\ref{fig:and-3trp-ff}), a unique 
solution is obtained by similar arguments.
The connection of the lower right part to the rest of the subpuzzle is 
the edge between tiles $w$ and~$g$. If the right input color is \emph{blue}, 
this edge will also be \emph{blue}, and if the right input color
is \emph{red}, this edge will be \emph{yellow}.

The heart of the AND
subpuzzle is its lower middle part, formed by the tiles $c$ and~$g$.
The colors at the joint edge between tiles $b$ and $c$ and at the joint edge 
between tiles $w$ and $g$ determine the orientation of the tiles $c$ and 
$g$ uniquely for all four possible combinations of input colors.
The output of this part is the color at the edge between $c$ and~$j$.
If both input colors are \emph{blue}, this edge will also be \emph{blue}, and 
otherwise this edge will always be \emph{yellow}.

The output of the whole AND subpuzzle
will be \emph{red} if the edge between $c$ and $j$ is \emph{yellow}, 
and if this edge is \emph{blue} then the output of the whole subpuzzle 
will also be
\emph{blue}.  If the input color for the upper part is \emph{blue}
(see Figure~\ref{fig:and-3trp-tt}), 
each of the
tiles $j$, $k$, $l$, $m$, and $n$ has a vertical \emph{blue} line.
Note that since the colors
\emph{red} and \emph{yellow} are symmetrical in these tiles, we would have
several possible solutions without tiles $o$, $u$, and~$v$.  However,
tile $v$ (which is of type~$t_9$)
contains neither ${\tt rr}$ nor ${\tt ry}$ for the
edges joint with tiles $k$ and~$j$, so the orientation of the tiles
$j$ through $n$ is fixed, except that tile $n$ without tiles $o$ and $u$
would still have two possible 
orientations. Tile $u$ (which is of type~$t_2$)
is fixed because of its color-sequence substring
${\tt yy}$ at the edges joint
with $l$ and~$m$, so due to tiles $o$ and $u$
the only color possible at the edge between $n$ and $o$
is \emph{yellow}, and we have a unique solution.  If the input color for
the upper part is \emph{yellow}
(see Figures~\ref{fig:and-3trp}(b)--(d)), we obtain unique solutions by similar
arguments.  Hence, this new AND subpuzzle uses only three colors and has
unique solutions for each of the four possible combinations of input colors.

\begin{figure}
  \centering
  \subfigure[Out: \emph{true}]{
    \label{fig:bool-4trp-t}
    \quad
    \quad
    \includegraphics[height=2.1cm]{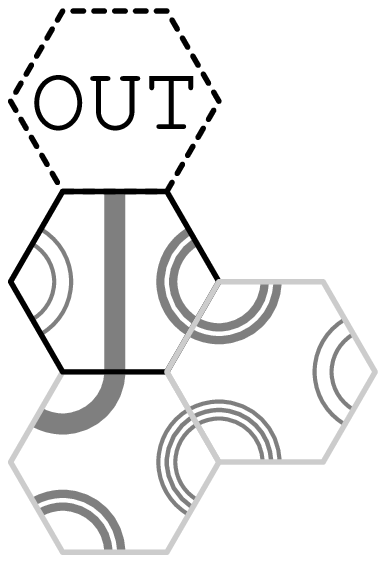}
    \quad
    \quad
  }
  \subfigure[Out: \emph{false}]{
    \label{fig:bool-4trp-f}
    \quad
    \quad
    \includegraphics[height=2.1cm]{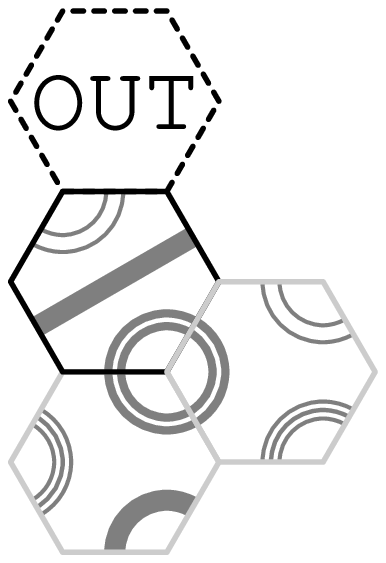}
    \quad
    \quad
  }
  \subfigure[Scheme]{
    \label{fig:bool-4trp-s}
    \quad
    \quad
    \includegraphics[height=2.1cm]{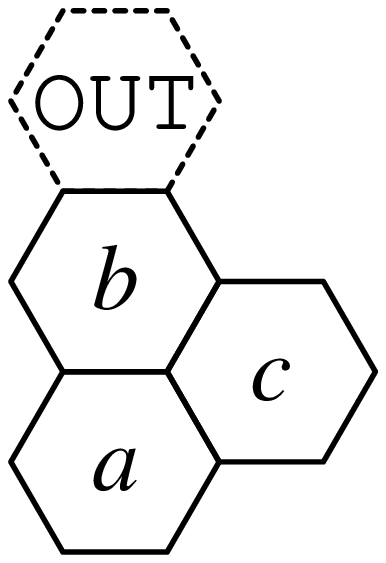}
    \quad
    \quad
  }
  \caption{Original BOOL subpuzzle, see~\cite{hol-hol:j:tantrix}}
\label{fig:bool-4trp}
\end{figure}

\begin{figure}[h!]
  \centering
  \subfigure[Out: \emph{true}]{
    \label{fig:bool-3trp-t}
    \quad
    \includegraphics[height=2.5cm]{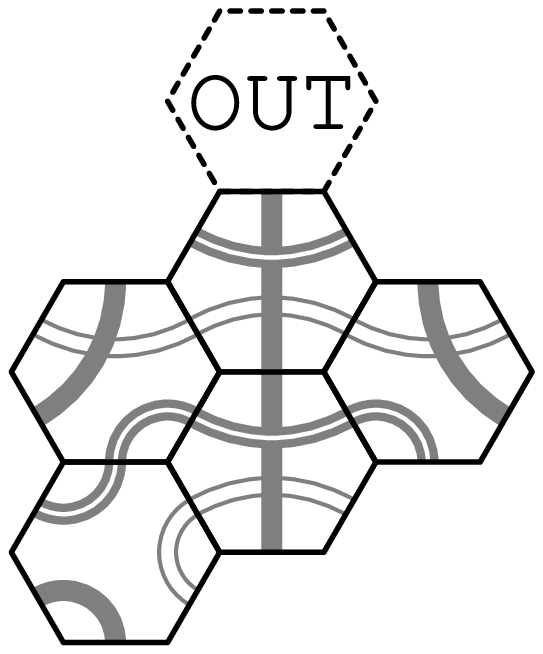}
    \quad
  }
  \subfigure[Out: \emph{false}]{
    \label{fig:bool-3trp-f}
    \quad
    \includegraphics[height=2.5cm]{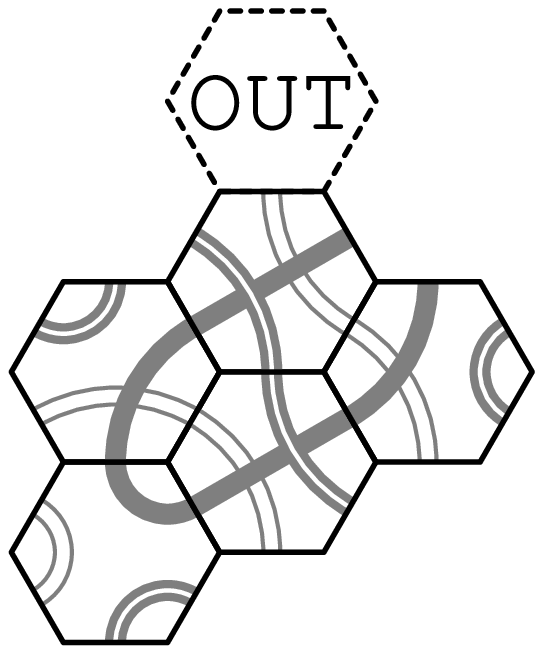}
    \quad
  }
  \subfigure[Scheme]{
    \label{fig:bool-3trp-s}
    \quad
    \includegraphics[height=2.5cm]{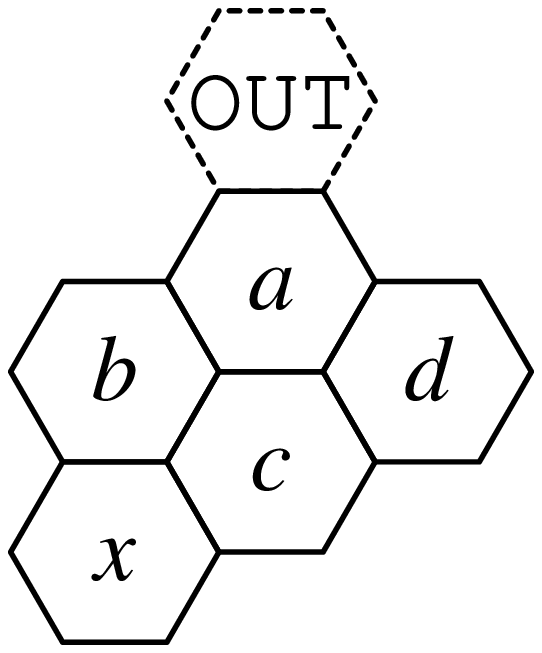}
    \quad
  }
  \caption{Three-color BOOL subpuzzle}
  \label{fig:bool-3trp}
\end{figure}

\paragraph{Input and output subpuzzles:}
The input variables of the boolean circuit are represented by the
subpuzzle {BOOL}.
The original four-color BOOL subpuzzle from~\cite{hol-hol:j:tantrix}
is shown in Figure~\ref{fig:bool-4trp}.
Our new three-color BOOL subpuzzle is
presented in Figure~\ref{fig:bool-3trp}, and since it is completely 
different from the original subpuzzle, no tiles are marked here. 
This subpuzzle has only two possible
solutions, one with the output color \emph{blue} (if the corresponding variable
is \emph{true}), and one with the output color \emph{red} (if the corresponding
variable is \emph{false}).  The original four-color BOOL subpuzzle
from~\cite{hol-hol:j:tantrix} (which was not modified
in~\cite{bau-rot:c:tantrix})
contains tiles with \emph{green} lines to
exclude certain rotations.  Our three-color BOOL subpuzzle
does not contain any \emph{green} lines, but it might not be that
obvious that there are only two possible solutions, one for each
output color.

First, we show that the output color \emph{yellow} is not
possible.  If the output color were \emph{yellow}, there would be
two possible orientations
for tile~$a$. In the first orientation, the joint edge between $a$ and $b$ is
\emph{blue}.  This is not possible, however, since $c$ (which is a \emph{Chin},
namely a tile of type~$t_8$) does not contain the color-sequence
substring~${\tt rr}$.
By a similar argument for tile~$d$, the other
orientation with the output color \emph{yellow} is not possible either.

Second, we show that tile $x$ makes the solution unique. For the output
color \emph{blue}, there are two possible orientations for each of 
the tiles~$a$, $b$, $c$, and~$d$.  In order to exclude one of these
orientations in each case, tile $x$ must contain
either of the color-sequence substrings ${\tt br}$ or
${\tt yr}$ at its edges joint with tiles $b$ and~$c$.
On the other hand, for the output color \emph{red},
tile $x$ must not contain the
color-sequence substring
${\tt ry}$ at its edges joint with $b$ and~$c$, because this
would leave two possible orientations for tile~$d$.  Tile $t_1$ satisfies
all these conditions and makes the solution of the BOOL subpuzzle unique,
while using only three colors.

\begin{figure}
  \centering
  \subfigure[TEST-true]{
    \label{fig:test-4trp-t}
    \quad
    \includegraphics[height=2.1cm]{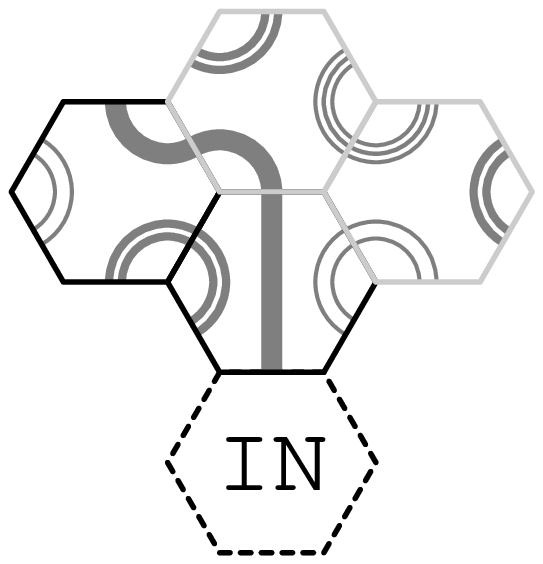}
    \quad
  }
  \subfigure[TEST-false]{
    \label{fig:test-4trp-f}
    \quad
    \includegraphics[height=2.1cm]{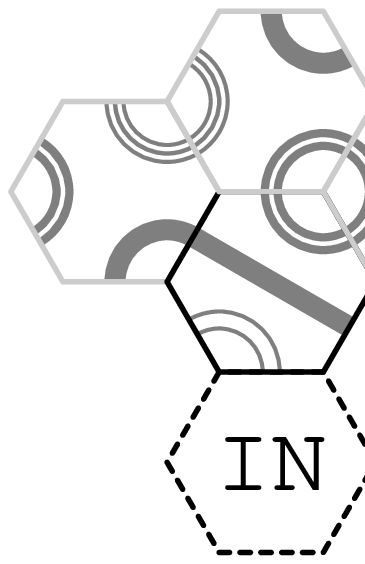}
    \quad
  }
  \subfigure[Scheme]{
    \label{fig:test-4trp-s}
    \quad
    \includegraphics[height=2.1cm]{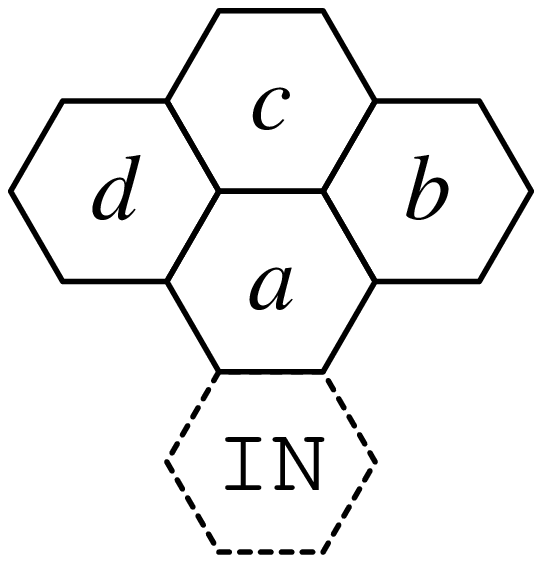}
    \quad
  }
  \caption{Original TEST subpuzzles, see~\cite{hol-hol:j:tantrix}}
  \label{fig:test-4trp}
\end{figure}

\begin{figure}[h!]
  \centering
  \subfigure[TEST-true]{
    \label{fig:test-3trp-t}
    \quad
    \includegraphics[height=2.5cm]{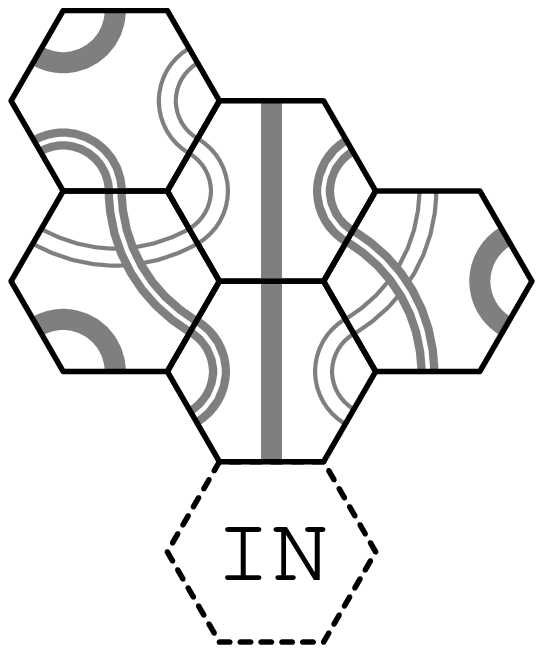}
    \quad
  }
  \subfigure[TEST-false]{
    \label{fig:test-3trp-f}
    \quad
    \includegraphics[height=2.5cm]{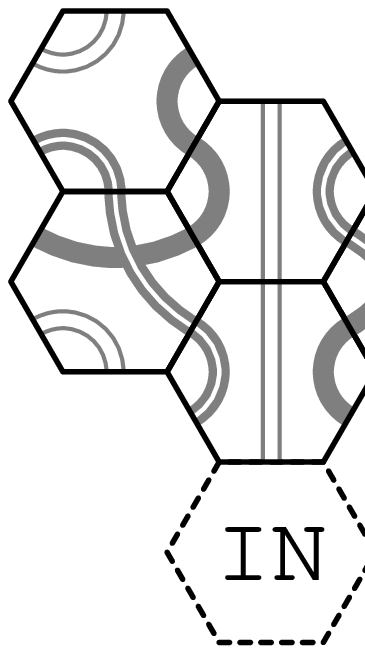}
    \quad
  }
  \subfigure[Scheme]{
    \label{fig:test-3trp-s}
    \quad
    \includegraphics[height=2.5cm]{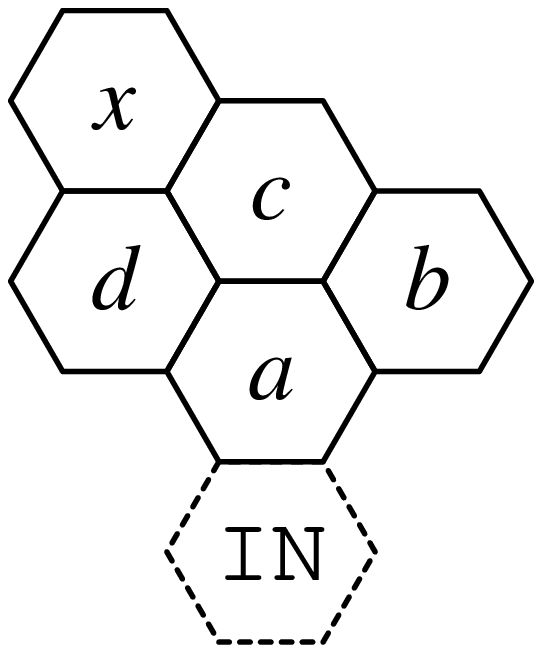}
    \quad
  }
  \caption{Three-color TEST subpuzzles}
  \label{fig:test-3trp}
\end{figure}

Finally, a subpuzzle is needed to check whether or not the circuit
evaluates to \emph{true}.  This is achieved by the subpuzzle TEST-true shown in
Figure~\ref{fig:test-3trp-t}.  It has only one valid solution, namely that its
input color is \emph{blue}.
Just like the subpuzzle BOOL, the original four-color TEST-true subpuzzle
from~\cite{hol-hol:j:tantrix}, which is shown in Figure~\ref{fig:test-4trp-t}
and which was not modified
in~\cite{bau-rot:c:tantrix},
uses \emph{green} lines to exclude certain rotations. 
Again, since the new TEST-true subpuzzle is completely different from 
the original subpuzzle, no tiles are marked here.
Note that in the three-color TEST-true subpuzzle of 
Figure~\ref{fig:test-3trp-t},
$a$ and $c$ are the same tiles as $a$ and $b$ in the WIRE subpuzzle of
Figure~\ref{fig:wire-3trp}.  To ensure that the input color is
\emph{blue}, we have to consider all possible color-sequence
substrings at
the edges of $d$ joint with $c$ and~$a$, and at the edges of $b$
joint with $a$ and~$c$.  For each input color, there are four
possibilities.

Assume that the input color is \emph{red}.  Then the
possible color-sequence substrings for tile $d$ at the edges
joint with $c$ and $a$ are: ${\tt bb}$, ${\tt yb}$, ${\tt yy}$, 
and~${\tt by}$.  
Similarly,  the possible color-sequence substrings for tile~$b$ at the edges
joint with $a$ and $c$ are: ${\tt yy}$, ${\tt yb}$, ${\tt bb}$, and ${\tt by}$.
Tile $t_{14}$ at position $d$ excludes ${\tt by}$ and~${\tt yy}$,
while tile $t_{11}$ at position $b$ excludes ${\tt yy}$ and~${\tt yb}$.
Thus, \emph{red} is not possible as the input color. The input color
\emph{yellow} can be excluded by similar arguments.  It follows that
\emph{blue}
is the only possible input color. It is clear that the tiles $a$ and $c$
have a vertical \emph{blue} line. Due to the fact that neither $t_{11}$ nor
$t_{14}$ contains the color-sequence substrings ${\tt rr}$ or 
${\tt yy}$ for the edges joint with
tiles $a$ and~$c$, two possible solutions are still left. The color-sequence
substrings for these solutions at the edges of $x$ joint with $c$ and $d$ are
${\tt ry}$ and~${\tt yr}$.  Since tile $t_2$ at position $x$
contains the former but not the 
latter sequence, the TEST-true subpuzzle uses only three
colors and has a unique solution.

(Note: The TEST-false subpuzzles in Figures~\ref{fig:test-3trp-f}
and~\ref{fig:test-2trp-f} will be needed for a circuit construction
in Section~\ref{sec:unique-infinite-variants},
see Figure~\ref{fig:circuit-inf}.
In particular, the three-color
TEST-false subpuzzle in Figure~\ref{fig:test-3trp-f} is identical to
the three-color TEST-true subpuzzle from Figure~\ref{fig:test-3trp-t},
except that the colors \emph{blue} and \emph{red} are exchanged.  By
the above argument, the TEST-false subpuzzle has only one valid
solution, namely that its input color is \emph{red}.)

The shapes of the subpuzzles constructed above
have changed slightly.
However, by Holzer and Holzer's
argument~\cite{hol-hol:j:tantrix} about
the minimal horizontal distance between two wires and/or gates being
at least four, unintended interactions between the subpuzzles do not
occur.
This concludes the proof of Theorem~\ref{thm:3trp-npc}.~\end{proofs}

%

%


Theorem~\ref{thm:3trp-npc} immediately gives the following corollary.

\begin{corollary}
\label{cor:3trp-npc}
$\threetrp$ is $\np$-complete.
\end{corollary}

Since the tile set $T_3$ is a subset of the tileset~$T_4$, we have
$\ktrp{3} \manyone \ktrp{4}$.  Thus, the hardness results for
$\ktrp{3}$ and its variants proven in this paper immediately are
inherited by $\ktrp{4}$ and its variants, which provides an
alternative proof of these hardness results for $\ktrp{4}$
and its variants established
in~\cite{hol-hol:j:tantrix,bau-rot:c:tantrix}.  In
particular, Corollary~\ref{cor:4trp-npc} follows from
Theorem~\ref{thm:3trp-npc} and Corollary~\ref{cor:3trp-npc}.

\begin{corollary}[\cite{hol-hol:j:tantrix,bau-rot:c:tantrix}]
\label{cor:4trp-npc}
$\ktrp{4}$ is $\np$-complete, via a parsimonious reduction from $\sat$.
\end{corollary}

\subsection{Parsimonious Reduction from SAT to 2-TRP}
\label{sec:two-trp-is-np-complete}

In contrast to the above-mentioned fact that $\ktrp{3} \manyone
\ktrp{4}$ holds trivially, the reduction $\ktrp{2} \manyone \ktrp{3}$
(which we will show to hold due to both problems being $\np$-complete,
see Corollaries~\ref{cor:3trp-npc}
and~\ref{cor:two-trp-is-np-complete}) is not immediatedly
straightforward, since the tile set $T_2$ is not a subset of the tile
set~$T_3$ (recall Figure~\ref{fig:tileset} in Section~\ref{sec:definitions}).
In this section, we study $\ktrp{2}$ and its variants.
Our main result here is Theorem~\ref{thm:two-trp-is-np-complete}
below.

\begin{theorem}
\label{thm:two-trp-is-np-complete}
$\sat$ parsimoniously reduces to $\ktrp{2}$.
\end{theorem}

\begin{proofs}
%
As in the proof of Theorem~\ref{thm:3trp-npc},
we again provide a reduction from $\circuitandnotsat$, but here we use
McColl's planar cross-over circuit~\cite{mcc:j:planar-crossovers} instead of 
a CROSS subpuzzle.\footnote{Whether there exists an analogous two-color
CROSS subpuzzle to simplify this
construction, is still an open question.}

We choose our color set $C_2$ to contain the colors \emph{blue}
and \emph{red} (corresponding to the truth values \emph{true} and
\emph{false}), and we use the tileset~$T_2$ shown in 
Figure~\ref{fig:tileset-two-color}.  
To simulate a boolean circuit with AND and NOT gates, we now
present the subpuzzles constructed only with tiles from~$T_2$.

\paragraph{Wire subpuzzles:}
We again use \emph{Brid} tiles with a straight \emph{blue} line to
construct the WIRE subpuzzle with the colors \emph{blue} and
\emph{red} as shown in Figure~\ref{fig:wire-2trp}.  If the input
color is \emph{blue}, then tiles $a$ and $b$ must have a vertical
\emph{blue} line, so the output color will be \emph{blue}.  If the
input color is \emph{red}, then the edge between $a$ and $b$ must be
\emph{red} too, and it follows that the ouput color will also be
\emph{red}.  Tile $x$ forces tiles $a$ and $b$ to fix the orientation
of the \emph{blue} line for the input color \emph{red}.  Since we care
only about distinct color sequences of the tiles (recall the remarks
made in 
Section~\ref{sec:tile-sets-color-sequences-orientations}),\footnote{By
contrast, if we were to count all distinct orientations of the
tiles even if they have identical color sequences, we would obtain
two solutions each for tiles $a$ and~$b$, and six solutions
for tile~$x$, which gives a total of $24$ solutions
for each input color in the WIRE
subpuzzle.  However, as argued in
Section~\ref{sec:tile-sets-color-sequences-orientations}, since our focus
is on the color sequences, we have unique solutions and thus a parsimonious
reduction from $\sat$ to $\ktrp{2}$.}
we have unique solutions for both input colors.

Note that this
construction allows wires of arbitrary height, unlike the WIRE
subpuzzle constructed in the proof of Theorem~\ref{thm:3trp-npc} or
the WIRE subpuzzles constructed
in~\cite{hol-hol:j:tantrix,bau-rot:c:tantrix},
which all are constructed so as to have
even height. To construct two-color WIRE subpuzzles of arbitrary height,
tile $x$ of type $t_8$ in Figure~\ref{fig:wire-2trp} would have to be 
placed on alternating sides of tiles~$a$, $b$, etc.\ in each level.

\begin{figure}[h!]
  \centering
  \subfigure[In: \emph{true}]{
    \label{fig:wire-2trp-t}
    \quad
    \includegraphics[height=2.8cm]{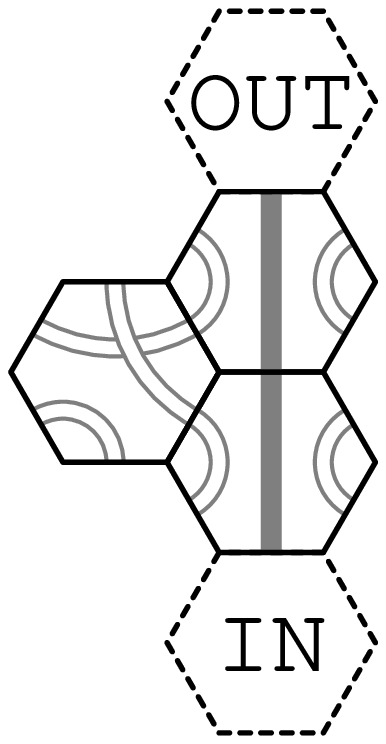}
    \quad
  }
  \subfigure[In: \emph{false}]{
    \label{fig:wire-2trp-f}
    \quad
    \includegraphics[height=2.8cm]{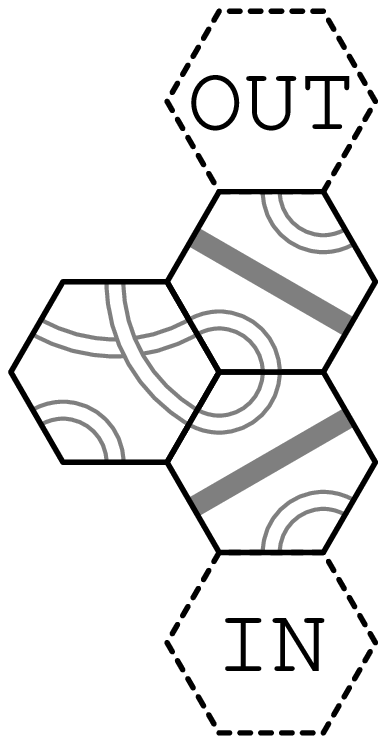}
    \quad
  }
  \subfigure[Scheme]{
    \label{fig:wire-2trp-s}
    \quad
    \includegraphics[height=2.8cm]{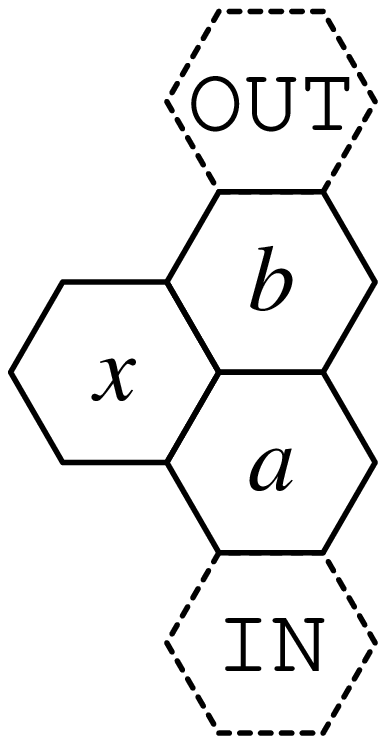}
    \quad
  }
  \caption{Two-color WIRE subpuzzle}
  \label{fig:wire-2trp}
\end{figure}

The two-color MOVE subpuzzle is shown in
Figure~\ref{fig:move-2trp}.  Just like the WIRE subpuzzle, it
consists only of tiles of types $t_3$ and~$t_8$ (see
Figure~\ref{fig:tileset-two-color}).  For the input color \emph{blue},
it is obvious that all tiles must have vertical \emph{blue} lines and
so the output color is also \emph{blue}. If the input color is
\emph{red}, then the edge between $a$ and $b$ is \emph{red}, too.  Since
neither $c$ nor $d$ contains the color-sequence substring~${\tt bb}$,
the blue lines of
these four tiles have all the same direction.  The same argument applies
to tiles $e$ and~$f$, and since tiles $f$, $g$, and $x$ behave like a
WIRE subpuzzle, the output color will be \emph{red} in this case.
As above, since we care only about the color sequences of the tiles,
we obtain unique solutions for both input colors.

Note that
Figure~\ref{fig:move-2trp} shows a move to the right.  A move to
the left can be made symmetrically, simply by mirroring this subpuzzle.

\begin{figure}[h!]
  \centering
  \subfigure[In: \emph{true}]{
    \label{fig:move-2trp-t}
    \quad
    \includegraphics[height=4.2cm]{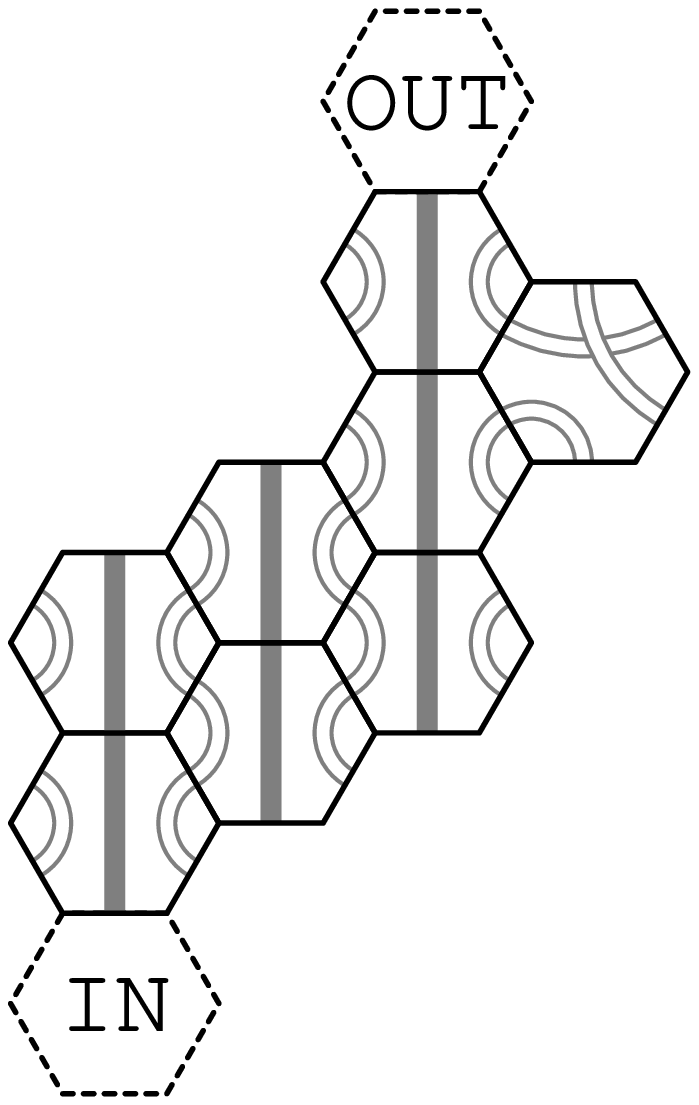}
    \quad
  }
    \subfigure[In: \emph{false}]{
    \label{fig:move-2trp-f}
    \quad
    \includegraphics[height=4.2cm]{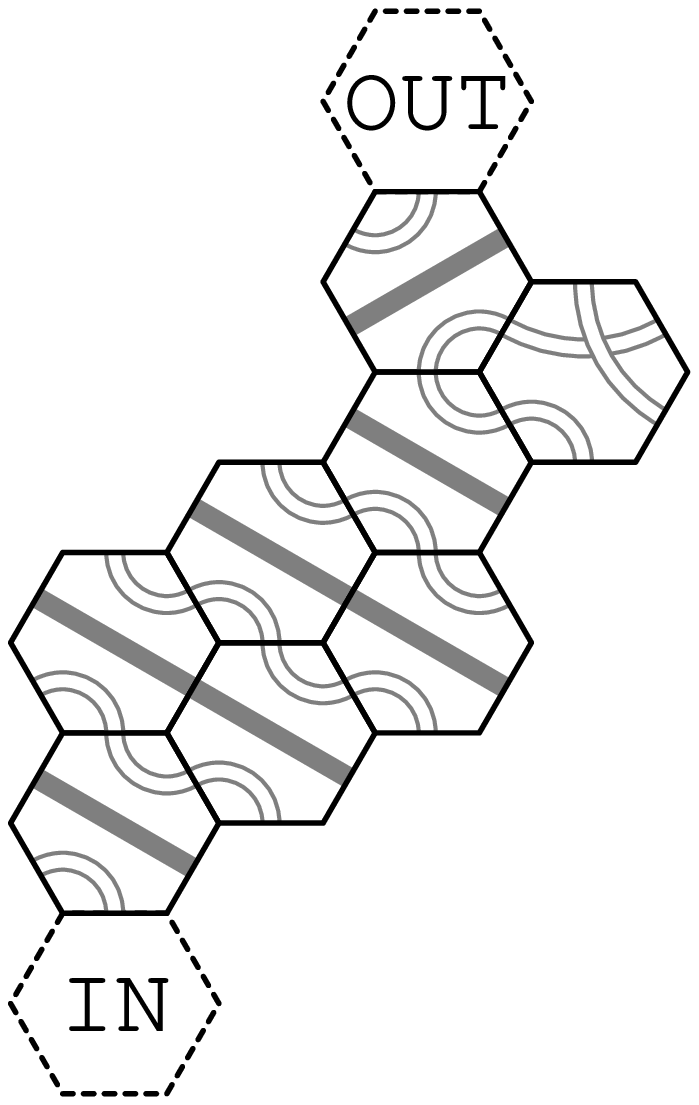}
    \quad
  }
  \subfigure[Scheme]{
    \label{fig:move-2trp-s}
    \quad 
    \includegraphics[height=4.2cm]{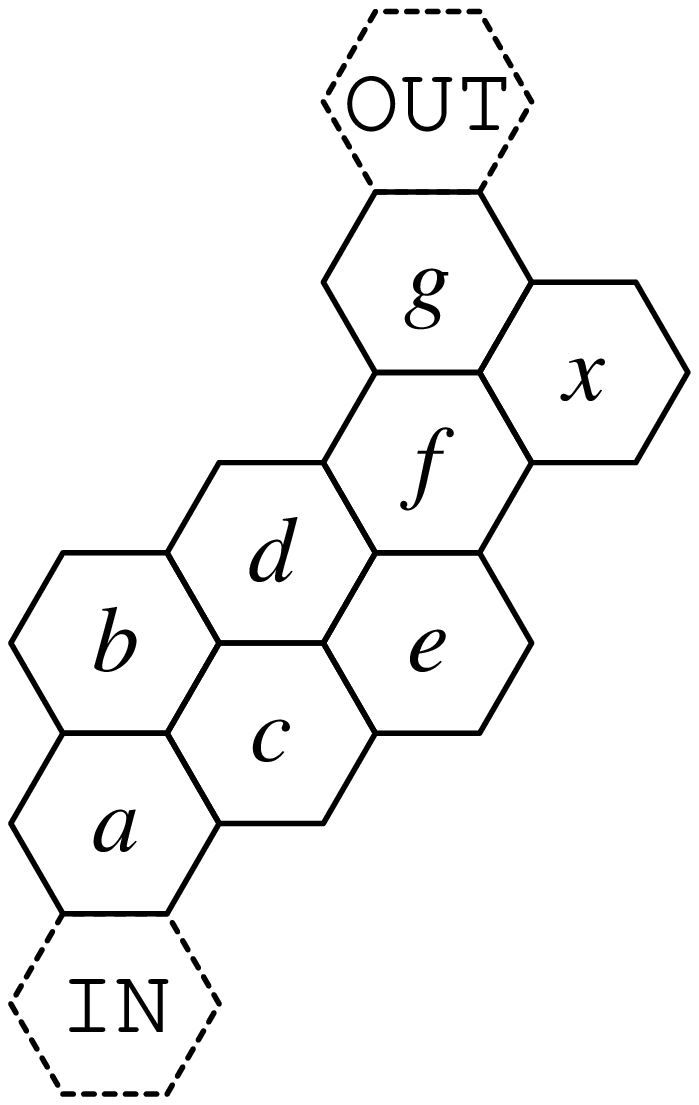}
    \quad 
  }
  \caption{Two-color MOVE subpuzzle}
  \label{fig:move-2trp}
\end{figure}

\begin{figure}[h!]
  \centering
  \subfigure[In: \emph{true}]{
    \label{fig:copy-2trp-t}
    \quad
    \includegraphics[height=3.6cm]{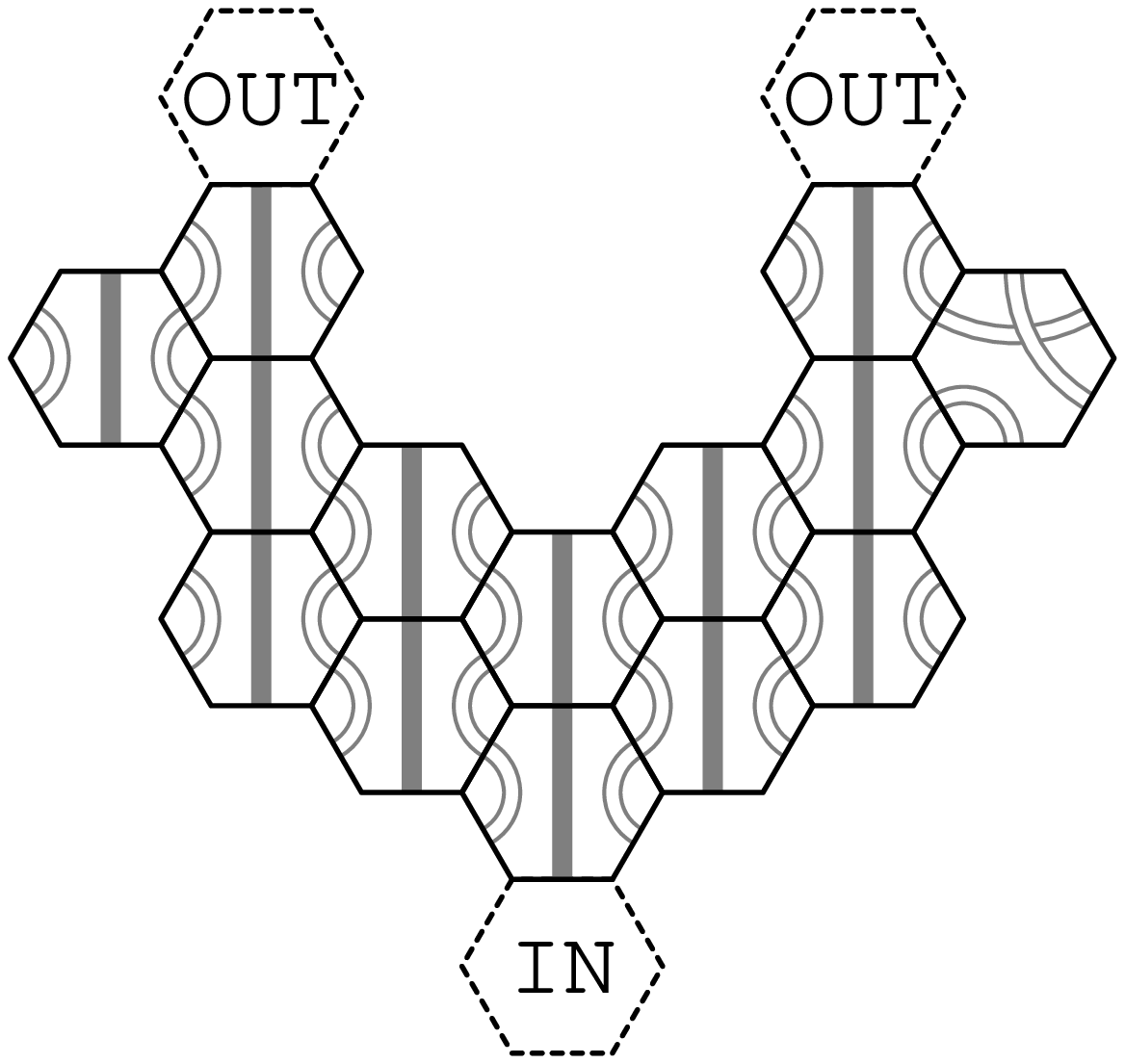}
    \quad
  }
  \subfigure[In: \emph{false}]{
    \label{fig:copy-2trp-f}
    \quad
    \includegraphics[height=3.6cm]{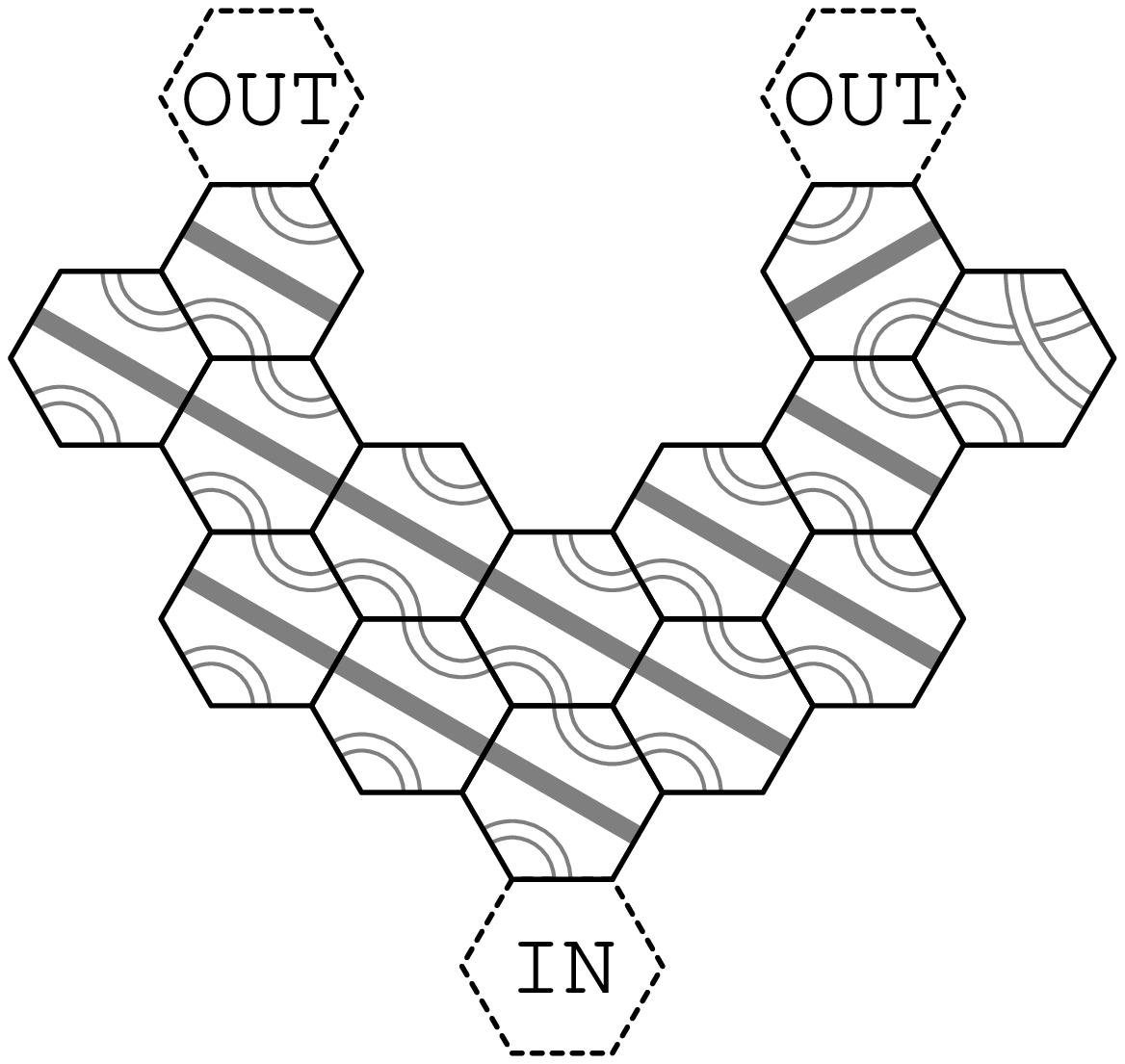}
    \quad
  }
  \subfigure[Scheme]{
    \label{fig:copy-2trp-s}
    \quad 
    \includegraphics[height=3.6cm]{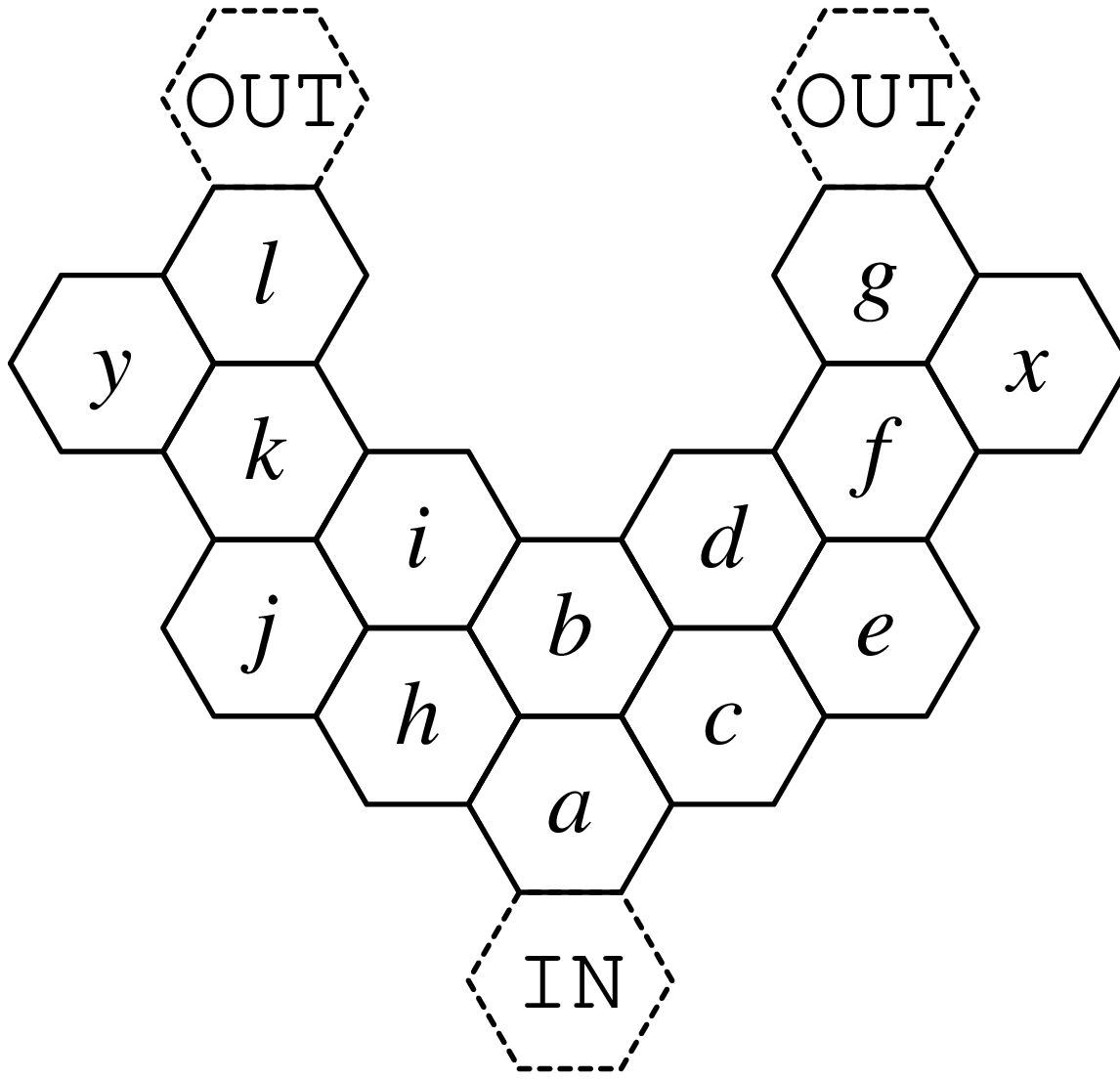}
    \quad 
  }
  \caption{Two-color COPY subpuzzle}
  \label{fig:copy-2trp}
\end{figure}

The last subpuzzle needed to simulate the wires of the boolean
circuit is the COPY subpuzzle
in Figure~\ref{fig:copy-2trp}.
This subpuzzle is akin to the subpuzzle obtained by mirroring the MOVE
subpuzzle in both directions,\footnote{We here say ``is akin to\ldots''
because the COPY subpuzzle in Figure~\ref{fig:copy-2trp} differs
from a true two-sided mirror version of MOVE by having a tile of type
$t_3$ at position $y$ instead of a $t_8$ as in position~$x$.  Why?
By the arguments for the MOVE subpuzzle,
tile $x$ already fixes the orientation of tiles $a$ through $k$ but
not of $l$ (if the input color is \emph{red}, see 
Figure~\ref{fig:copy-2trp-f}).
The orientation of tile $l$ is then fixed by a $t_3$ tile at position~$y$, 
since obviously a $t_8$ would not lead to a solution.
However, it is clear that an argument analogous  to that for the MOVE
subpuzzle shows that all \emph{blue} lines (except that of $g$ in
Figure~\ref{fig:copy-2trp-f}) have the same direction.}
so similar arguments as above work.
Again, since we disregard the repetitions of color sequences, we have
unique solutions for both input colors.

\paragraph{Gate subpuzzles:}
The construction of the NOT subpuzzle presented in 
Figure~\ref{fig:not-2trp} is similar to the corresponding
subpuzzle with
three colors (see Figure~\ref{fig:not-3trp}).  
Tiles $b$ and $d$ in the two-color version 
allow only two possible orientations
of tile~$c$, one for each input color.
The first one has \emph{blue} at the edge joint with $a$
and, consequently, \emph{red} at the edge joint with~$e$; the second
possible orientation has the
same colors exchanged.  Since tiles $e$, $f$, and $x$
behave like a WIRE subpuzzle, the output color will ``negate'' the
input color, i.e., the output color will be \emph{blue}
if the input color is \emph{red}, and it will be
\emph{red} if the input color is \emph{blue}.
Tile $x$ fixes the orientation of tiles $f$ and $e$ and the orientation 
of tile $a$ is fixed by tile $b$. We again obtain unique solutions, 
since we focus on color sequences.

\begin{figure}[h!]
  \centering
  \subfigure[In: \emph{true}]{
    \label{fig:not-2trp-t}
    \quad
    \includegraphics[height=4.2cm]{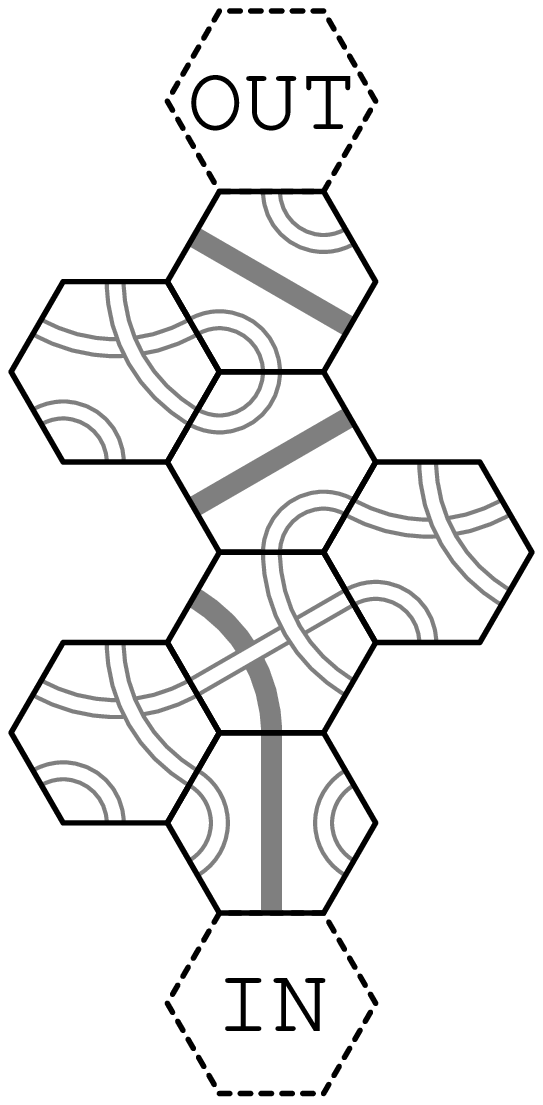}
    \quad
  }
    \subfigure[In: \emph{false}]{
    \label{fig:not-2trp-f}
    \quad
    \includegraphics[height=4.2cm]{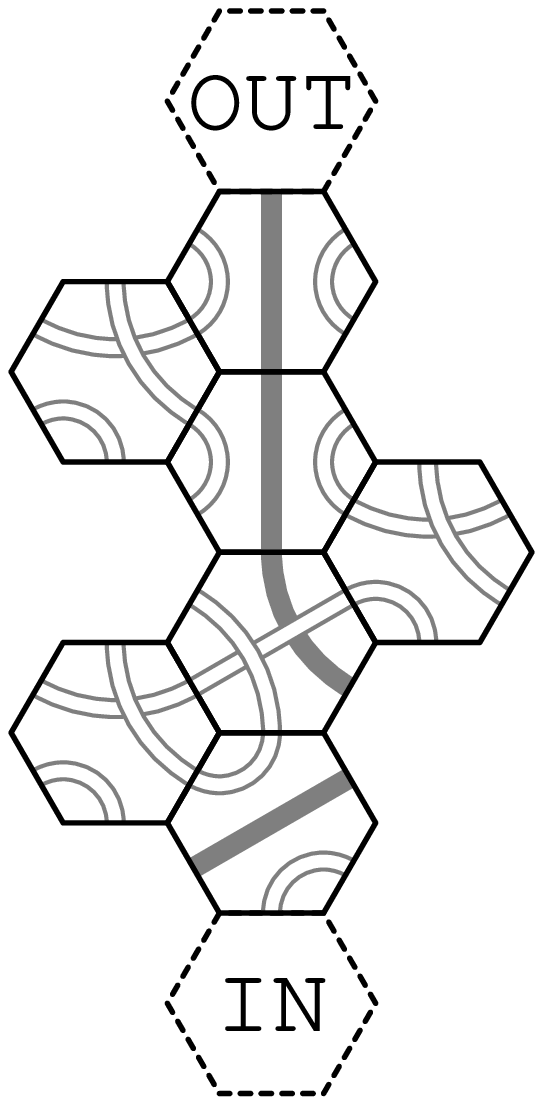}
    \quad
  }
  \subfigure[Scheme]{
    \label{fig:not-2trp-s}
    \quad 
    \includegraphics[height=4.2cm]{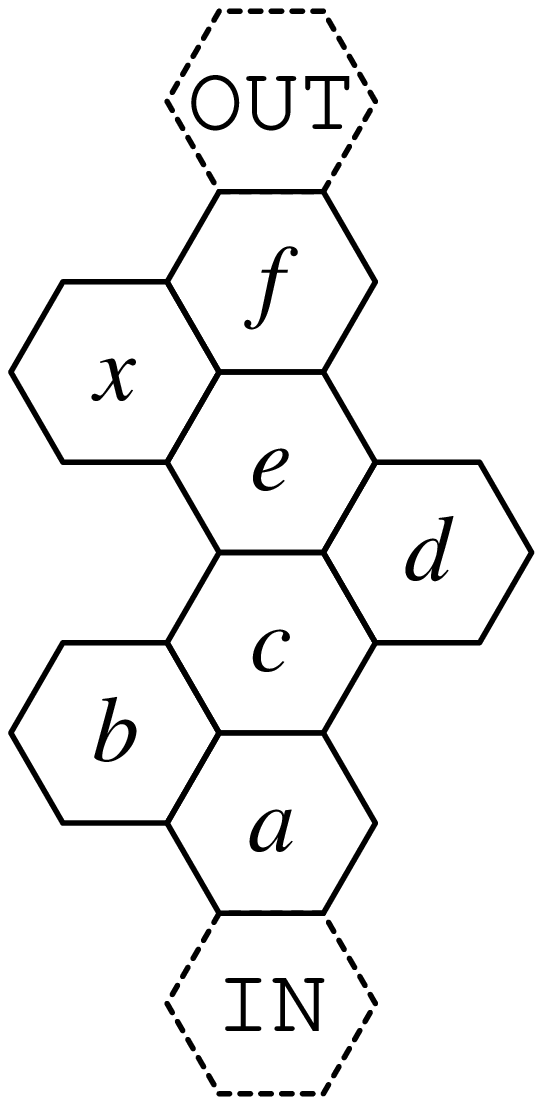}
    \quad 
  }
  \caption{Two-color NOT subpuzzle}
  \label{fig:not-2trp}
\end{figure}

\begin{figure}[h!]
  \centering
  \subfigure[In: \emph{true, true}]{
    \label{fig:and-2trp-tt}
    \quad
    \includegraphics[height=5.6cm]{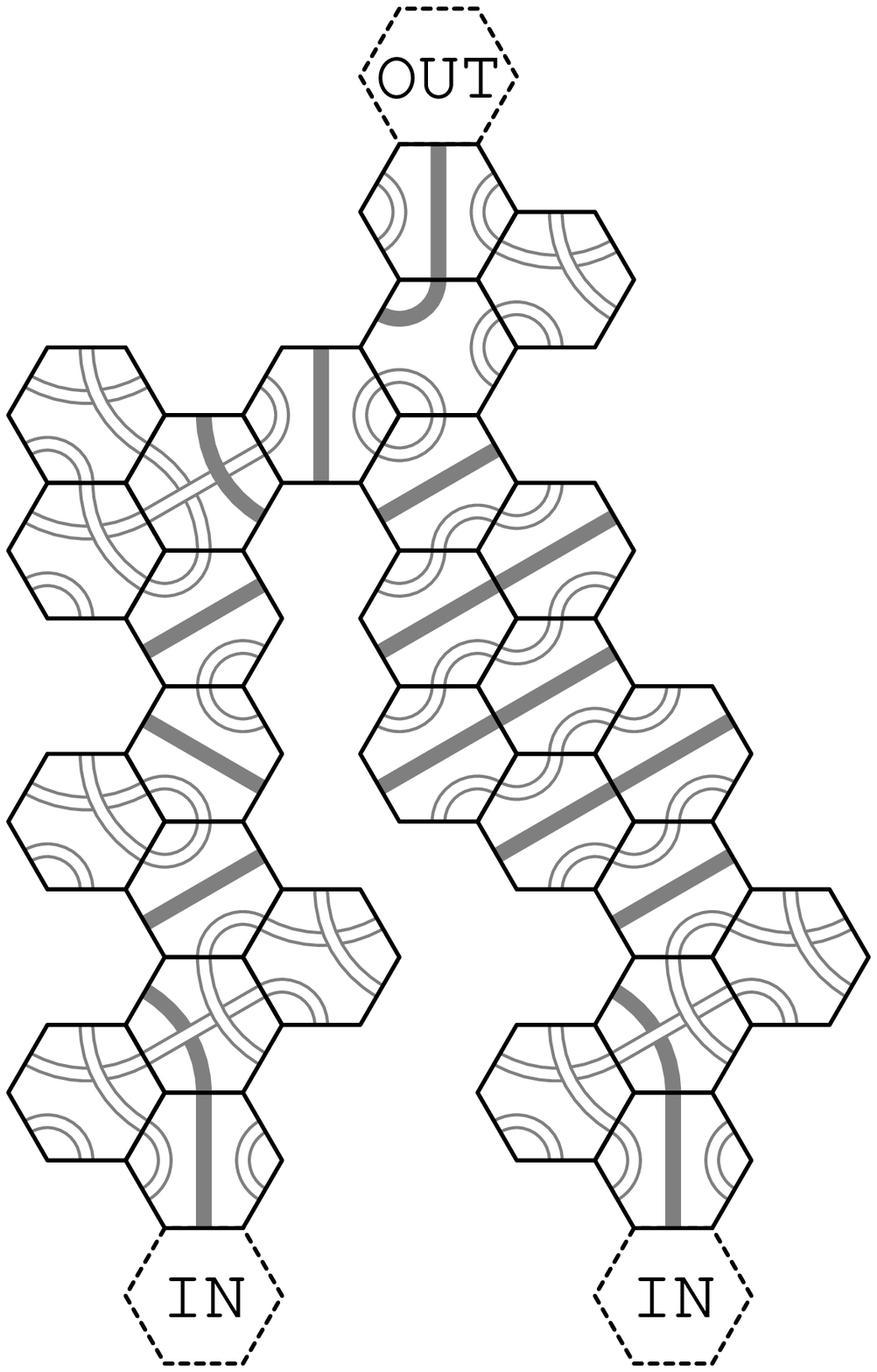}
    \quad
  }
  \subfigure[In: \emph{true, false}]{
    \label{fig:and-2trp-tf}
    \quad
    \includegraphics[height=5.6cm]{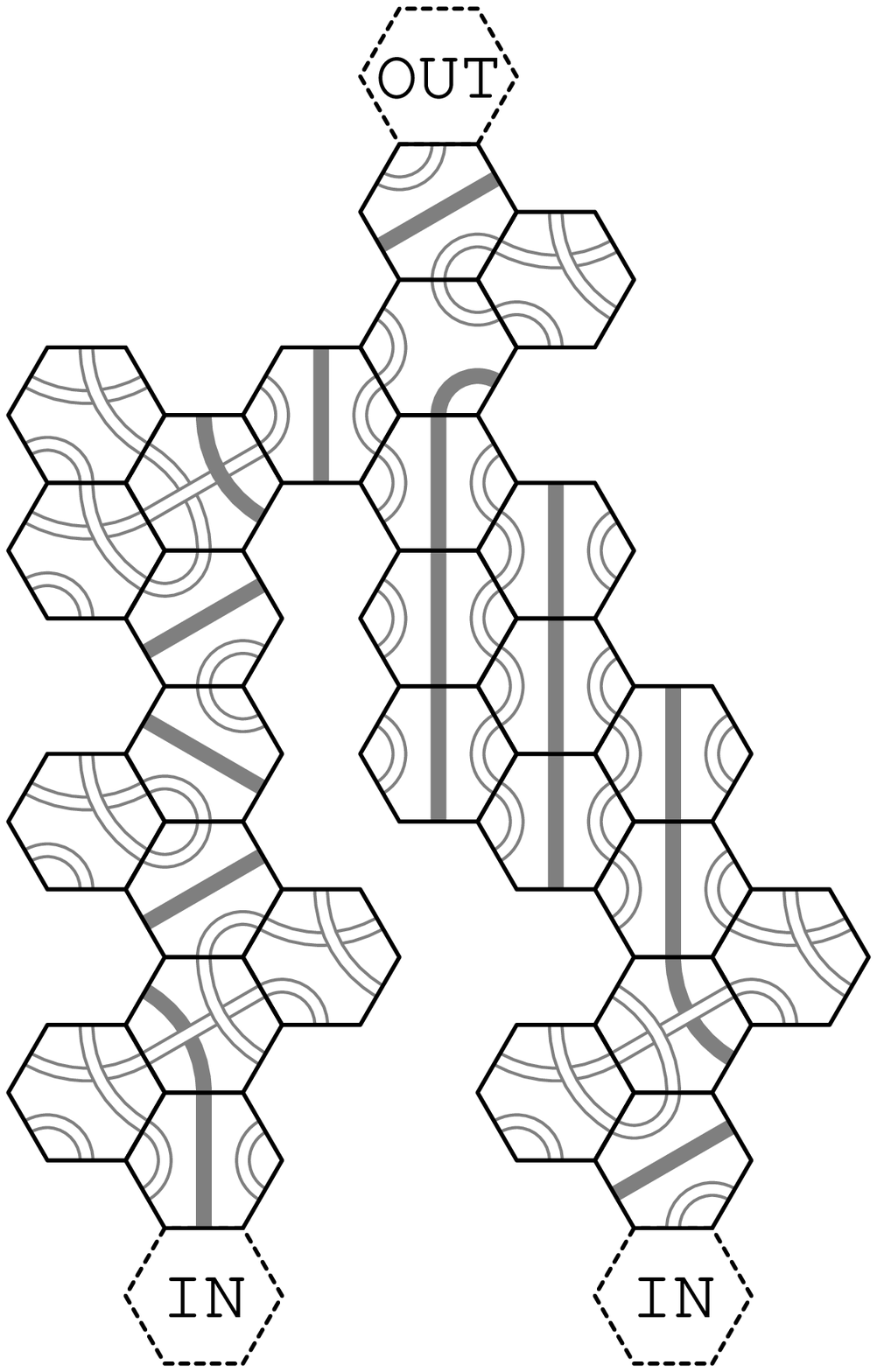}
  }
  \subfigure[In: \emph{false, true}]{
    \label{fig:and-2trp-ft}
    \quad
    \includegraphics[height=5.6cm]{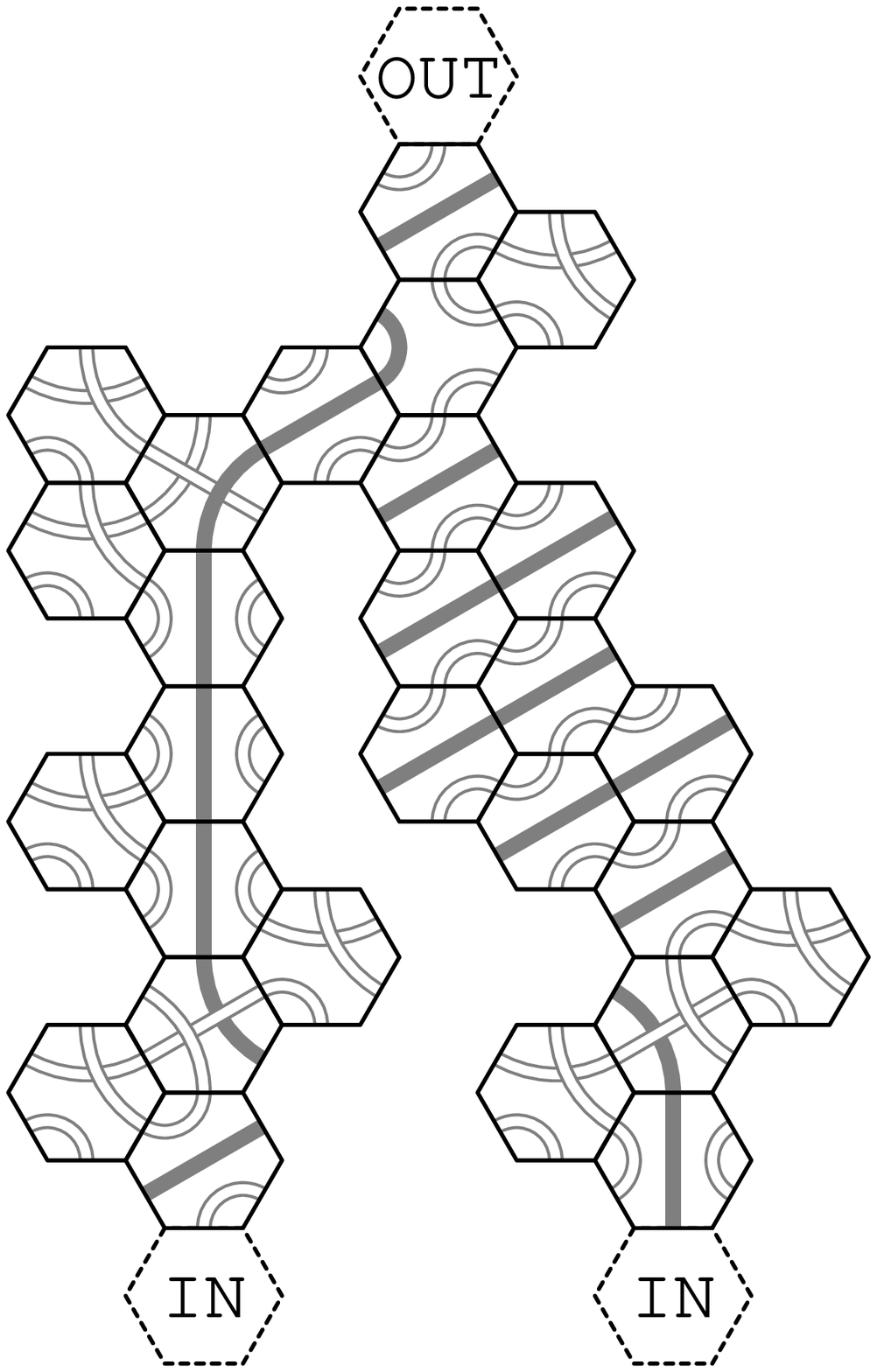}
    \quad
  }
  \subfigure[In: \emph{false, false}]{
    \label{fig:and-2trp-ff}
    \quad
    \includegraphics[height=5.6cm]{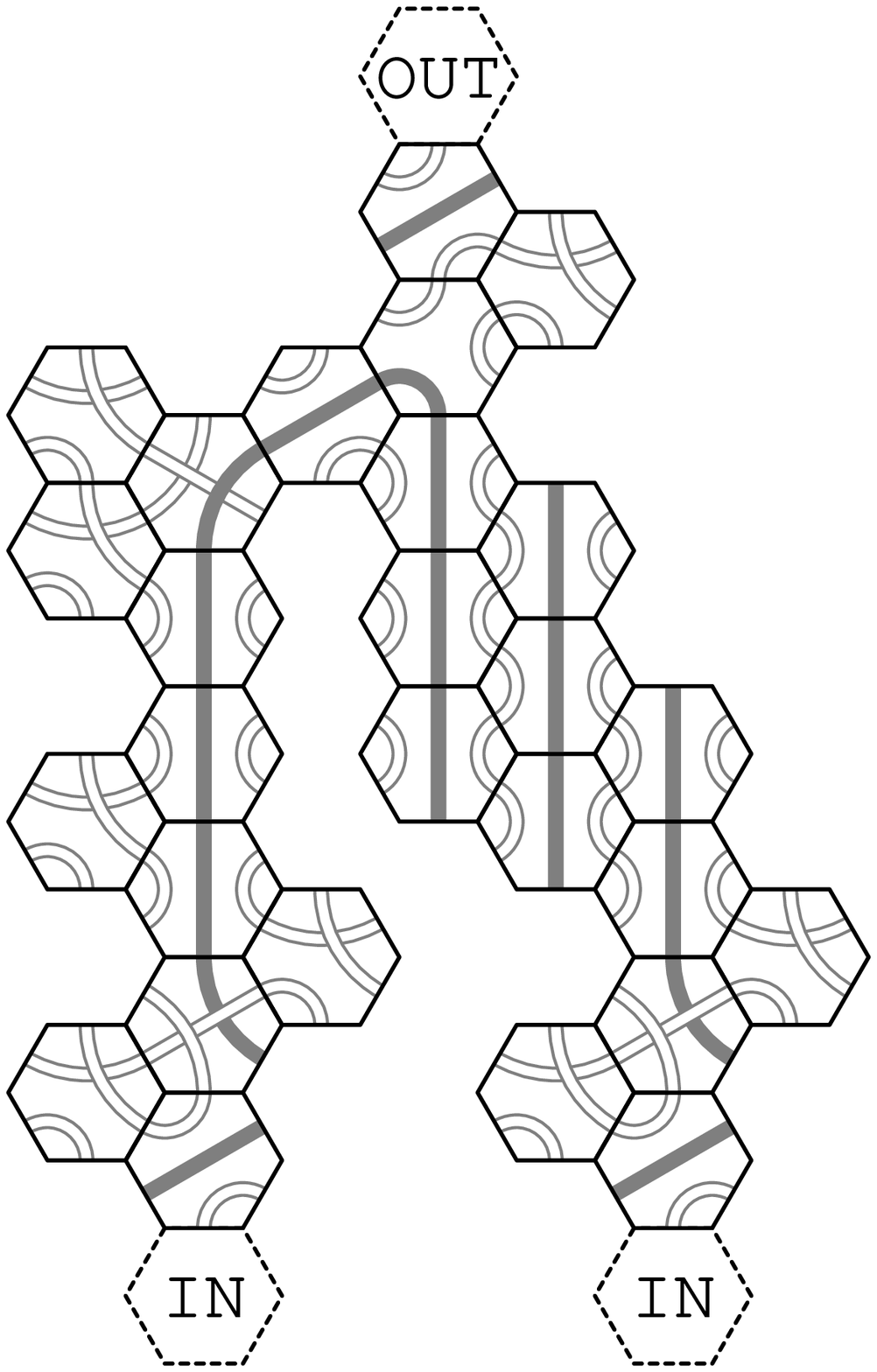}
    \quad
  }
  \subfigure[Scheme]{
    \label{fig:and-2trp-s}
    \quad 
    \psfrag{z1}[lB]{$z_1$}
    \psfrag{z2}{$z_2$}
    \psfrag{z3}{$z_3$}
    \includegraphics[height=5.6cm]{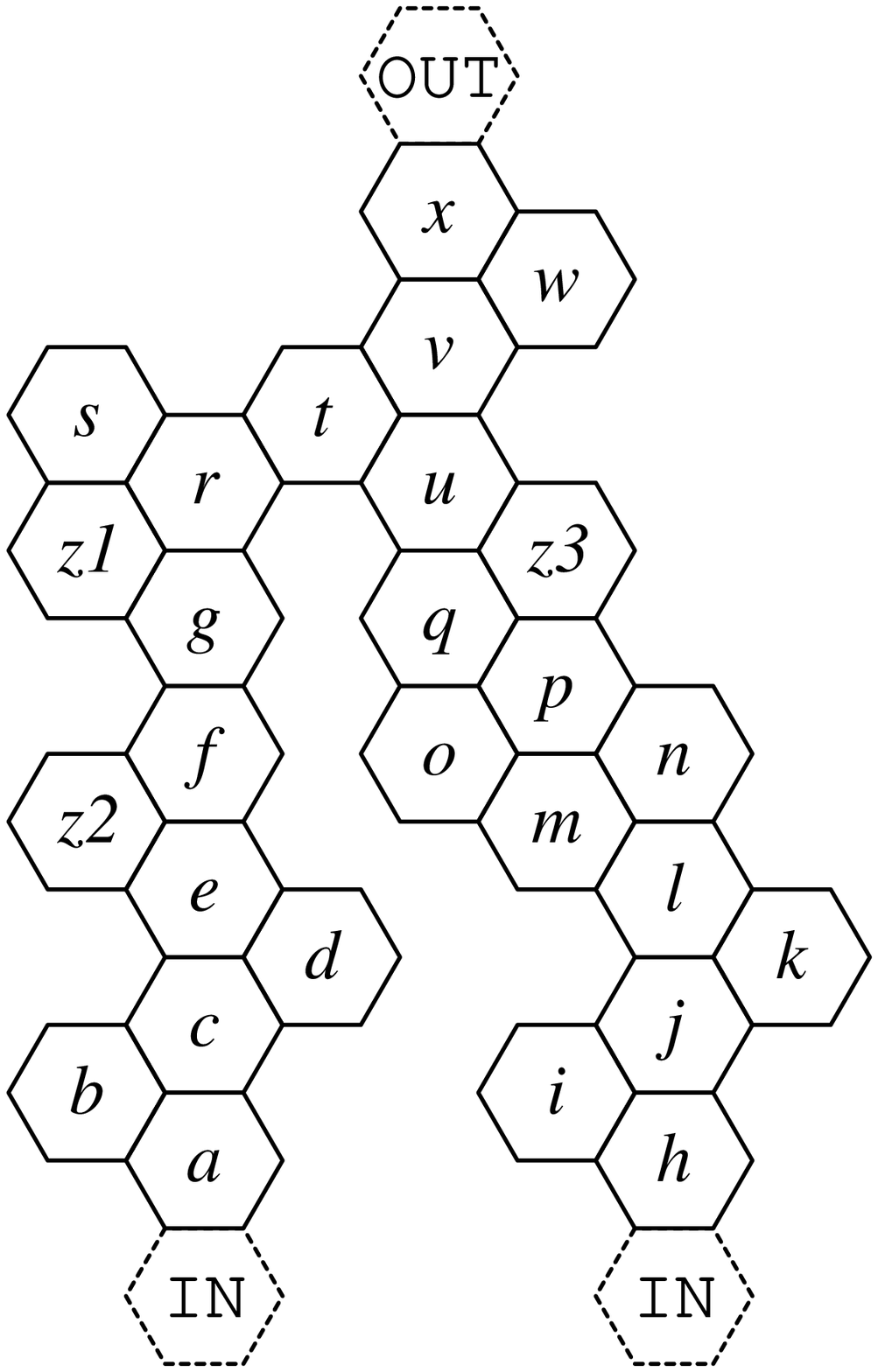}
    \quad 
  }
  \caption{Two-color AND subpuzzle}
  \label{fig:and-2trp}
\end{figure}

The AND subpuzzle is again the most complicated one.  To analyze this
subpuzzle, we subdivide it into three disjoint parts:
\begin{enumerate}
\item The first part consists of the tiles $a$ through~$g$, $z_1$,
  and~$z_2$. Tiles $a$ through $f$ and $z_2$ form a two-color NOT 
  subpuzzle, and tile $g$ passes the color at the edge between tiles 
  $f$ and $g$ on to the edge between tiles $g$ and~$r$. So the negated  
  left input color will be at the edge between tiles $g$ and~$r$. Tile $z_1$ 
  fixes the orientation of tile $g$ to obtain a unique solution for 
  this part of the subpuzzle.

\item The second part is formed by the tiles $h$ through~$q$,
  and~$z_3$.  This part is made from a two-color NOT and a two-color
  MOVE subpuzzle to negate the right input and move it by two
  positions to the left, which both are slightly modified with respect
  to the NOT in Figure~\ref{fig:not-2trp} and the MOVE in
  Figure~\ref{fig:move-2trp}.

  First, the minor differences between the move-to-the-left analog of
  the MOVE subpuzzle from Figure~\ref{fig:move-2trp} and this
  modified MOVE subpuzzle as part of the AND subpuzzle are the
  following: (a)~tile $z_3$ is positioned to the right of tiles $q$
  and $u$ and not to their left, and (b)~$z_3$ is a $t_3$ tile,
  whereas the tile at position $x$ in Figure~\ref{fig:move-2trp}
  is of type~$t_8$.  However, it is clear that
  the orientation of the \emph{blue} lines of tiles $l$ through $q$ is
  fixed by tile~$k$, and $z_3$ enforces $u$ and $q$ to have the same
  direction of \emph{blue} lines.

  Second, the minor difference between the NOT from
  Figure~\ref{fig:not-2trp} and this modified NOT subpuzzle as
  part of the AND subpuzzle is that tile $m$ is not of type $t_8$ (as
  is the $x$ in Figure~\ref{fig:not-2trp}) but of type~$t_3$,
  since the modified NOT and MOVE subpuzzles have been merged.  These
  changes are needed to ensure that we get a suitable height for this
  part of the AND subpuzzle.  However, it is again clear that
  the orientation of the \emph{blue} lines of tiles $l$ through $q$ is
  fixed by tile~$k$.

\item Finally, the third part, formed by the tiles $r$ through~$x$,
  behaves like a two-color subpuzzle simulating a boolean NOR gate,
  which is defined as $\neg (\alpha \vee \beta) \equiv \neg \alpha
  \wedge \neg \beta$.  The two inputs to the NOR subpuzzle come from
  the edges between $g$ and $r$ and between $q$ and~$u$.

  If the left input color (at the edge between $g$ and~$r$) is
  \emph{red}, then tiles $s$ and $z_1$ ensure that the edge between
  $r$ and $t$ will also be \emph{red}.  If the left input color is
  \emph{blue}, then the edge between $r$ and $t$ will be \emph{blue}
  by similar arguments, and since tile $t$ is of type~$t_3$, it passes
  this input color on to its joint edge with $v$ in both cases.  The
  right input to the upper part (at the edge between $q$ and~$u$) is
  passed on by tile $u$ to the edge between $u$ and~$v$.

  Now, we have both input colors at the edges between $t$ and $v$ and
  between $u$ and~$v$.  If both of these edges are \emph{red} (see
  Figure~\ref{fig:and-2trp-tt}), then tile $w$ enforces that the edge
  between $v$ and $x$ will be \emph{blue}.  On the other hand, if one
  or both of $v$'s edges with $t$ and~$u$ are \emph{blue}, then $v$'s
  short \emph{blue} arc must be at these edges, which enforces that
  the color at the edge between $v$ and $x$ will be \emph{red}.
  Finally, tile $x$ passes the color at the edge joint with tile $v$
  to the output.  With the negated inputs of the first and second
  part, this subpuzzle behaves like an AND gate, i.e., as a whole this
  subpuzzle simulates the computation of the boolean function AND:
  $\neg (\neg \alpha \vee \neg \beta) \equiv \neg \neg \alpha \wedge \neg
  \neg \beta \equiv \alpha \wedge \beta$.
\end{enumerate}
Again, since we care only about the color sequences of the tiles, we
obtain unique solutions for each pair of input colors.

\begin{figure}[h!]
  \centering
  \subfigure[BOOL Out: \emph{true}]{
    \label{fig:bool-2trp-t}
    \quad
    \quad
    \quad
    \includegraphics[height=1.8cm]{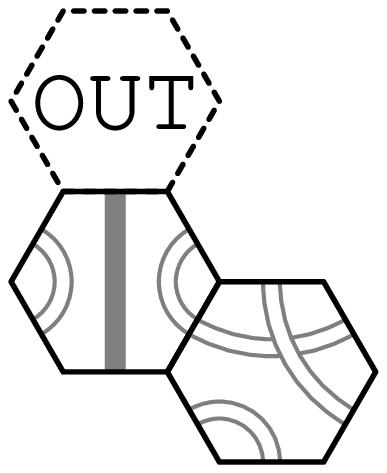}
    \quad
    \quad
    \quad
  }
  \subfigure[BOOL Out: \emph{false}]{
    \label{fig:bool-2trp-f}
    \quad
    \quad
    \quad
    \includegraphics[height=1.8cm]{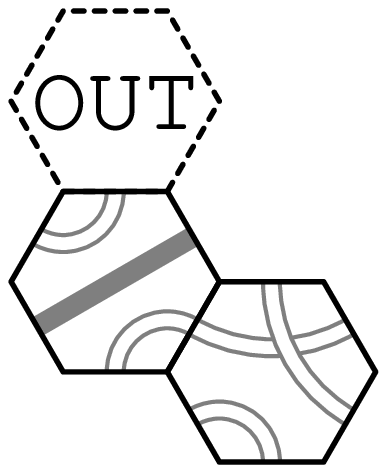}
    \quad
    \quad
    \quad
  }
  \subfigure[BOOL Scheme]{
    \label{fig:bool-2trp-s}
    \quad
    \quad
    \quad
    \includegraphics[height=1.8cm]{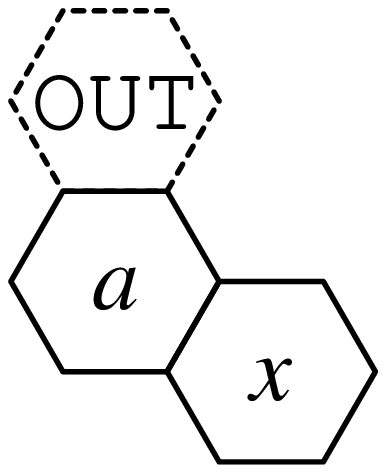}
    \quad
    \quad
    \quad
  }

  \subfigure[TEST-true]{
    \label{fig:test-2trp-t}
    \quad
    \quad
    \quad
    \includegraphics[height=1.4cm]{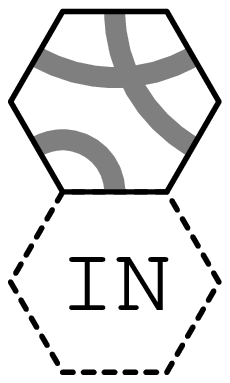}
    \quad
    \quad
    \quad
  }
  \subfigure[TEST-false]{
    \label{fig:test-2trp-f}
    \quad
    \quad
    \quad
    \includegraphics[height=1.4cm]{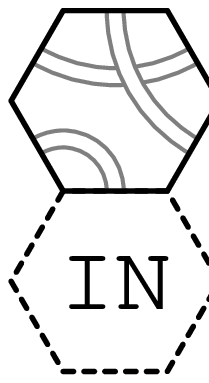}
    \quad
    \quad
    \quad
  }
  \subfigure[TEST Scheme]{
    \label{fig:test-2trp-s}
    \quad
    \quad
    \quad
    \includegraphics[height=1.4cm]{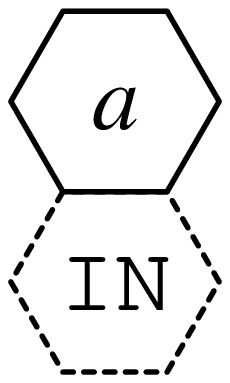}
    \quad
    \quad
    \quad
  }
  \caption{Two-color BOOL and TEST subpuzzles}
  \label{fig:bool-test-2trp}
\end{figure}

\paragraph{Input and output subpuzzles:}
The input variables of the circuit are simulated by the subpuzzle
BOOL.  Constructing a subpuzzle with the only possible outputs
\emph{blue} or \emph{red} is quite easy, since all tiles except $t_7$
and $t_8$ satisfy this condition.
Figures~\ref{fig:bool-test-2trp}(a)--(c) show our two-color BOOL subpuzzle.
Note that tile $x$ ensures the uniqueness of the solutions.

The last step is to check if the output of the whole circuit is
\emph{true}.  This is done by the subpuzzle TEST-true shown in
Figure~\ref{fig:bool-test-2trp}(d), which sits on top of the
subpuzzle simulating the circuit's output gate.  Since tile $t_7$
contains only \emph{blue} lines,
the solution is unique.

(Note: The subpuzzle TEST-false in Figure~\ref{fig:test-2trp-f}
will again be needed
in Section~\ref{sec:unique-infinite-variants},
see Figure~\ref{fig:circuit-inf}.
It has only \emph{red} lines, so the
input is always \emph{red} and the solution is unique.)~\end{proofs}

Theorem~\ref{thm:two-trp-is-np-complete} immediately gives the
following corollary.

\begin{corollary}
\label{cor:two-trp-is-np-complete}
$\ktrp{2}$ is $\np$-complete.
\end{corollary}

\subsection{Complexity of the Unique, Another-Solution, and Infinite
Variants of 3-TRP and 2-TRP}
\label{sec:unique-infinite-variants}

Parsimonious reductions preserve the number of solutions and, in
particular, the uniqueness of solutions.  Thus,
Theorems~\ref{thm:3trp-npc} and~\ref{thm:two-trp-is-np-complete} imply
Corollary~\ref{cor:3-uniquetrp-is-DP-randomized-complete} below that
also employs Valiant and Vazirani's results on the $\DP$-hardness of
$\usat$ under $\randomized$-reductions (which were defined in
Section~\ref{sec:definitions}).  The proof of
Corollary~\ref{cor:3-uniquetrp-is-DP-randomized-complete} follows the
lines of the proof of~\cite[Theorem~6]{bau-rot:c:tantrix}, which
states the analogous result for $\uktrp{4}$ in place of $\uktrp{3}$
and $\uktrp{2}$.

\begin{corollary}
\label{cor:3-uniquetrp-is-DP-randomized-complete}
\begin{enumerate}
\item $\usat$ parsimoniously reduces to the problems $\uktrp{3}$ and
$\uktrp{2}$.
\item Both $\uktrp{3}$ and $\uktrp{2}$ are $\DP$-complete under
$\randomized$-reductions.
\end{enumerate}
\end{corollary}

We now turn to the another-solution problems for $\ktrp{k}$.

\begin{corollary}
\label{cor:another-solution-trp-is-NP-complete}
\begin{enumerate}
\item For each $k \in \{2,3,4\}$, $\sat \aspred \ktrp{k}$.
\item For $k \in \{2,3,4\}$, $\asktrp{k}$ is $\np$-complete.
\end{enumerate}
\end{corollary}

\begin{proofs}
In Sections~\ref{sec:sharpsat-parsimoniously-reduces-to-sharp3trp} 
and~\ref{sec:two-trp-is-np-complete}, we showed a parsimonious reduction
from $\circuitandnotsat$ to $\ktrp{3}$ and $\ktrp{2}$.  To prove the
first part of
this corollary,
we have to show (see Section~\ref{sec:definitions-complexity})
that there is a polynomial-time computable
function bijectively mapping the solutions of any given
$\circuitandnotsat$ instance $C$ to the
solutions of the $\ktrp{k}$ instance corresponding to~$C$,
for each $k\in \{2,3,4\}$.  However, note that a satisfying
assignment to the variables of the circuit $C$ immediately gives the
solution for the BOOL subpuzzles according to our reduction for
$\ktrp{k}$, see the
proof of
Theorem~\ref{thm:two-trp-is-np-complete} (for $k = 2$), of
Theorem~\ref{thm:3trp-npc} (for $k = 3$), and 
of the result presented for $\ktrp{4}$ in~\cite{bau-rot:c:tantrix}
(for $k = 4$).

In each case, our circuit is constructed as a
sequence of steps, so the solutions for the BOOL subpuzzles determine
the color at the input for all subpuzzles at the next step, and so on.
Since all
subpuzzles have unique solutions we can construct a solution to our
puzzle in polynomial time from bottom to top using the parsimonious
reductions mentioned above.  Now, given the assignment of the variables,
we just have to place the tiles of the single subpuzzles according to
the determined solution and so specify their orientation.  
Conversely, if we have a solution of a
resulting $\ktrp{k}$ instance for $k\in \{2,3,4\}$, the output colors at
the BOOL subpuzzles gives the corresponding satisfying assignment to
the variables of the circuit.

To prove the second part of
Corollary~\ref{cor:another-solution-trp-is-NP-complete}, note that
AS-$\sat$ is $\np$-complete~\cite{yat-set:j:finding-another-solution},
and since the parsimonious reduction from $\sat$ to
$\circuitandnotsat$ provides a bijective transformation between these
problems' solution sets,
AS-$\circuitandnotsat$ is also $\np$-complete.  It follows
immediately, that the problems $\asktrp{3}$ and $\asktrp{2}$ are
$\np$-complete.  Furthermore, $\asktrp{4}$ inherits the
$\np$-completeness result from $\asktrp{3}$.~\end{proofs}


Holzer and Holzer~\cite{hol-hol:j:tantrix} proved that $\infktrp{4}$,
the infinite Tantrix\texttrademark\ rotation puzzle problem with four
colors, is undecidable, via a reduction from (the complement of)
the empty-word problem for Turing machines. The proof of
Theorem~\ref{thm:inf-3-trp-and-inf-2-trp-are-undecidable} below uses
essentially the same argument but is based on our modified three-color
and two-color constructions.

\begin{theorem}
\label{thm:inf-3-trp-and-inf-2-trp-are-undecidable}
Both $\infktrp{2}$ and $\infktrp{3}$ are undecidable.
\end{theorem}

\begin{figure}
  \centering
  \subfigure[Empty word not accepted]{
    \label{fig:circuit-inft}
    \quad
    \quad
    \quad
    \quad
    \quad
    \includegraphics[height=4cm]{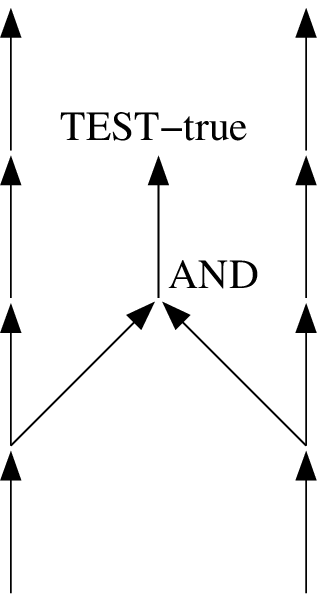}
    \quad
    \quad
    \quad
    \quad
    \quad
  }
  \subfigure[Empty word accepted]{
    \label{fig:circuit-inff}
    \quad
    \quad
    \quad
    \quad
    \quad
    \includegraphics[height=4cm]{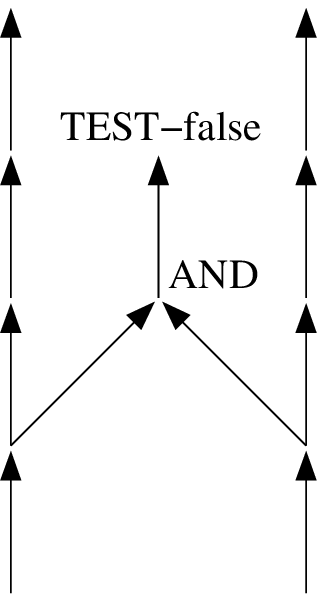}
    \quad
    \quad
    \quad
    \quad
    \quad
  }
\caption{Two choices for the $i$th layer 
of the infinite circuit for
 $\infktrp{2}$ and $\infktrp{3}$}
\label{fig:circuit-inf}
\end{figure}

\begin{proofs}
The empty-word problem for Turing machines asks whether the empty word,
$\lambda$, belongs to the language $L(M)$ accepted by a given Turing
machine~$M$.  By Rice's 
Theorem~\cite{ric:j:nontrivial-properties-of-re-sets}, both this problem
and its complement are undecidable.  To reduce the latter
problem to either $\infktrp{2}$ or $\infktrp{3}$,
we do the following.  Let $M_i$
denote the simulation of a Turing machine $M$ for exactly $i$
steps. Then, $M_i$ accepts its input
if and only if $M$ accepts the input within $i$ steps.

We employ another circuit construction that will be simulated
by a Tantrix\texttrademark\ rotation puzzle.
First, two wires are initialized with the
boolean value \emph{true}.  Then, in each step, we use either the
circuit shown in Figure~\ref{fig:circuit-inft} or the one shown in
Figure~\ref{fig:circuit-inff}. The former circuit is chosen in step~$i$
if $\lambda \notin L(M_i)$, and the latter one is chosen in step~$i$ 
if $\lambda \in L(M_i)$.
To transform this circuit into an $\infktrp{k}$ instance,
where $k$ is either two or three,
we use the TEST-true subpuzzle from either Figure~\ref{fig:test-3trp-t}
or Figure~\ref{fig:test-2trp-t}, rotated by $180$ degrees and with
the ``IN'' tile becoming an ``OUT'' tile, in order to
initialize both wires with the input \emph{true}.  Then we substitute the
single layers of the circuit by the subpuzzles described above, step by step,
always choosing either the circuit from Figure~\ref{fig:circuit-inft}
(where TEST-true is the subpuzzle from Figure~\ref{fig:test-3trp-t}
if $k = 3$, or from Figure~\ref{fig:test-2trp-t} if $k = 2$),
or the circuit from Figure~\ref{fig:circuit-inff}
(where TEST-false is the subpuzzle from Figure~\ref{fig:test-3trp-f}
if $k = 3$, or from Figure~\ref{fig:test-2trp-f} if $k = 2$).

Since
both wires are initialized with the value \emph{true}, it is obvious
that the constructed subpuzzle has a
solution if and only if
$\lambda \notin L(M)$.  Note that the layout of the circuit is computable,
and our reduction will output the encoding of a Turing machine computing
first this circuit layout and then the transformation to the 
Tantrix\texttrademark\ rotation puzzle as described above.
By this reduction, both $\infktrp{2}$
and $\infktrp{3}$ are shown to be undecidable.
\end{proofs}

\section{Conclusions}
\label{sec:conclusions}

This paper studied the three-color and two-color
Tantrix\texttrademark\ rotation puzzle problems, $\ktrp{3}$ and
$\ktrp{2}$, and their unique, another-solution, and infinite variants.
Our main contribution is that both $\ktrp{3}$ and $\ktrp{2}$ are
$\np$-complete via a parsimonious reduction from $\sat$, which in
particular solves a question raised by Holzer and
Holzer~\cite{hol-hol:j:tantrix}.  Since restricting the number of
colors to three and two, respectively, drastically reduces the number
of Tantrix\texttrademark\ tiles available, our constructions as well
as our correctness arguments substantially differ from those
in~\cite{hol-hol:j:tantrix,bau-rot:c:tantrix}.
Table~\ref{tab:results} in Section~\ref{sec:introduction} shows that
our results give a complete picture of the complexity of $\ktrp{k}$,
$1 \leq k \leq 4$.  An interesting question still remaining open is
whether the analogs of $\ktrp{k}$ \emph{without holes} still are
$\np$-complete.

\bigskip

\noindent
{\bf Acknowledgments:} We are grateful to Markus Holzer and Piotr
Faliszewski for inspiring discussions on Tantrix\texttrademark\
rotation puzzles, and we thank Thomas Baumeister for his help with
producing reasonably small figures.  We thank the anonymous
LATA~2008 referees for helpful comments, and in particular the
referee who let us know that he or she has also written a program
for verifying the correctness of our constructions.

\bibliographystyle{alpha}

\bibliography{/home/rothe/BIGBIB/joergbib}

\newcommand{\etalchar}[1]{$^{#1}$}
\begin{thebibliography}{CGH{\etalchar{+}}89}

\bibitem[BR]{bau-rot:c-toappear:tantrix-three-two-color}
D.~Baumeister and J.~Rothe.
\newblock The three-color and two-color {T}antrix\texttrademark\ rotation
  puzzle problems are {NP}-complete via parsimonious reductions.
\newblock In {\em Proceedings of the 2nd International Conference on Language
  and Automata Theory and Applications}. Springer-Verlag {\it Lecture Notes in
  Computer Science}.
\newblock To appear.

\bibitem[BR07]{bau-rot:c:tantrix}
D.~Baumeister and J.~Rothe.
\newblock Satisfiability parsimoniously reduces to the {Tantrix}\texttrademark\
  rotation puzzle problem.
\newblock In {\em Proceedings of the 5th Conference on Machines, Computations
  and Universality}, pages 134--145. Springer-Verlag {\it Lecture Notes in
  Computer Science \#4664}, September 2007.

\bibitem[CGH{\etalchar{+}}88]{cai-gun-har-hem-sew-wag-wec:j:bh1}
J.~Cai, T.~Gundermann, J.~Hartmanis, L.~Hemachandra, V.~Sewelson, K.~Wagner,
  and G.~Wechsung.
\newblock The boolean hier\-archy {I}: {S}truc\-tural proper\-ties.
\newblock {\em SIAM Jour\-nal on Com\-pu\-ting}, 17(6):1232--1252, 1988.

\bibitem[CGH{\etalchar{+}}89]{cai-gun-har-hem-sew-wag-wec:j:bh2}
J.~Cai, T.~Gundermann, J.~Hartmanis, L.~Hemachandra, V.~Sewelson, K.~Wagner,
  and G.~Wechsung.
\newblock The boolean hierarchy {II}: Applications.
\newblock {\em SIAM Journal on Computing}, 18(1):95--111, 1989.

\bibitem[CKR95]{cha-kad-roh:uniquesat-randomized-reductions}
R.~Chang, J.~Kadin, and P.~Rohatgi.
\newblock On unique satisfiability and the threshold behavior of randomized
  reductions.
\newblock {\em Journal of Computer and System Sciences}, 50(3):359--373, 1995.

\bibitem[Coo71]{coo:c:theorem-proving}
S.~Cook.
\newblock The complexity of theorem-proving procedures.
\newblock In {\em Proceedings of the 3rd ACM Symposium on Theory of Computing},
  pages 151--158. ACM Press, 1971.

\bibitem[Dow05]{dow:j:tantrix}
K.~Downing.
\newblock Tantrix: {A} minute to learn, 100 (genetic algorithm) generations to
  master.
\newblock {\em Genetic Programming and Evolvable Machines}, 6(4):381--406,
  2005.

\bibitem[Gol77]{gol:j:monotone-planar-circuits}
L.~Goldschlager.
\newblock The monotone and planar circuit value problems are log space complete
  for {P}.
\newblock {\em SIGACT News}, 9(2):25--29, 1977.

\bibitem[Gr{\"{a}}90]{gra:j:domino}
E.~Gr{\"{a}}del.
\newblock Domino games and complexity.
\newblock {\em SIAM Journal on Computing}, 19(5):787--804, 1990.

\bibitem[HH04]{hol-hol:j:tantrix}
M.~Holzer and W.~Holzer.
\newblock Tantrix\texttrademark\ rotation puzzles are intractable.
\newblock {\em Discrete Applied Mathematics}, 144(3):345--358, 2004.

\bibitem[McC81]{mcc:j:planar-crossovers}
W.~McColl.
\newblock Planar crossovers.
\newblock {\em IEEE Transactions on Computers}, C-30(3):223--225, 1981.

\bibitem[Pap94]{pap:b-1994:complexity}
C.~Papadimitriou.
\newblock {\em Computational Complexity}.
\newblock Addison-Wesley, 1994.

\bibitem[PY84]{pap-yan:j:dp}
C.~Papadimitriou and M.~Yannakakis.
\newblock The complexity of facets (and some facets of complexity).
\newblock {\em Journal of Computer and System Sciences}, 28(2):244--259, 1984.

\bibitem[Ric53]{ric:j:nontrivial-properties-of-re-sets}
H.~Rice.
\newblock Classes of recursively enumerable sets and their decision problems.
\newblock {\em Transactions of the American Mathematical Society}, 74:358--366,
  1953.

\bibitem[Rot05]{rot:b:cryptocomplexity}
J.~Rothe.
\newblock {\em Complexity Theory and Cryptology. An Introduction to
  Cryptocomplexity}.
\newblock EATCS Texts in Theoretical Computer Science. Springer-Verlag, Berlin,
  Heidelberg, New York, 2005.

\bibitem[UN96]{ued-nag:t:nonogramm}
N.~Ueda and T.~Nagao.
\newblock {NP}-completeness results for {NONOGRAM} via parsimonious reductions.
\newblock Technical Report TR96-0008, Tokyo Institute of Technology, Department
  of Information Science, Tokyo, Japan, May 1996.

\bibitem[Val79]{val:j:permanent}
L.~Valiant.
\newblock The complexity of computing the permanent.
\newblock {\em Theoretical Computer Science}, 8(2):189--201, 1979.

\bibitem[VV86]{val-vaz:j:np-unique}
L.~Valiant and V.~Vazirani.
\newblock {N}{P} is as easy as detecting unique solutions.
\newblock {\em Theoretical Computer Science}, 47:85--93, 1986.

\bibitem[YS02]{yat-set:j:finding-another-solution}
T.~Yato and T.~Seta.
\newblock Complexity and completeness of finding another solution and its
  application to puzzles.
\newblock {\em Joho Shori Gakkai Kenkyu Hokoku}, 2002(103(AL-87)):9--16, 2002.

\end{thebibliography}

\clearpage

\end{document}